\documentclass[twocolumn,showpacs,preprintnumbers,amsmath,amssymb,nofootinbib]{revtex4-1}
\usepackage{epsfig}

\usepackage{comment}
\usepackage[table]{xcolor}
\usepackage{graphicx}
\usepackage{dcolumn}
\usepackage{bm}

\usepackage{hyperref}
\hypersetup{colorlinks=true,linkcolor=magenta,anchorcolor=green,citecolor=cyan,filecolor=black,menucolor=black,urlcolor=brown}
\usepackage{slashed}

\newcommand{\be}{\begin{equation}}
\newcommand{\ee}{\end{equation}}
\newcommand{\ba}{\begin{eqnarray}}
\newcommand{\ea}{\end{eqnarray}}

\newcommand{\bea}{\begin{eqnarray}}
\newcommand{\eea}{\end{eqnarray}}

\begin{document}

\title{Solitons and halos for self-interacting scalar dark matter}

\author{Raquel Galazo Garc\'ia}
\affiliation{Universit\'{e} Paris-Saclay, CNRS, CEA, Institut de physique th\'{e}orique, 91191, Gif-sur-Yvette, France}
\author{Philippe Brax}
\affiliation{Universit\'{e} Paris-Saclay, CNRS, CEA, Institut de physique th\'{e}orique, 91191, Gif-sur-Yvette, France}
\author{Patrick Valageas}
\affiliation{Universit\'{e} Paris-Saclay, CNRS, CEA, Institut de physique th\'{e}orique, 91191, Gif-sur-Yvette, France}

\begin{abstract}

We study the formation and evolution of solitons supported by repulsive self-interactions inside extended halos,
for scalar-field dark matter scenarios. We focus on the semiclassical regime where the quantum pressure
is typically much smaller than the self-interactions. We present numerical simulations, with initial conditions
where the halo is described by the WKB approximation for its eigenfunction coefficients.
We find that when the size of the system is of the order of the Jeans length associated with the
self-interactions, a central soliton quickly forms and makes about $50\%$ of the total mass.
However, if the halo is ten times greater than this self-interaction scale, a soliton only quickly forms 
in cuspy halos where the central density is large enough to trigger the self-interactions. If the halo has a flat core, it takes
a longer time for a soliton to appear, after small random fluctuations on the de Broglie wavelength size
build up to reach a large enough density.
In some cases, we observe the co-existence of several narrow density spikes inside the larger
self-interaction-supported soliton.
All solitons appear robust and slowly grow, unless they already make up $40\%$ of the total mass. 
We develop a kinetic theory, valid for an inhomogeneous background, to estimate the soliton growth
rate for low masses. It explains the fast falloff of the growth rate as resonances between the ground
state and halo excited states disappear. 
Our results suggest that cosmological halos would show a large scatter for their
soliton mass, depending on their assembly history.

\end{abstract}

\date{\today}

\maketitle

\section{Introduction}
\label{sec:introduction}

Scalar dark matter \citep{Matos2000} is a fruitful alternative to the WIMP paradigm \citep{Jungman1995}. Originating
from the first axion models \citep{Peccei1977,Wilczek1978,Weinberg1978}, it has now been extended to a wide range of scenarios where the mass
of the scalar can be between $10^{-22}$ eV and 1 eV \citep{  PRESKILL1983127,Hu2000,Ferreira2020,Hui_2021}. 
In particular these models allow for a modification of the
distribution of dark matter inside galaxies where the core-cusp problem \citep{Navarro1996,Moore1999,Block2010} of $\Lambda$-CDM can be addressed without any baryonic
feedback \citep{Burkert_1995,Flores1994,Moore1994}. Scalar dark matter models can lead to different types of stable configurations of dark matter, often called solitons \citep{Goodman:2000tg,Chavanis:2011zi,Chavanis:2011zm,Luu:2018afg,Mocz2017}, which
would render the distribution of dark matter smooth on short scales and evade the cuspy profiles found in the $\Lambda$-CDM model. These solitons result
from the balance between different types of pressures and interactions acting on dark matter. Gravity is always active but two other sources of pressure play also a role. The first one is the quantum pressure whose origin springs from the Schr\"odinger description of the non-relativistic scalar (i.e., from the Heisenberg inequalities that follow from the wave-like properties of the system). This quantum pressure is always repulsive and can form
solitons by compensating gravity \citep{Schive2014a}. These solitons have a radius that decreases with their mass as $R \propto 1/M$. They are a manifestation of the wave nature of models such as fuzzy dark matter (FDM) \citep{Hu2000,Hui2017}, where the de Broglie wavelength reaches galactic scales when the mass of the scalar is $m\simeq 10^{-22}$ eV. Fuzzy dark matter is described by a non-interacting massive scalar field. When self-interactions are present, such as a quartic term in the Lagrangian of the scalar model \citep{Chavanis2016a,Chavanis2018,Shapiro2013}, this can generate a repulsive self-interaction that can balance gravity on large scales and again create smooth solitons of finite size  \citep{Chavanis:2011zi,Chavanis:2011zm,Chavanis2016,Guth2015,Brax:2019fzb}. Their size is directly related to the mass and the self-interaction of the scalar field and no longer depends on the soliton mass. 
The cosmological evolution of such models has been studied for instance in
\cite{Li:2013nal,Suarez:2017mav,Brax:2019fzb}.

In this paper we will be interested in the formation of self-interacting solitons inside a larger dark matter halo. 
The 1D spherical collapse was studied with a fluid approximation in \cite{Dawoodbhoy:2021beb}.
In contrast, in this paper we use the 3D Schr\"odinger-Poisson equations with stochastic
initial conditions around a mean halo target profile that can have either a flat or cuspy core.
We also consider the cases where there is initially a small soliton or not superimposed
on this stochastic halo.
Thus, we assume that a large halo has formed by Jeans' instability \citep{Guzman2002,Chavanis2018,Harko2019,Brax:2019fzb} and study whether solitons can emerge dynamically from the time evolution of the dark matter inside the halo. In particular, we will show that due to the initial density fluctuations around the initial halo profile, self-interacting solitons always emerge and swallow a significant portion of the halo mass. This takes place whether a small soliton already exists or even when no soliton is present initially. We also compare what happens when the halo profile is flat or cuspy. In the case of cuspy profiles, we find that the formation of a soliton from no initial one is very fast and happens in a few dynamical times. In all cases, a soliton forms, grows and reaches a significant size of the initial halo. This behaviour differs from what happens for fuzzy dark matter where solitons are only stable if initially present and of a large enough mass \citep{Schive2014a,Schive2014b,Veltmaat2018,Chan:2022bkz}. In the self-interacting case, the solitons are spontaneously created.

We confirm these results by deriving a kinetic equation from the nonlinear Schr\"odinger equation,
which also applies to a non-homogeneous background.
With the addition of a simple energy-cutoff ansatz for the occupation numbers of the halo excited
states, we obtain that the growth rate of solitons is positive but shows a fast falloff with the 
increase of the soliton mass.

Our results would have relevant consequences for astrophysical situations as they suggest that if dark matter happened to be a scalar with self-interactions, then dark matter halos would be composed of a mix between a diffuse halo and a smaller soliton whose size would depend on the formation history of the halo. This could have phenomenological consequences for the dynamics of stars in dark matter halos that we leave for future studies.

The paper is arranged as follows. In Sec.~\ref{sec:Equations-of-motion}
we describe the model and the region in the parameter space that we focus on, where
self-interactions are important.
We present the choice of initial conditions and the numerical procedure in 
Sec.~\ref{sec:initial-conditions}.
In the next Sec.~\ref{sec:flat-core}, we describe the emergence of solitons in flat halos 
and then in Sec.~\ref{sec:cuspy} in cuspy  halos. 
Finally, we develop a kinetic theory in section \ref{sec:kine} to analyse the growth rate of solitons. We then conclude in Sec.~\ref{sec:Conclusion}.

\section{Equations of motion}
\label{sec:Equations-of-motion}
 
\subsection{The scalar field action}

We consider scalar dark matter models described by the action
\be
S= \int d^4x \sqrt{-g} \left[\frac{R}{16\pi {\cal G}_N} -\frac{1}{2}(\partial \phi)^2 -\frac{m^2\phi^2}{2} -V_I(\phi) \right] ,
\ee
where the quartic self-interaction potential $V_I$ is small as compared with the quadratic term
and given by
\be
V_{I}(\phi)= \frac{\lambda_4}{4} \phi^4 ,
\label{eq:V_I-def}
\ee
where $\lambda_4$ dictates the strength of the self-interactions. 
At late times, when the Hubble expansion rate is smaller than the scalar mass, $H \ll m$, 
the scalar field oscillates inside its potential well with a period $2\pi/m$ and can be described by the 
nonrelativistic ansatz
\be
\phi= \frac{1}{\sqrt{2m}} ( \psi e^{-imt} +  \psi^* e^{imt}) ,
\ee
where the complex wavefunction $\psi$ varies on time scales and lengths much larger than $1/m$.
In this nonrelativistic approximation \citep{Brax:2019fzb}, the wavefunction satisfies the nonlinear 
Schr\"odinger equation
\be
i \frac{\partial\psi}{\partial t} = - \frac{\nabla^2\psi}{2m} + m (\Phi_N+\Phi_I) \psi .
\label{eq:Schrodinger-m}
\ee
Here and in the following, we consider time scales that are much shorter than the Hubble time
and we neglect the expansion of the Universe.
The Newtonian gravitational potential $\Phi_N$ is given by the Poisson equation
\be
\nabla^2 \Phi_N = 4\pi {\cal G}_N m \vert \psi\vert^2 ,
\ee
and the self-interacting potential is
\be
\Phi_I= \frac{3\lambda_4}{4m^3} \vert \psi \vert^2.
\ee
This a coupled system where the non-linear Schr\"odinger equation couples to the Poisson equation. This can also be reduced to a single integro-differential equation, which will be analysed in section \ref{sec:kine},
\be
i \frac{\partial\psi}{\partial t} = - \frac{\nabla^2\psi}{2m} + m^2 \psi \left( 4\pi {\cal G}_N \nabla^{-2} 
+ \frac{1}{\rho_a} \right) \vert \psi\vert ^2 ,
\label{SNL}
\ee
with
\be
\rho_a = \frac{4m^4}{3\lambda_4} .
\ee

Simple configurations can be understood from the hydrodynamical picture that follows from
the Madel\"ung transform \citep{Madelung1926}
\be 
\psi= \sqrt{\frac{\rho}{m}} e^{iS} , \;\;\; \mbox{whence} \;\; \rho = m |\psi|^2 , 
\ee
where the dark matter velocity is identified as $\vec v = \nabla S/m$. The real and imaginary parts of
the Schr\"odinger equation give the continuity and Euler equations
\bea
&& \partial_t \rho + \nabla \cdot (\rho \vec v ) = 0 , \nonumber \\
&& \partial_t \vec v + (\vec v \cdot \nabla) \vec v= - \nabla (\Phi_N + \Phi_I + \Phi_Q) ,
\label{eq:Euler-1}
\eea
with 
\be 
\Phi_I= \frac{\rho}{\rho_a} , \;\;\; \Phi_Q =  - \frac {\nabla^2 \sqrt \rho}{2m^2 \sqrt \rho} ,
\label{eq:PhiI-PhiQ}
\ee
where $\Phi_Q$ is the so-called quantum pressure.

\subsection{Hydrostatic equilibrium and Thomas-Fermi limit}
\label{sec:hydrostatic}

As seen from Eq.(\ref{eq:Euler-1}), such scalar field models admit hydrostatic equilibria 
given by $\vec v=0$ and $\Phi_N+\Phi_I+\Phi_Q = {\rm constant}$.
The spherically symmetric ground state is also called a soliton or boson star.
In the Thomas-Fermi regime that we will consider in this paper, this soliton
is governed by the balance between gravity and the repulsive force associated with the 
self-interactions (for $\lambda_4 > 0$). This means that $\Phi_Q \ll \Phi_I$ over most of the
extent of the soliton and the Laplacian term $- \frac{\nabla^2\psi}{2m}$ can be neglected
in Eq.(\ref{eq:Schrodinger-m}). Then, the wavefunction reads 
$\psi(r,t) = e^{-iE t} \hat\psi(r)$ with
\be 
\Phi_N +\Phi_I=\frac{E}{m} .
\label{eq:bal-TF}
\ee
The soliton density profile is given by \cite{Chavanis:2011zi,Harko:2011jy,Brax:2019fzb}

\be
\rho_{\rm sol}(r) = \rho_{0{\rm sol}} \frac{\sin(\pi r/R_{\rm sol})}{\pi r/R_{\rm sol}} , 
\label{eq:rho-soliton-1}
\ee
with the radius
\be
R_{\rm sol} = \pi r_a , \;\; \mbox{with} \;\; r_a^2 = \frac{3 \lambda_4}{16\pi {\cal G}_N m^4}
= \frac{1}{4\pi {\cal G}_N \rho_a} .
\label{eq:Rsol-1}
\ee
In fact, outside of the radius $r_a$ where Eq.(\ref{eq:rho-soliton-1}) would give
a zero density we can no longer neglect $\Phi_Q$ and the exact solution develops
an exponential tail at large radii.
Nevertheless, from Eq.(\ref{eq:PhiI-PhiQ}) we can see that the approximation
(\ref{eq:rho-soliton-1}) is valid up to $r \lesssim R_{\rm sol}$ for
\be
\Phi_Q \ll \Phi_I : \;\;\; \frac{\rho_{0\rm sol}}{\rho_a} \gg \frac{1}{r_a^2 m^2} .
\label{eq:TF-1}
\ee

\subsection{Outer halo and semi-classical limit}

In this paper, we will study the emergence and the evolution of these solitons within 
a larger halo of radius $R_{\rm halo} > R_{\rm sol}$. 
As seen above, the self-interactions can only support an hydrostatic equilibrium 
within the radius $R_{\rm sol}$ of Eq.(\ref{eq:Rsol-1}), independently of the soliton mass.
Therefore, while inside $R_{\rm sol}$ the self-interactions can balance gravity and build a flat core
when the condition (\ref{eq:TF-1}) is satisfied, outside of $R_{\rm sol}$ the 
self-interactions are negligible. There, as for FDM and CDM models, gravity is balanced
by the velocity dispersion or the angular momentum of the system. Thus, in cosmological numerical simulations of FDM halos, one finds a flat core governed by the quantum pressure
inside an NFW halo 
 that is similar to the halos found in CDM simulations \citep{Navarro:1995iw}.
The halo is made of granules that are stochastic fluctuations with a size of the order of the
de Broglie wavelength.
A similar configuration would then apply to our case, except that the flat core is now 
supported by the self-interactions instead of the quantum pressure.

We will consider the semi-classical limit (i.e., large scalar mass $m$), where
the de Broglie wavelength is much smaller than both the core and halo radii.
Then, the granules also correspond to temporary wave packets that play the role of particules
\citep{Hui2017} 
with a velocity dispersion or an angular momentum that balances gravity and supports a virialized halo.
This means that $\Phi_Q \ll \Phi_N$. For a system of size $L_\star$ and density $\rho_\star$,
this gives
\be
\Phi_Q \ll \Phi_N : \;\;\; \epsilon \ll 1 \;\; \mbox{with} \;\; 
\epsilon = \frac{1}{\sqrt{{\cal G}_N \rho_\star} m L_\star^2} .
\label{eq:epsilon-1}
\ee
For a virialized system governed by gravity, the gravitational dynamical time $t_\star$ and the
virial velocity are
\be
t_\star = \frac{1}{\sqrt{{\cal G}_N \rho_\star}} \;\; \mbox{and} \;\;
v_\star = \frac{L_\star}{t_\star} . 
\label{eq:t*-1}
\ee
Therefore, the de Broglie wavelength $\lambda_{\rm dB}$ reads
\be
\lambda_{\rm dB} = \frac{2\pi}{m v_\star} = \frac{2\pi t_\star}{m L_\star} = \frac{2\pi}{\sqrt{{\cal G}_N \rho_\star} m L_\star} = \epsilon 2\pi L_\star .
\label{eq:lambda-dB-epsilon}
\ee
Thus, the limit $\epsilon \to 0$ corresponds to the semiclassical limit, where the de Broglie wavelength
is much smaller than the size of the system
In this paper, we focus on the semiclassical regime $\epsilon = 0.01 \ll 1$.
Then, the halo is composed of incoherent stochastic fluctuations of size $\lambda_{\rm dB}$,
with a velocity dispersion set by the virial velocity, whereas a coherent static soliton can appear at
the center.

\subsection{Parameter space}
\label{sec:parameter-space}

\begin{figure}[ht]
\centering
\includegraphics[height=6.5cm,width=0.5\textwidth]{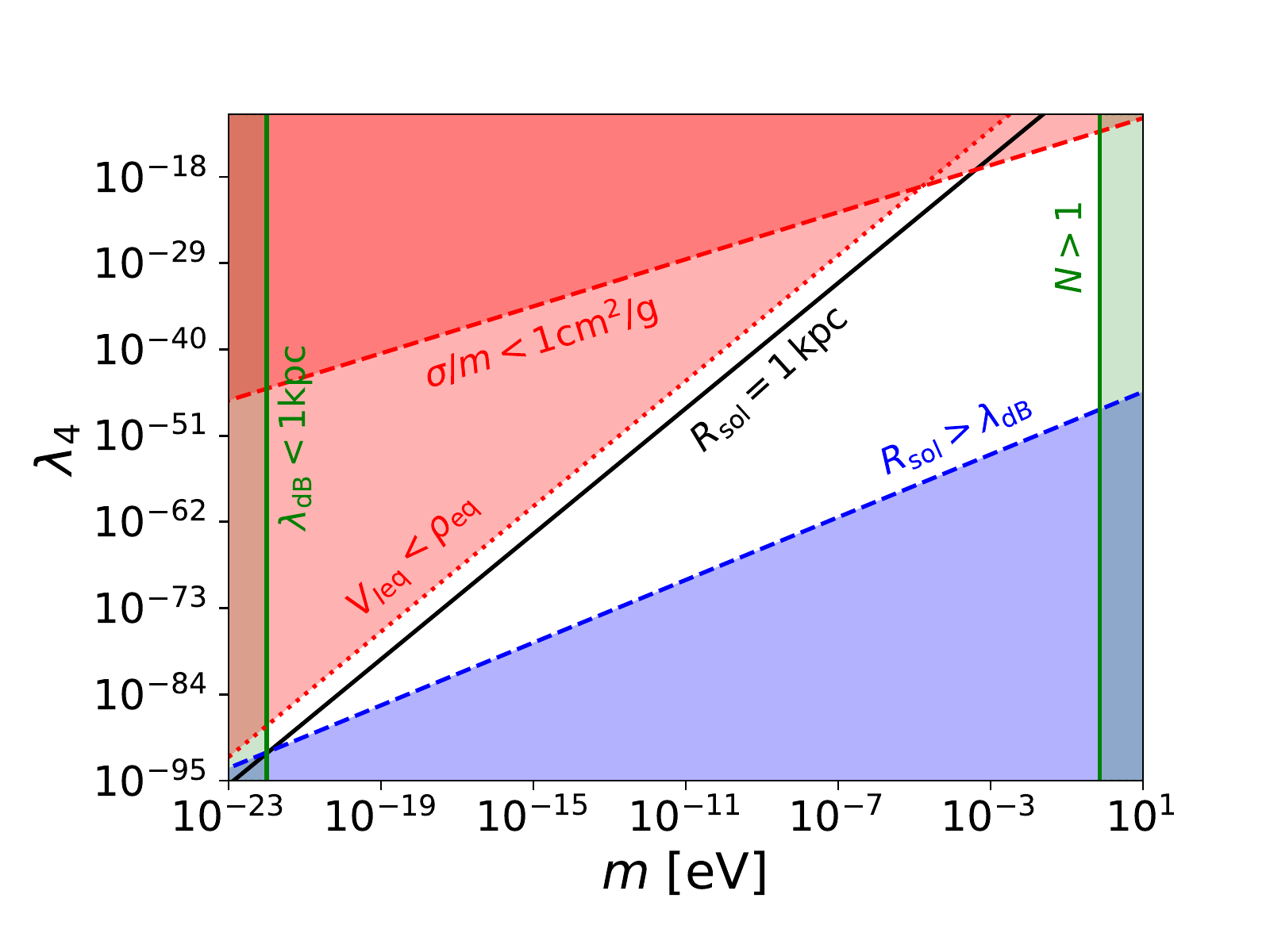}
\includegraphics[height=6.5cm,width=0.5\textwidth]{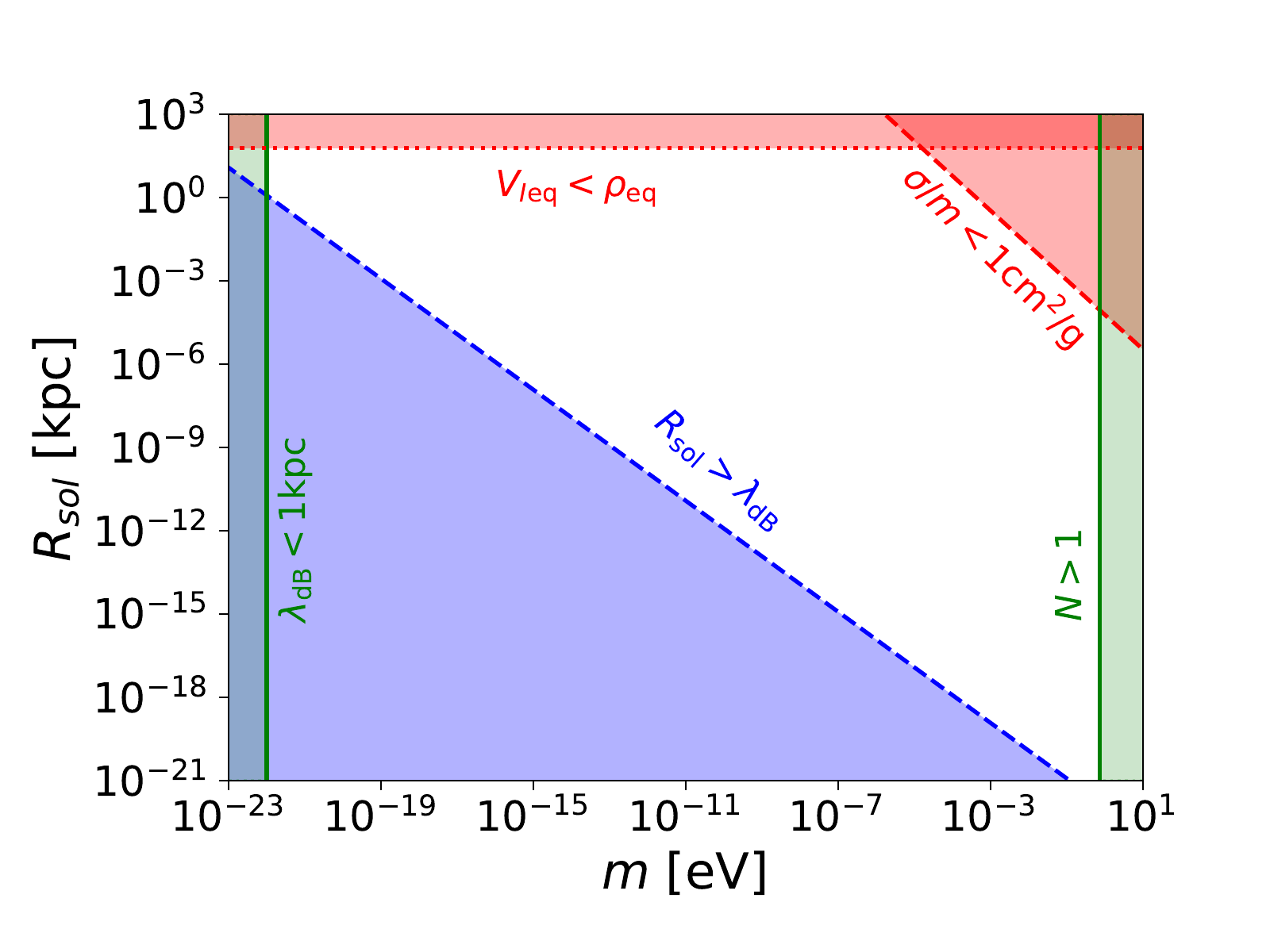}
\caption{
Domains in the planes $(m,\lambda_4)$ (upper panel) and $(m,R_{\rm sol})$ (lower panel)
where our computations are relevant. See the main text for the explanation of the shaded exclusion
regions. 
}
\label{fig:constraints}
\end{figure}

To set our study within the context of dark matter models and the formation of large-scale
structures, we show in Fig.~\ref{fig:constraints} the range of parameters allowed by
astrophysical and cosmological constraints and where our computations are relevant.
The exclusion regions, shown by the shaded domains, are displayed in the planes
$(m,\lambda_4)$ (upper panel) and $(m,R_{\rm sol})$ (lower panel).
At fixed $m$, the soliton radius (\ref{eq:Rsol-1}) can be used in place of $\lambda_4$ as the second
dark matter parameter and it can be more convenient for astrophysical and cosmological purposes.
In the upper panel, we also show the black solid line $R_{\rm sol}=1$ kpc for reference.

A first observational constraint is associated with cluster mergers, such as the bullet cluster, that provide
an upper bound on the dark matter cross-section, 
$\sigma/m_{\rm DM}  \lesssim 1$ cm$^2$/g \cite{Randall_2008}.
This gives the upper bound \cite{Brax:2019fzb}
\be
\sigma/m \lesssim 1 \; {\rm cm}^2/{\rm g} : \;\;\;   
\lambda_4 \lesssim 10^{-12} \left(\frac{m}{1 \,{\rm eV}}\right)^{\frac{3}{2}} ,
\label{eq:cross_section_constraint}
\ee
shown by the dashed red line.
A second constraint is the requirement that the quartic term $V_I$ of Eq.(\ref{eq:V_I-def}) be much smaller
than the quadratic term in the scalar field potential since at least the time of matter-radiation equality
$t_{\rm eq}$. This ensures that the scalar field behaves as dark matter with a mean cosmological density
that decreases as $\bar\rho \propto a^{-3}$, where $a(t)$ is the scale factor.
This gives the condition $\lambda_4 \bar\rho_{\rm eq}/m^4 \ll1$, which reads 
\be
V_{I {\rm eq}} \ll \bar\rho_{\rm eq} : \;\;\; \lambda_4 \ll \left( \frac{m}{1 \, {\rm eV}} \right)^4 ,
\label{eq:V-I-small}
\ee
and is shown by the dotted red line.
For the same reason, we also require that the scalar field has started oscillating in its almost quadratic
potential before the matter-radiation equality. This gives the constraint
\be
m \gg H_{\rm eq} \sim 10^{-28} {\rm eV} .
\ee
This lower bound does not appear in Fig.~\ref{fig:constraints} because it is less stringent than the
lower bound (\ref{eq:m-FDM}) considered below.

The de Broglie wavelength (\ref{eq:lambda-dB-epsilon}) sets the scale where the quantum pressure
alone and wave-like effects lead to strong departure from CDM. We require $\lambda_{\rm dB} < L$
with $L \sim 1$ kpc, which gives
\be
\lambda_{\rm dB} < L : \;\;\; m > \left( \frac{L}{1 \, {\rm kpc}} \right)^{-1} 
\left( \frac{v}{100 \, {\rm km/s}} \right)^{-1} 10^{-22} \, {\rm eV} .
\label{eq:m-FDM}
\ee
It is shown by the left vertical green line for $L=1$ kpc and $v=100$ km/s.
The lower limit $m \sim 10^{-22} \, {\rm eV}$ corresponds to Fuzzy Dark Matter scenarios,
where galactic cores could be partly explained by the solitons formed at the center of galaxies,
although this is disfavoured by observations of the Lyman-$\alpha$ forest power spectrum.
Thus, in this paper we consider models with masses above $10^{-22}$ eV.
More generally, we focus on scenarios where the soliton is governed by the self-interactions rather
than by the quantum pressure.
This corresponds to the condition $R_{\rm sol} > \lambda_{\rm dB}$, which reads
\be
R_{\rm sol} > \lambda_{\rm dB} : \;\;\; \lambda_4 > 4 \times 10^{-48} \left( \frac{v}{100 \, {\rm km/s}} \right)^{-2}
\left( \frac{m}{1 \, {\rm eV}} \right)^2 ,
\ee
and is shown by the blue dashed line for $v=100$ km/s.

Finally, to treat the system as a classical field, the occupation number 
$N \sim (\rho/m) \lambda_{\rm dB}^3 \sim \rho/(m^4 v^3)$ must be much greater than unity.
At the soliton scale, with $v^2 \sim \Phi_N \sim \Phi_I$ this gives $\lambda_4 v \ll 1$,
which reads
\be
\mbox{soliton}, \;\; N \gg 1 : \;\;\;  \lambda_4 \ll 3 \times 10^3  \left( \frac{v}{100 \, {\rm km/s}} \right)^{-1} .
\ee
This upper bound does not appear in Fig.~\ref{fig:constraints} because it is less stringent than the
upper bounds (\ref{eq:cross_section_constraint}) and (\ref{eq:V-I-small}) considered above.
We also require that the field be classical on cosmological scales, that is,
$\bar N \sim \bar\rho_0/(m^4 v^3) \gg 1$. For the cosmological density today this gives the condition
\be
\mbox{cosmology}, \;\; N \gg 1 : \;\;\; m \ll \left( \frac{v}{100 \, {\rm km/s}} \right)^{-3/4} 0.7 \, {\rm eV} ,
\ee
shown by the right vertical green line.

Thus, the white area in Fig.~\ref{fig:constraints} shows the domain of parameters
that we consider, where the scalar field is a viable dark matter candidate and self-interactions
dominate over the quantum pressure.
In our simulations we take $\epsilon=0.01$. This typically corresponds to models that are parallel
to the line labelled $R_{\rm sol} > \lambda_{\rm dB}$ in Fig.~\ref{fig:constraints}, with a coupling
$\lambda_4$ or a radius $R_{\rm sol}$ that is greater by a factor $100$. However, we expect similar
behaviours to hold for smaller $\epsilon$, that is, further into the allowed domain.
The soliton radius can range from a few kpc down to the meter, depending on $m$ and $\lambda_4$.
The larger value could allow these scenarios to play the same role as Fuzzy Dark Matter models, where
the soliton may cure some of the small-scale tensions of CDM.
Smaller solitons would not play such a role and would behave like CDM on galactic scales.
However, they could be distinguished from CDM by astrophysical probes, such as their impact
on the orbital dynamics and the emission of gravitational waves of Black Hole binary systems
\citep{Boudon:2023aa}.

\section{Initial conditions and numerical procedure}
\label{sec:initial-conditions}

\subsection{Dimensionless variables}

Going back to the Schr\"odinger equation, it is convenient to work with dimensionless quantities 
denoted with a tilde,
\be
\psi = \psi_\star \tilde\psi, \;\;\; t = t_\star \tilde t, \;\;\;
\vec x = L_\star \tilde{\vec x} , \;\;\;
\Phi = \frac{L_\star^2}{t_*^2} \tilde\Phi,
\ee
where $t_\star$ and $L_\star$ are the characteristic time and length scales of
the system (in our case the halo that may contain a smaller soliton at the center). 
This gives the dimensionless Schr\"odinger equation
\be
i \epsilon \frac{\partial\tilde\psi}{\partial\tilde t} =
- \frac{\epsilon^2}{2} \tilde\nabla^2\tilde\psi
+ (\tilde\Phi_N+\tilde\Phi_I) \tilde\psi ,
\label{eq:Schrod-eps}
\ee
with
\be
\epsilon = \frac{t_\star}{m L_\star^2} .
\label{eq:epsilon-def}
\ee
We have already introduced in Eq.(\ref{eq:epsilon-1}) the parameter $\epsilon$, 
which plays the role of $\hbar$ and  measures the relevance
of wave effects, such as interferences or the quantum pressure.
The Poisson equation takes the dimensionless form
\be
\tilde\nabla^2 \tilde\Phi_N = 4 \pi \tilde\rho, \;\;\; \mbox{with} \;\;\;
t_\star = \frac{1}{\sqrt{{\cal G}_N \rho_\star}} , \;\;\; \rho = \rho_\star \tilde\rho ,
\label{eq:PhiN-epsilon}
\ee
As in Eq.(\ref{eq:t*-1}), $t_\star$ is the gravitational dynamical time associated with 
the characteristic density $\rho_\star$ of the system.
We also define the characteristic mass $M_\star$,
\be
\tilde M = \int d \tilde {\vec x} \, \tilde\rho , \;\;\;  \mbox{with} \;\;\;
M = M_\star \tilde M , \;\; M_\star = \rho_\star L_\star^3
\ee
and the characteristic wavefunction amplitude $\psi_\star$,
\be
\tilde\rho = \tilde\psi\tilde\psi^* , \;\;\; \mbox{with} \;\;\;
\psi_\star = \sqrt{\rho_\star /m} .
\ee
Then, the self-interaction potential reads
\be
\tilde\Phi_I = \lambda \tilde \rho , \;\; \mbox{with} \;\;
\lambda = \frac{4\pi r_a^2}{L_\star^2} = \frac{1}{{\cal G}_N \rho_a L_\star^2}
= \frac{6\pi \lambda_4 M_{\rm Pl}^2}{m^4 L_\star^2} .
\label{eq:PhiI-tilde}
\ee
In the following, we remove the tildes for simplicity, as we always work with
the dimensionless variables.
We will choose $L_\star$ as the radius of our initial spherical halo, so
that in dimensionless coordinates we have $R_{\rm halo}= 1$.

\subsection{Initial conditions and central soliton}

In this paper, we study the evolution of solitons inside self-gravitating halos.
As initial conditions of our numerical simulations, we write the wavefunction as
\be
\psi_{\rm initial} = \psi_{\rm sol} + \psi_{\rm halo} .
\label{eq:soliton+halo}
\ee
The first term $\psi_{\rm sol}$ corresponds to a solitonic core, where gravity
is balanced by the self-interactions, whereas the second term $\psi_{\rm halo}$
corresponds to the halo that makes up most of the volume and mass of the object,
where quantum pressure and self-interactions are negligible and the scalar field
behaves like cold dark matter.

As seen in Sec.~\ref{sec:hydrostatic}, in the Thomas-Fermi limit the spherically symmetric 
soliton is given by the hydrostatic equilibrium
\be
\Phi_N(r)+\Phi_I(r) = E_{\rm sol} ,
\label{eq:hydrostatic-alpha}
\ee
where we used the dimensionless variables and $E_{\rm sol}$ is a constant with
\be
\psi_{\rm sol}(\vec x,t) = e^{-i E_{\rm sol} t/\epsilon}
\hat\psi_{\rm sol}(r) .
\ee
For a quartic self-interaction $\lambda_4\phi^4$, which gives $\Phi_I = \lambda \rho$,
this yields a linear Helmholtz equation in $\rho$, with the solution
\be
\rho_{\rm sol}(r) = \rho_{0{\rm sol}}
\frac{\sin(\pi r/R_{\rm sol})}{\pi r/R_{\rm sol}} , \;\;\;
\hat\psi_{\rm sol}(r) = \sqrt{\rho_{\rm sol}(r)} ,
\label{eq:rho-soliton}
\ee
over $r \leq R_{\rm sol}$, and $\rho_{\rm sol}=0$ for $r>R_{\rm sol}$, as in
Eq.(\ref{eq:rho-soliton-1}).
This is a compact object of dimensionless radius and mass
\be
R_{\rm sol} = \frac{\sqrt{\lambda \pi}}{2} , \;\;\;
M_{\rm sol} = \frac{4}{\pi} \rho_{0{\rm sol}} R_{\rm sol}^3 .
\label{eq:R-soliton}
\ee
In practice, we define our system by $R_{\rm sol}$, and the self-interaction coupling
$\lambda$ follows from Eq.(\ref{eq:R-soliton}) as
$\lambda = 4 R_{\rm sol}^2/\pi$.
As the size of the halo is $R_{\rm halo}=1$, we consider cases with
$R_{\rm sol} \lesssim 1$, whence $\lambda \lesssim 1$.

In our numerical computations, we focus on the semiclassical regime $\epsilon=0.01 \ll 1$.
The central soliton is governed by the balance between gravity and self-interactions 
if the condition (\ref{eq:TF-1}) is satisfied. This reads
\be
\rho_{0\rm sol} \gg \frac{4\pi \epsilon^2}{\lambda^2} , \;\;\; 
\rho_{0\rm sol} \gg \frac{\pi^3\epsilon^2}{4 R_{\rm sol}^4} .
\label{eq:Thomas-Fermi-2}
\ee
We will consider the cases $R_{\rm sol}=0.5$ and $0.1$. In the former case the soliton
is always dominated by the self-interactions as $\rho \gtrsim 1$, whereas in the latter
case the self-interactions dominate over the quantum pressure for $\rho \gtrsim 10$.

\subsection{Decomposition of the halo in eigenfunctions}
\label{sec:halo-eigenfunctions}

\subsubsection{Eigenmodes}
\label{sec:Eigenmodes}

For a given time-independent potential $\Phi_N+\Phi_I = \bar\Phi$, Eq.(\ref{eq:Schrod-eps}) 
takes the form of the usual linear Schr\"odinger equation, which can be solved in terms 
of the energy eigenmodes $e^{-i E t/\epsilon} \hat\psi_E(\vec x)$ that obey
\be
- \frac{\epsilon^2}{2} \nabla^2 \hat\psi_E + \bar\Phi \hat\psi_E = E \hat\psi_E .
\ee
For a spherically symmetric potential $\bar\Phi$, we can expand these eigenmodes
in spherical harmonics,
\be
\hat\psi_{n\ell m}(\vec x) = {\cal R}_{n\ell}(r) Y_\ell^m(\theta,\varphi) ,
\ee
where the radial parts obey the usual radial time-independent Schr\"odinger equation
\be
\left[ - \frac{\epsilon^2}{2} \frac{1}{r^2} \frac{d}{dr} \left( r^2 \frac{d}{dr}
\right) + \frac{\epsilon^2}{2} \frac{\ell(\ell+1)}{r^2} + \bar\Phi \right]
{\cal R}_{n\ell} = E_{n\ell} {\cal R}_{n\ell}
\label{eq:Rnl-Enl}
\ee
and form an orthonormal basis
\be
\int dr \, r^2 \, {\cal R}_{n_1 \ell} {\cal R}_{n_2 \ell} = \delta_{n_1,n_2} \; .
\label{eq:Rn1-Rn2-orthonormal}
\ee
The energy levels $E_{n\ell}$ depend on the radial and orbital quantum numbers
$n$ and $\ell$ and are independent of the azimuthal number $m$.
As initial condition for the halo, we take a semiclassical equilibrium solution
defined by a target spherical density profile $\bar\rho(r)$,
and hence the associated target gravitational potential $\bar\Phi_N(r)$,
where we neglect the self-interactions and the central soliton,
\be
\bar\Phi(r) = \bar\Phi_N(r) , \;\;\;
\nabla^2\bar\Phi_N = 4\pi \bar\rho .
\ee
More precisely, in a fashion similar to \cite{Lin2018,2022PhRvD.105b3512Y},
we take for the initial halo wavefunction
\be
\psi_{\rm halo}(\vec x,t) = \sum_{n\ell m} a_{n\ell m} \hat\psi_{n \ell m}(\vec x)
e^{-i E_{n\ell} t/\epsilon} ,
\label{eq:psi-halo-eigenmodes}
\ee
where we choose the coefficients $a_{n\ell m}$ of the eigenmodes as
\be
a_{n\ell m} = a(E_{n\ell}) e^{i \Theta_{n\ell m}} ,
\label{eq:a-E}
\ee
where the amplitude $|a_{n\ell m}| = a(E_{n\ell})\geq 0$ is a deterministic
function $a(E)$ of the energy while the phases $\Theta_{n\ell m}$ are uncorrelated
random variables with a uniform distribution over $0 \leq \Theta < 2\pi$.

This gives a stochastic halo density $\rho_{\rm halo} = |\psi_{\rm halo}|^2$,
which fluctuates between different realizations of the phases $\Theta_{n\ell m}$.
Defining the average $\langle \dots \rangle$ over these random realizations, that is,
over the uncorrelated phases $\Theta_{n\ell m}$, we obtain the averaged density
\be
\langle \rho_{\rm halo} \rangle = \sum_{n\ell m} a(E_{n\ell})^2
|\hat\psi_{n\ell m}|^2 = \sum_{n\ell} \frac{2\ell+1}{4\pi} a(E_{n\ell})^2 \,
{\cal R}_{n\ell}^2 ,
\label{eq:rho-average-nl}
\ee
where we used $\sum_m |Y_\ell^m|^2= (2\ell+1)/(4\pi)$.
Then, the function $a(E_{n\ell})$ that determines the occupation numbers is chosen
so that $\langle \rho_{\rm halo} \rangle = \bar\rho$, i.e. we recover
the target density profile $\bar\rho(r)$ as the averaged profile over
the random realizations.
In the classical case of discrete particles, this corresponds to the construction
of the phase space distribution function $f(\vec x,\vec v)$ from the density profile,
and the choice (\ref{eq:a-E}) corresponds to an isotropic distribution $f(E)$.

Here we must point out that the average $\langle \dots \rangle$ is neither a spatial
or temporal average, but a statistical average over the random initial condition,
defined by the random phases $\Theta_{n\ell m}$ in Eq.(\ref{eq:a-E}).
As explained in the next section, we are looking for a set of coefficients $a_{n\ell m}$ so that
the initial density profile approximates a target spherical profile $\rho_{\rm target}(r)$.
As in \cite{Lin2018,2022PhRvD.105b3512Y}, a simple procedure is to choose these coefficients
of the form (\ref{eq:a-E}) with $a(E)$ given by Eq.(\ref{eq:a(E)-f(E)}) below. 
This ensures that the statistical average (\ref{eq:rho-average-nl}) recovers $\rho_{\rm target}(r)$.
In the limit $\epsilon \ll 1$ that we consider in this paper, many modes contribute to a fixed radial bin,
which is why the weight of each mode decreases as $\epsilon^3$ in Eq.(\ref{eq:a(E)-f(E)}). 
Therefore, a coarse graining of the initial condition will also recover $\rho_{\rm target}(r)$.
However, as seen in Eq.(\ref{eq:Proba-rho-halo}) below and in the initial conditions shown in
Figs.~\ref{fig:rho-plane-sinc-R0p5-rho0}-\ref{fig:rho-plane-sinc-R0p1-rho5} and
\ref{fig:rho-plane-cos-R0p5-rho0}-\ref{fig:rho-plane-cos-R0p1-rho100}, pointwise the initial density
does not converge to $\rho_{\rm target}$ as $\epsilon \to 0$. It always shows relative fluctuations of order
unity, but their spatial width decreases as $\epsilon$, so that the coarse-grained density converges
to $\rho_{\rm target}$.

These fluctuations are thus the minimal fluctuations that are always shown by the scalar field dark matter
scenario. They arise from the wave-like nature of the system and are set by the de Broglie wavelength
(\ref{eq:lambda-dB-epsilon}).
Thus, $\epsilon^3$ or $\lambda_{\rm dB}^3$ plays the role of $1/N$ for a continuous classical system
that is approximated by a finite number $N$ of particles.
Because of these fluctuations, the initial density profile is not exactly $\rho_{\rm target}(r)$ 
nor an equilibrium solution. As found in the numerical simulations and understood in the kinetic theory
developed in section~\ref{sec:kine} below, these fluctuations drive the growth of the soliton.
They would only cease when all the matter is within the soliton, that is, when there is only one
eigenmode of the Schr\"odinger equation left and no more interference terms.

\subsubsection{WKB approximation}

As we consider the semiclassical regime $\epsilon \ll 1$, we can expect the
Wentzel-Kramers-Brillouin (WKB) approximation \citep{landau1977,merzbacher1998,2022PhRvD.105b3512Y} to be valid. This gives for the radial part ${\cal R}_{n\ell}(r)$ the
form
\be
r_1 \!<\! r \!<\! r_2 \!: \;  {\cal R}_{n\ell}(r) \simeq
\frac{N_{n\ell}}{r \sqrt{k_{n\ell}(r)}}
\sin\left[ \frac{1}{\epsilon} \int_{r_1}^r \! dr' k_{n\ell}(r')
\!+\! \frac{\pi}{4} \right]
\label{eq:WKB-Rnl}
\ee
where $N_{n\ell}$ is the normalization factor, $k_{n\ell}(r)$ is defined by
\be
k_{n\ell}(r) = \sqrt{2 \left(E_{n\ell} - \bar\Phi_N(r) - \frac{\epsilon^2}{2}
\frac{\ell(\ell+1)}{r^2} \right)} ,
\label{eq:kr-def}
\ee
and $r_1<r_2$ are the two turning points of the classical trajectory, where
$k_{n\ell}(r)=0$.
The lower bound $r_1$ is due to the centrifugal barrier and the upper bound $r_2$
to the confining gravitational potential $\bar\Phi_N$.
For radial trajectories, associated with $\ell=0$, we have $r_1=0$.
Outside of the interval $[r_1,r_2]$ the wavefunction shows a fast decrease
as this corresponds to the forbidden region in the classical limit and we consider
the semiclassical regime $\epsilon \ll 1$.
The normalization condition (\ref{eq:Rn1-Rn2-orthonormal}) gives
\be
N_{n\ell} = \left( \int_{r_1}^{r_2} \frac{dr}{2 k_{n\ell}(r)} \right)^{-1/2} ,
\ee
where we neglected the contributions from the classically forbidden regions and
took the average over the fast oscillations of the wavefunction.
Finally, the quantization condition of the energy levels is given in this WKB
approximation by
\be
\frac{1}{\epsilon} \int_{r_1}^{r_2} dr \, k_{n\ell}(r) =
\left( n+\frac{1}{2} \right) \pi ,
\label{eq:quanta-k-n}
\ee
where $n=0, 1, 2, \dots$ is a non-negative integer.
We can see that in the semiclassical regime, $\epsilon \ll 1$, the quantum numbers
become large as
\be
n \sim 1/\epsilon , \;\;\; \ell \sim 1/\epsilon ,
\label{eq:n-ell-epsilon}
\ee
and the difference between energy levels decreases as $\Delta E \sim \epsilon$.
In particular, at fixed $\ell$ we obtain from Eq.(\ref{eq:quanta-k-n})
\be
\frac{\partial n}{\partial E} = \frac{1}{\pi \epsilon} \int_{r_1}^{r_2} \frac{dr}{k_{n\ell}(r)} .
\ee
In this continuum limit, we can replace the sums in Eq.(\ref{eq:rho-average-nl}) by integrals
and we obtain
\be
\langle\rho_{\rm halo}(r)\rangle = \frac{1}{2\pi^2\epsilon^3} \int dE \, a(E)^2
\sqrt{2[E-\bar\Phi_N(r)]} ,
\ee
where we used the WKB approximation (\ref{eq:WKB-Rnl}).
Comparing this expression with the classical result that expresses the density
in terms of the particle phase-space distribution \citep{Binney2008},
\be
\rho_{\rm classical}(r) = 4 \pi \int_{\bar\Phi_N(r)}^0 dE \, f(E) \sqrt{2 [E-\bar\Phi_N(r)]} ,
\label{eq:rho-fE}
\ee
where we normalized the potential so that bound orbits correspond to $E<0$,
we obtain
\be
a(E)^2 = (2 \pi \epsilon)^3 f(E) .
\label{eq:a(E)-f(E)}
\ee
The classical phase-space distribution can be obtained from the density by
Eddington's formula \citep{Binney2008},
\be
f(E) = \frac{1}{2\sqrt{2}\pi^2} \frac{d}{dE} \int_E^0
\frac{d\Phi_N}{\sqrt{\Phi_N-E}} \frac{d\rho_{\rm classical}}{d\Phi_N} .
\label{eq:Eddington}
\ee

In practice, choosing a target halo density profile
$\rho_{\rm target}(r)$, we obtain the classical phase-space
distribution $f(E)$ from Eddington's formula (\ref{eq:Eddington}), the eigenmode
coefficients $a_{n\ell m}$ from Eqs.(\ref{eq:a-E}) and (\ref{eq:a(E)-f(E)}), and
the initial halo wavefunction from Eq.(\ref{eq:psi-halo-eigenmodes}).
However, to avoid the singularity of the WKB approximation at the turning points,
we do not use the WKB expression (\ref{eq:WKB-Rnl}) for the eigenmodes.
Instead, we explicitly solve the linear eigenmode problem associated with the
radial Schr\"odinger equation (\ref{eq:Rnl-Enl}).
Therefore, the WKB approximation is only used for the determination of
the initial coefficients $a_{n\ell m}$.
This is sufficient for our purpose, which is to build random initial conditions
with a target radial density profile.

\subsection{Numerical methods}

The system in dimensionless units is fully described by the Schrödinger equation (\ref{eq:Schrod-eps}), supplemented by the Poisson equation (\ref{eq:PhiN-epsilon})
and the self-interaction potential (\ref{eq:PhiI-tilde}). 
We consider periodic boundary conditions, which allows us to use Fourier transforms to compute
the Poisson equation and the Laplacian in the Schrödinger equation.
In this setting, the gravitational potential is obtained as
\be
\Phi_N = -4\pi \mathcal{F}^{-1} k^{-2} \mathcal{F} (\rho - \bar\rho) ,
\label{eq:Poisson-num}
\ee
where $\mathcal{F}$ and $\mathcal{F}^{-1}$ are the Fourier 
transform and its inverse, $\vec k$ is the wavenumber in Fourier space and $k=|\vec k|$.
This means that in real space, as in cosmological codes, we solve the Poisson equation defined
by $\nabla^2\Phi_N = 4 \pi (\rho-\bar\rho)$, where $\bar\rho$ is the mean density in the
simulation box. This is the appropriate form for periodic boundary conditions, in contrast with
the Poisson equation (\ref{eq:PhiN-epsilon}) without the $\bar\rho$ term, which is appropriate
for isolated objects. For large simulation box $\bar\rho \to 0$ and one recovers the isolated
case.

As found in \cite{Alvarez_Rios_2023}, the choice of the boundary conditions can have
some impact on the dynamics, especially on the tails of halos around solitons.
This is less so in our case because we consider the relaxation of single compact halos.
As shown by the density profiles found in the simulations presented in the next
sections, the density outside of the initial halo is very small and negligible amounts of
matter can reach the boundaries of the simulation box.
Therefore, the periodic boundary conditions mostly differ from truly isolated simulations
by the impact on the gravitational potential of the neighboring halos in the periodic chain.
As found in Fig.~\ref{fig:rho-plane-sinc-R0p1-rho0}, this only affects our results in the
case of the formation of a small-size soliton inside a flat halo. We find that the formation
time is lower in a bigger simulation box, but the main result on the formation of a soliton
remains true. This is because the formation of this soliton is triggered by the growth of 
small perturbations until one of them reaches a density that is large enough to show strong
self-interactions. This stochastic process is sensitive to the details of the dynamics.
In the other cases, we find that our results are stable when we compare with a run that has
a twice bigger simulation box.

We have developed a numerical code to compute the 3D dynamics, using a symmetrized 
split-step Fourier technique as in \citep{Edwards2018,Glennon2020}. 
Thus, the wavefunction is advanced by a timestep $\Delta t$ as
\bea
&& \psi(\vec{x},t+ \Delta t) = \exp\left[-\frac{i \Delta t}{2\epsilon} 
\Phi(\vec{x},t+ \Delta t)\right] \times \nonumber \\
&&  \mathcal{F}^{-1} \exp\left[-\frac{i \epsilon\Delta t}{2} k^2\right]\mathcal{F} 
\exp\left[-\frac{i \Delta t }{2\epsilon}\Phi(\vec{x},t)\right]\psi(\vec{x},t) , \nonumber \\
&&
\label{eq:step}
\eea
where $\Phi = \Phi_N+\Phi_I$.
This operator splitting scheme is based on the fact that in the Schrödinger equation 
(\ref{eq:Schrod-eps}) the operator $\Phi \psi$ is diagonal in configuration space whereas
the operator $\nabla^2\psi$ is diagonal in Fourier space. 
Thus, we proceed in three steps. First, from $\psi$ we compute $\rho = |\psi|^2$,
$\Phi_I = \lambda\rho$ and $\Phi_N$ by Fourier transforms as explained in (\ref{eq:Poisson-num}).
Then, we apply the first real-space operator $e^{-i \Delta t \Phi/(2\epsilon)}$.
Second, going to Fourier space, we apply the Fourier-space operator $e^{-i \epsilon \Delta t k^2/2}$.
Third, we compute again $\rho$, $\Phi_I$ and $\Phi_N$, which have been modified by the Laplacian
operator on $\psi$, and we perform the last multiplication by the real-space operator 
$e^{-i \Delta t \Phi/(2\epsilon)}$. Note that this last operation does not modify $\rho$, $\Phi_I$ and $\Phi_N$,
because it only modifies the local phase of $\psi$. Thus, the total potential $\Phi$ used in this step is also
the potential at the end of the timestep $\Delta t$.
We iterate this process until we reach the final time of the simulation.
The timestep satisfies the conditions $\Delta t < 2\pi / (\epsilon k_{\rm max}^2)$ and
$\Delta t < 2 \pi \epsilon / | \Phi_{\rm max} |$. In practice, we take a somewhat smaller timestep
to ensure that mass and energy are well conserved over the full simulation time.
This numerical procedure has already been used in many works on Fuzzy Dark Matter
\citep{Edwards2018,Glennon2020,Chan:2021aa}.

We have employed the \textsc{FFTW3} libraries \citep{FFTW05} to compute the discrete Fourier transform (DFT). These libraries adapt the DFT algorithm to details of the underlying hardware to maximize performance. In addition, we have taken advantage of the \textsc{OpenMP} tools to parallelize the multi-threaded routines \citep{openmp,openmp1997}.

\section{Halo with a flat-core density profile}
\label{sec:flat-core}

\begin{figure*}[ht]
\centering
\includegraphics[height=6.5cm,width=0.39\textwidth]{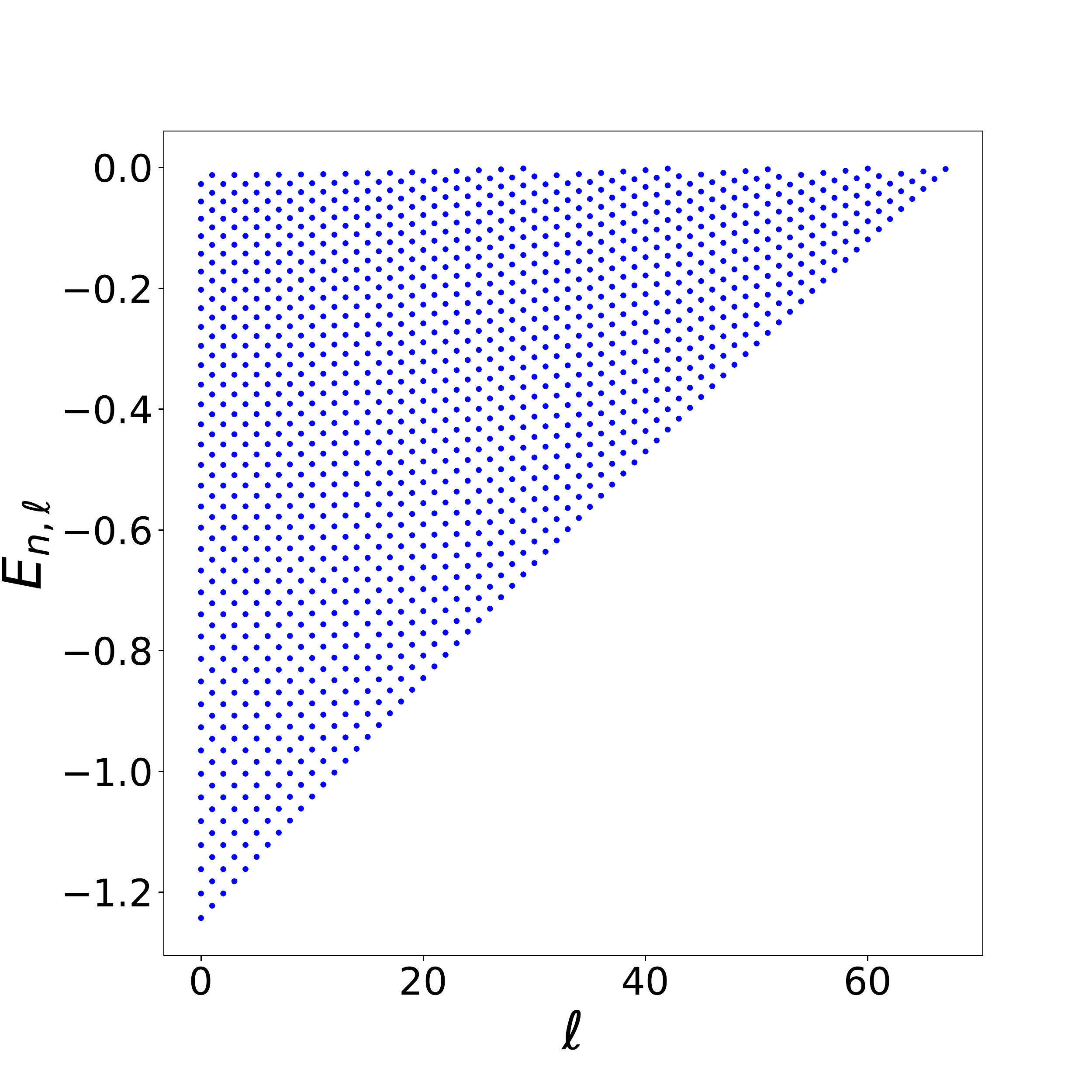}
\includegraphics[height=6.5cm,width=0.39\textwidth]{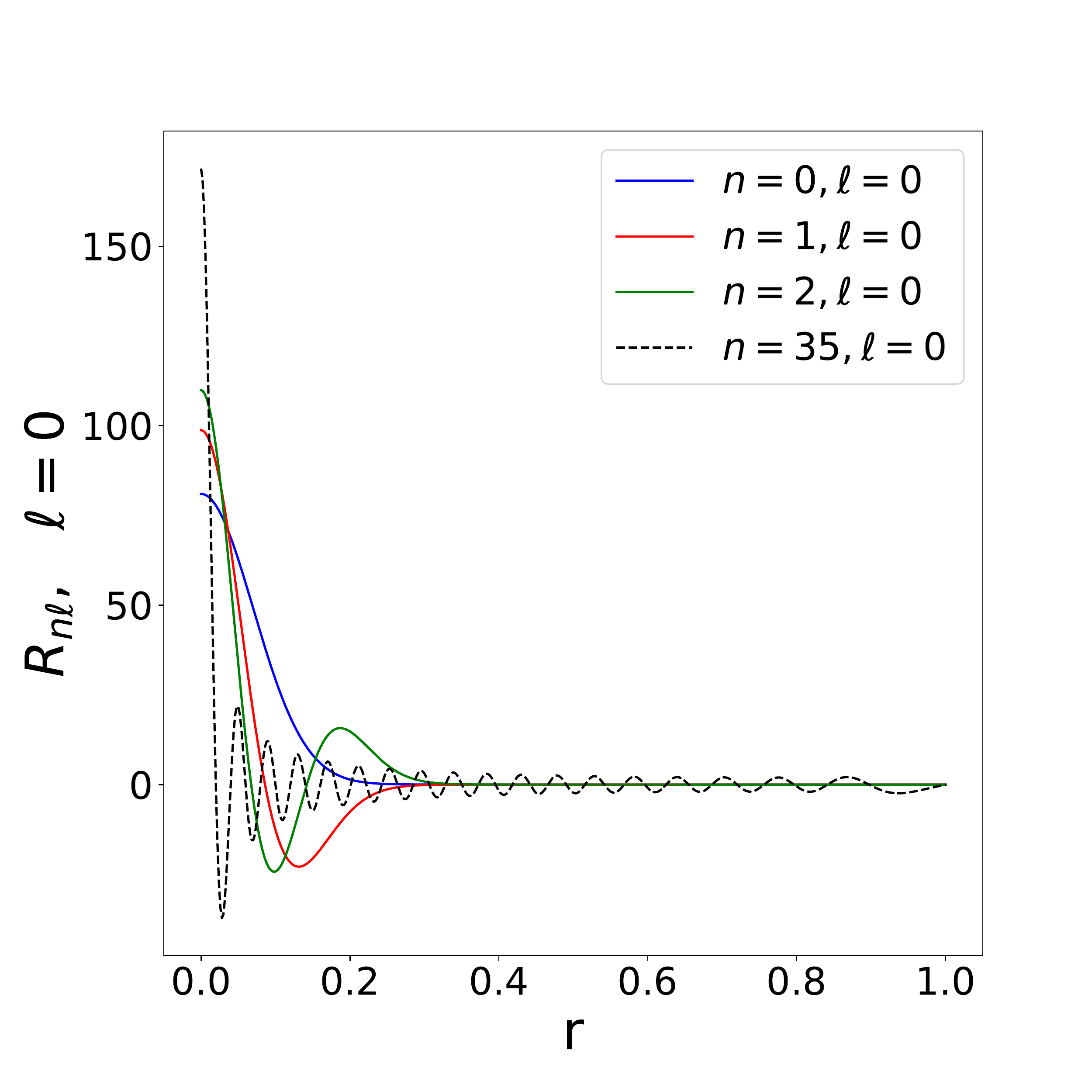}\hfill
\includegraphics[height=6.5cm,width=0.39\textwidth]{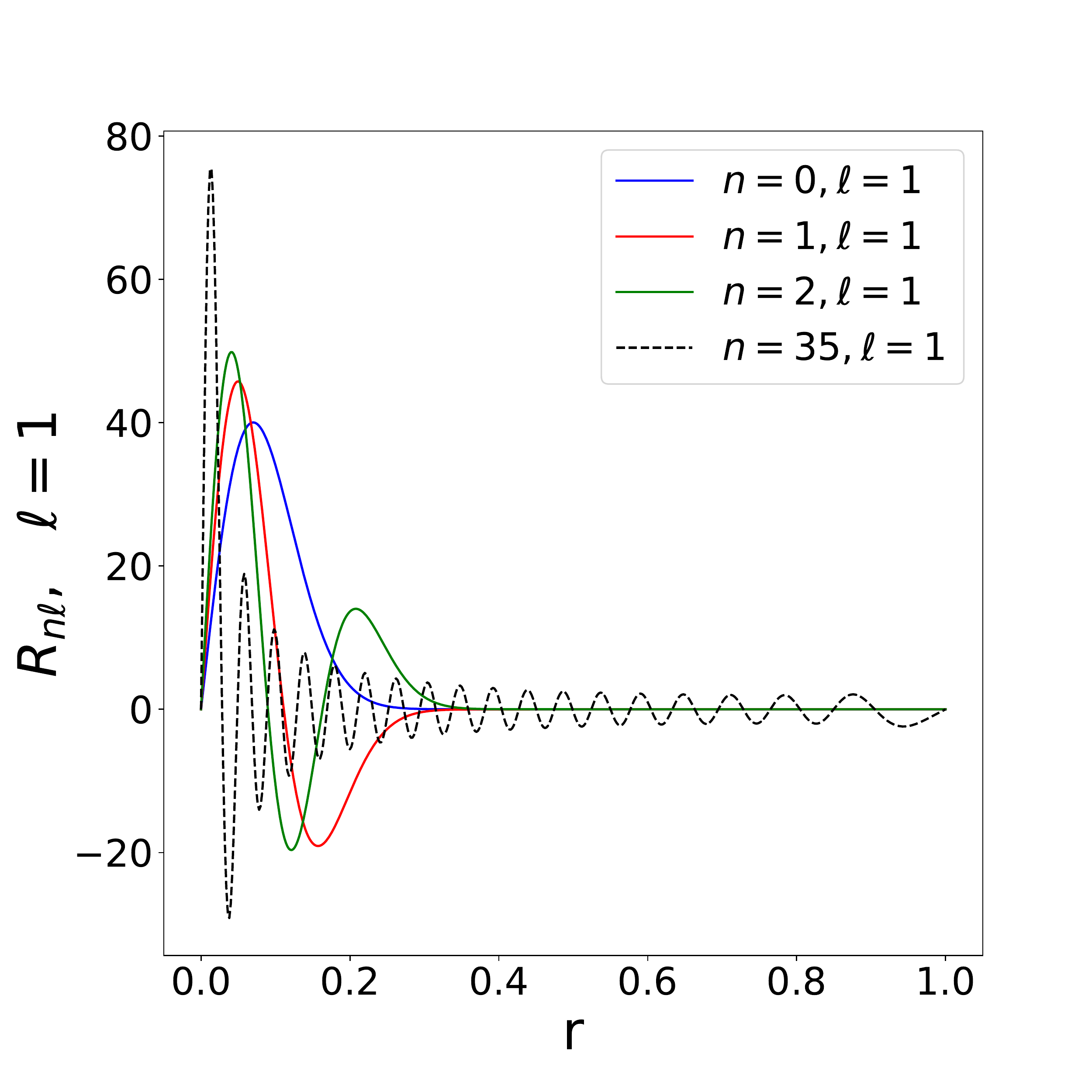}
\includegraphics[height=6.5cm,width=0.39\textwidth]{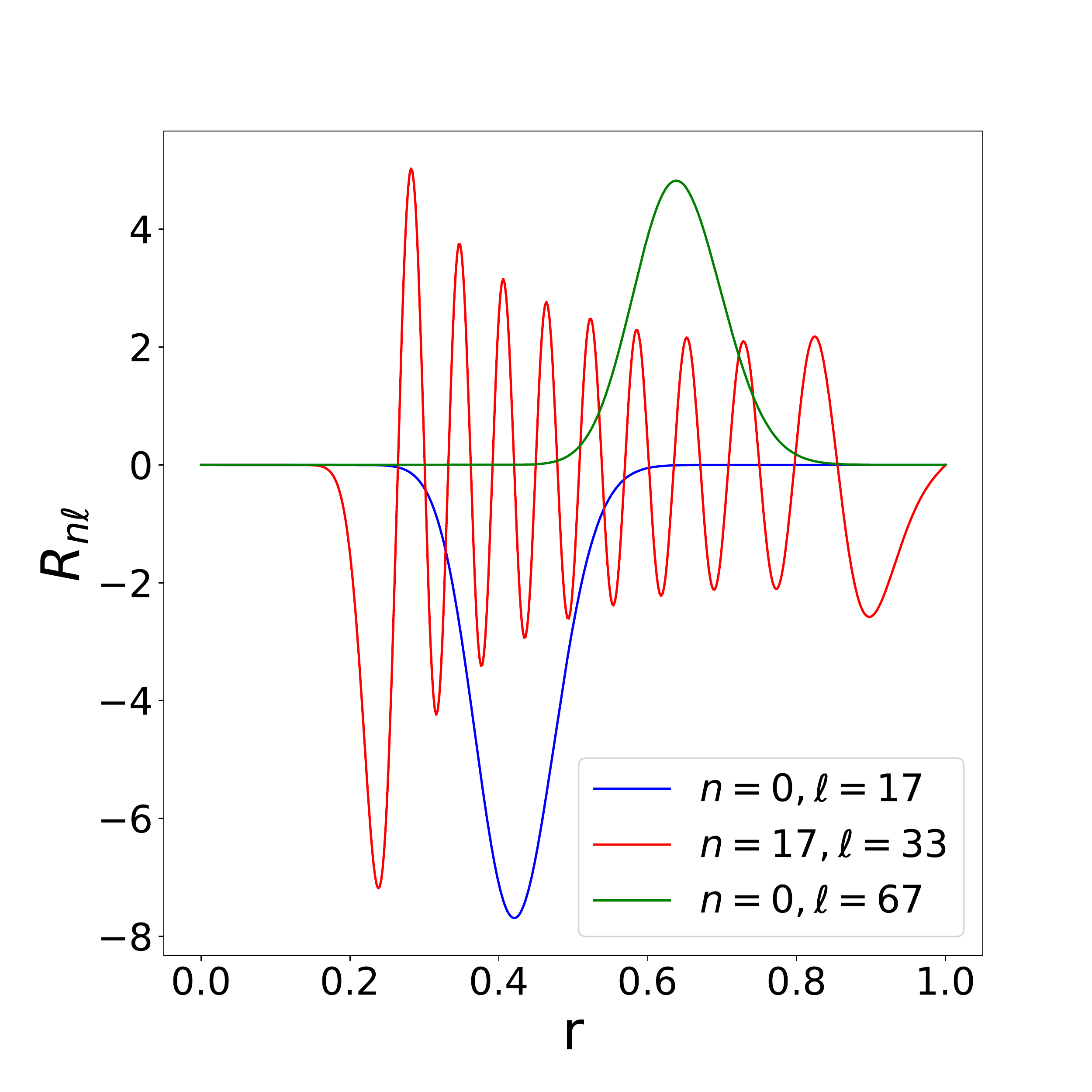}
\caption{
Energy levels $E_{n\ell}$ in the $(\ell,E_{n\ell})$ plane ({\it upper left panel}),  for the gravitational
potential (\ref{eq:rho-halo-sinc}).
Eigenmodes ${\cal R}_{n\ell}(r)$ for $\ell=0$ ({\it upper right panel}), $\ell=1$ ({\it lower left panel})
and some large values of $n$ or $\ell$ ({\it lower right panel}).
These eigenmodes are normalized to unity as in Eq.(\ref{eq:Rn1-Rn2-orthonormal}).
}
\label{fig:E-levels-sinc}
\end{figure*}

\subsection{Halo eigenmodes}
\label{sec:flat-core-eigenmodes}

We first investigate the dynamics of systems with a flat halo density core.
Thus, we consider a Lane-Emden profile with a polytropic index $n=1$,
\be
0 \!\leq\! r \!\leq\! 1 \!: \; \bar\rho(r) = \rho_0 \frac{\sin(\pi r)}{\pi r} ,
\; \bar\Phi_{N}(r) = - \frac{4\rho_0\sin(\pi r)}{\pi^2 r} ,
\label{eq:rho-halo-sinc}
\ee
which corresponds to the phase-space distribution
\be
- \frac{4\rho_0}{\pi} < E < 0 : \;\;\;  f(E) = \frac{1}{8\pi\sqrt{- 2 E}} .
\label{eq:f(E)-sinc}
\ee
Although this halo profile happens to take the same form as the
hydrostatic soliton (\ref{eq:rho-soliton}), its physics is quite different.
Indeed, here gravity is not balanced by self-interactions but by the velocity
dispersion, as for collisionless particles.
With $\rho_0=1$, this is a simple model for a halo
with a flat-core density profile and  $\rho_{\rm halo} \sim 1$ within the radius
$R_{\rm halo}=1$.

We solve the eigenvalue problem (\ref{eq:Rnl-Enl}) with a numerical spectral method.
For each orbital quantum number $\ell$, we expand the radial wavefunctions
${\cal R}_{n \ell}$ on the basis defined by the eigenvectors of the spherical flat potential
well with infinite walls at $r=1$ (they are given by the spherical Bessel functions
$j_\ell(k_n r)$ where $k_n$ is a zero of $j_\ell$).
This automatically satisfies the boundary condition at $r=0$,
${\cal R}_{n\ell} \propto r^\ell$.
This also gives ${\cal R}_{n\ell}(r=1)=0$, which is a good approximation in the semiclassical
regime $\epsilon \ll 1$, as we only include bound eigenmodes with $E<0$ that
are classically forbidden beyond $r=1$.
Truncating the basis at the first $100$ eigenvectors, we obtain a finite linear
eigenvalue problem associated with a real symmetric matrix of size $100\times 100$.
Then, we obtain the $n_{\rm max}(\ell)$ energy levels with $E<0$ and their associated
bound-state eigenvector.
Starting from $\ell=0$ we increase $\ell$ with unit step until there are no
more negative eigenvalues.

We show in Fig.~\ref{fig:E-levels-sinc} the energy levels and some radial eigenmodes
associated with the gravitational potential (\ref{eq:rho-halo-sinc}), with
$\epsilon=0.01$.
These eigenmodes are normalized to unity as in Eq.(\ref{eq:Rn1-Rn2-orthonormal}).
We find that there exist bound states until $\ell_{\max}=67$. The number $n$ of bound
radial modes decreases as $\ell$ increases and we find $n_{\max}=35$ at $\ell=0$.
In agreement with (\ref{eq:n-ell-epsilon}), because $\epsilon \ll 1$ there are many eigenmodes
inside the potential well $\Phi_N$, which has a depth of the order of unity.
As seen in Fig.~\ref{fig:E-levels-sinc}, high-energy modes with large $n$
can probe small scales, down to $\Delta r \sim \epsilon = 0.01$, while
high orbital momentum modes with large $\ell$ probe large radii.
The modes $\ell=0$ correspond to radial trajectories in the classical limit.

\begin{figure*}[ht]
\centering
\includegraphics[height=4.cm,width=0.24\textwidth]{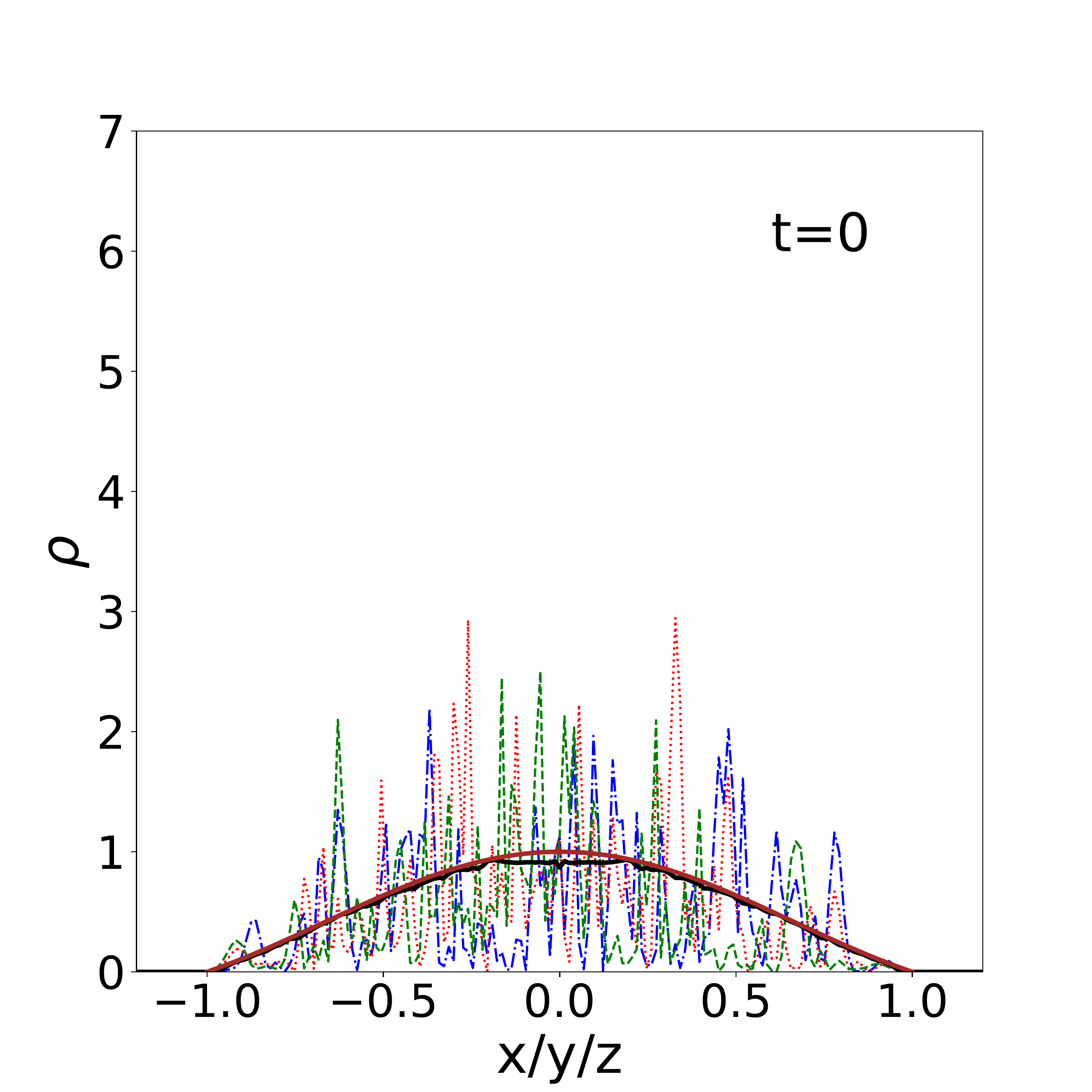}
\includegraphics[height=4.cm,width=0.25\textwidth]{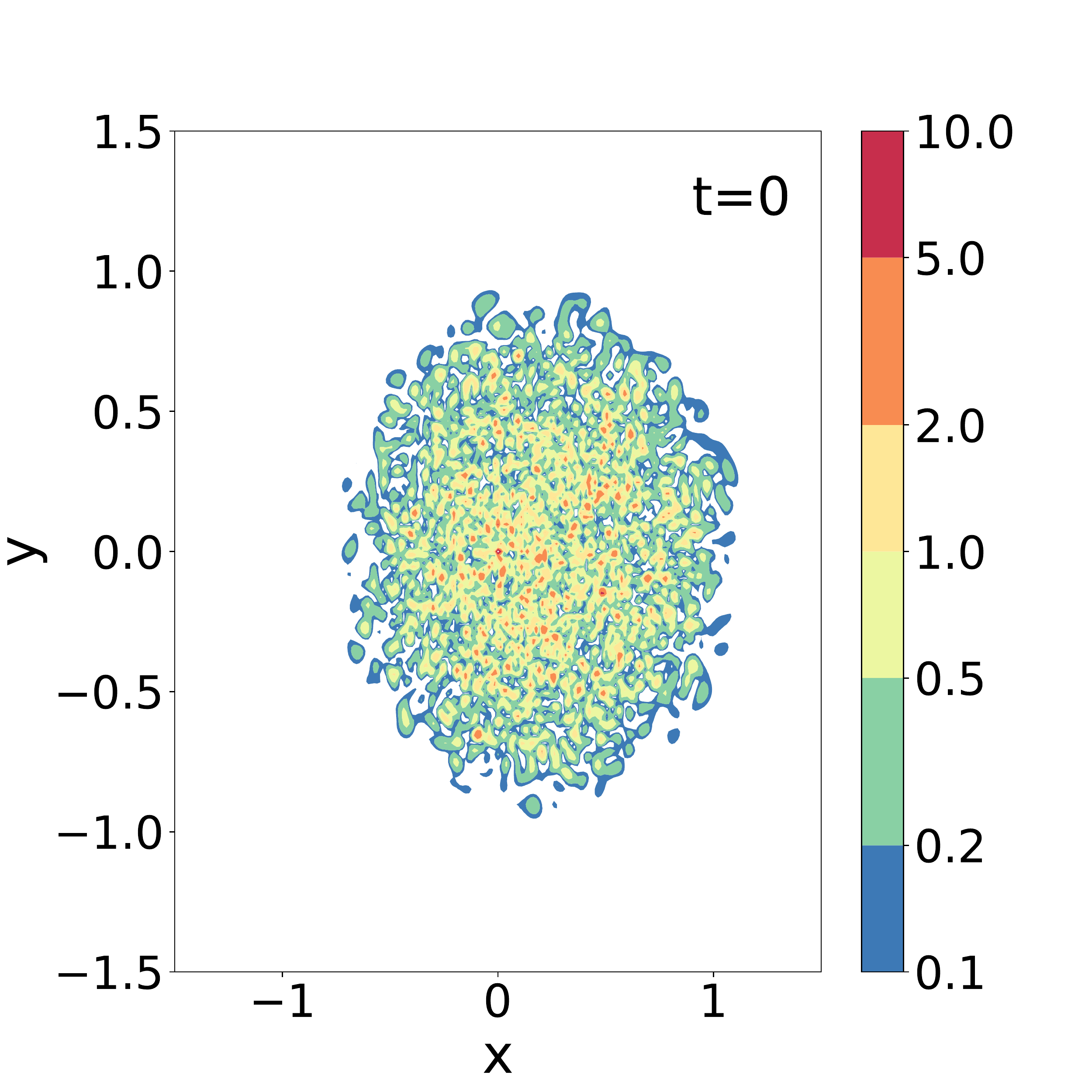}
\includegraphics[height=4.cm,width=0.24\textwidth]{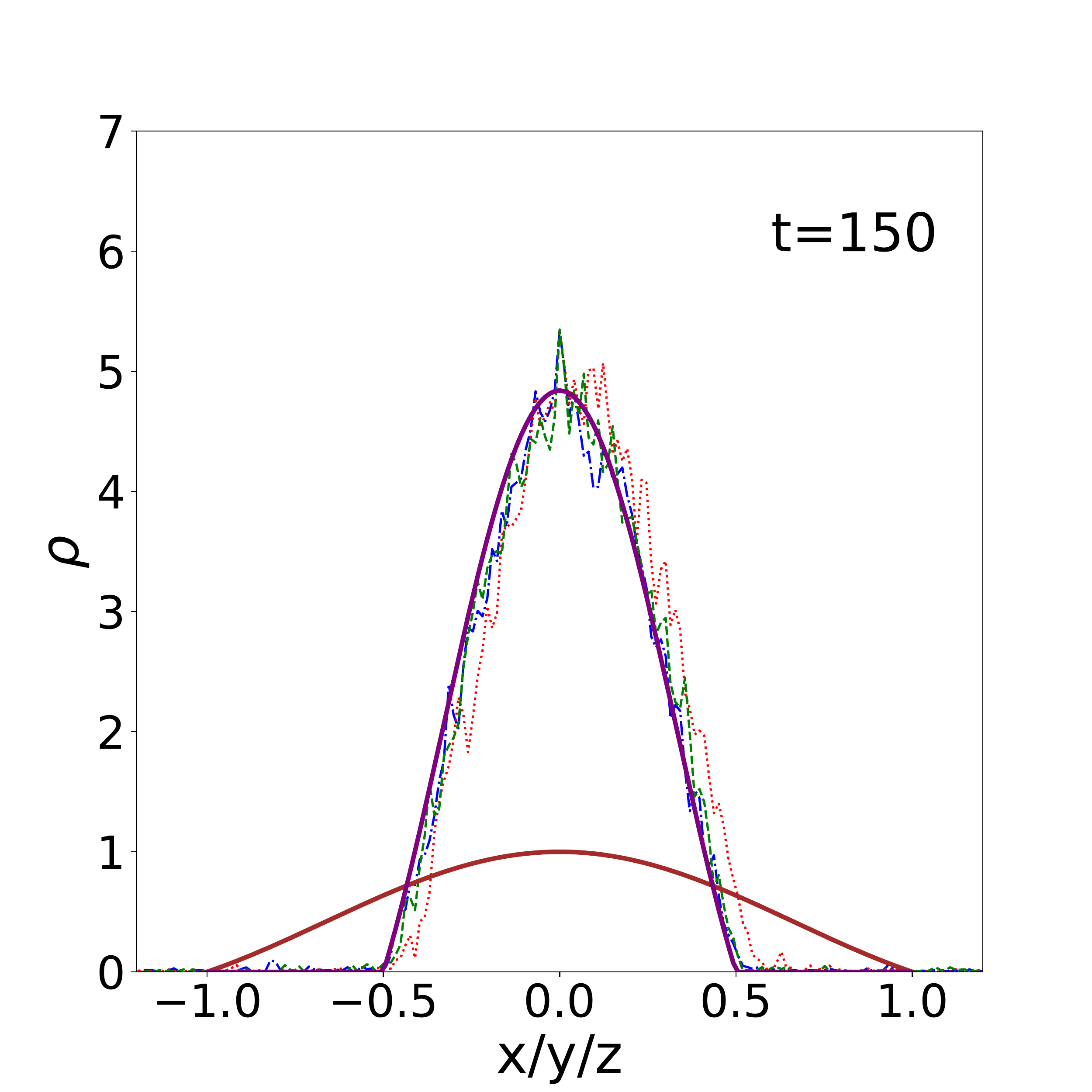}
\includegraphics[height=4.cm,width=0.25\textwidth]{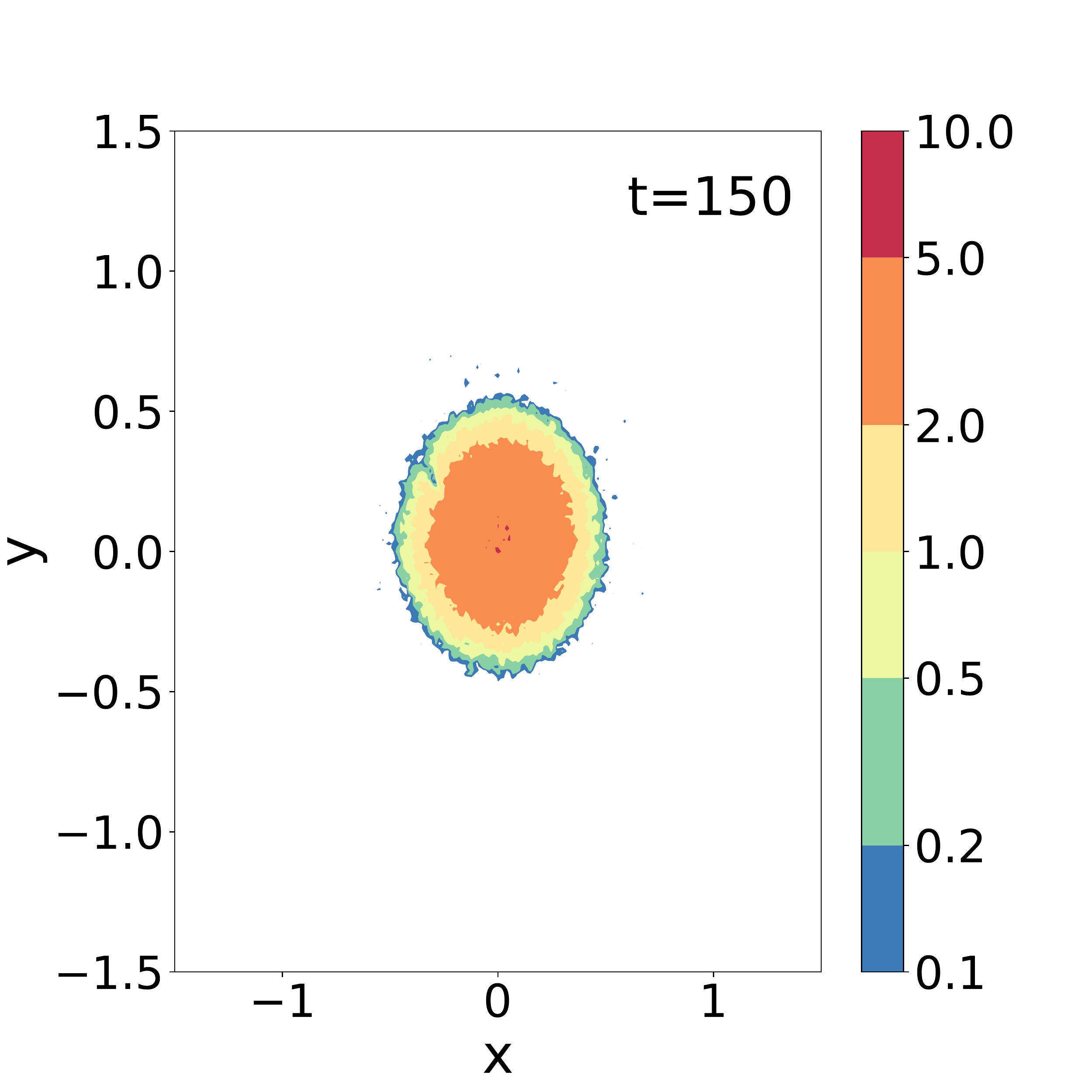}\\
\includegraphics[height=4.2cm,width=0.32\textwidth]{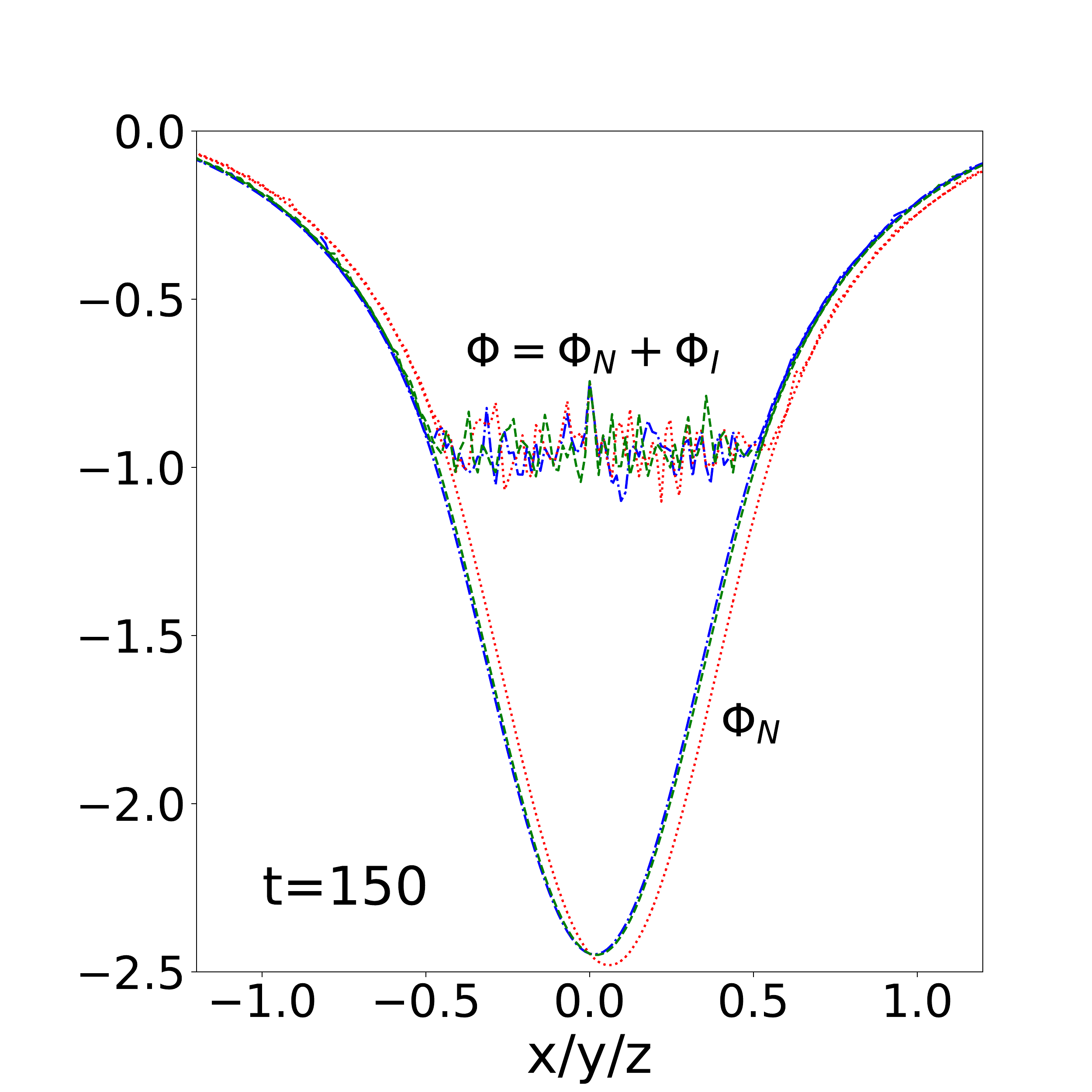}
\includegraphics[height=4.2cm,width=0.28\textwidth]{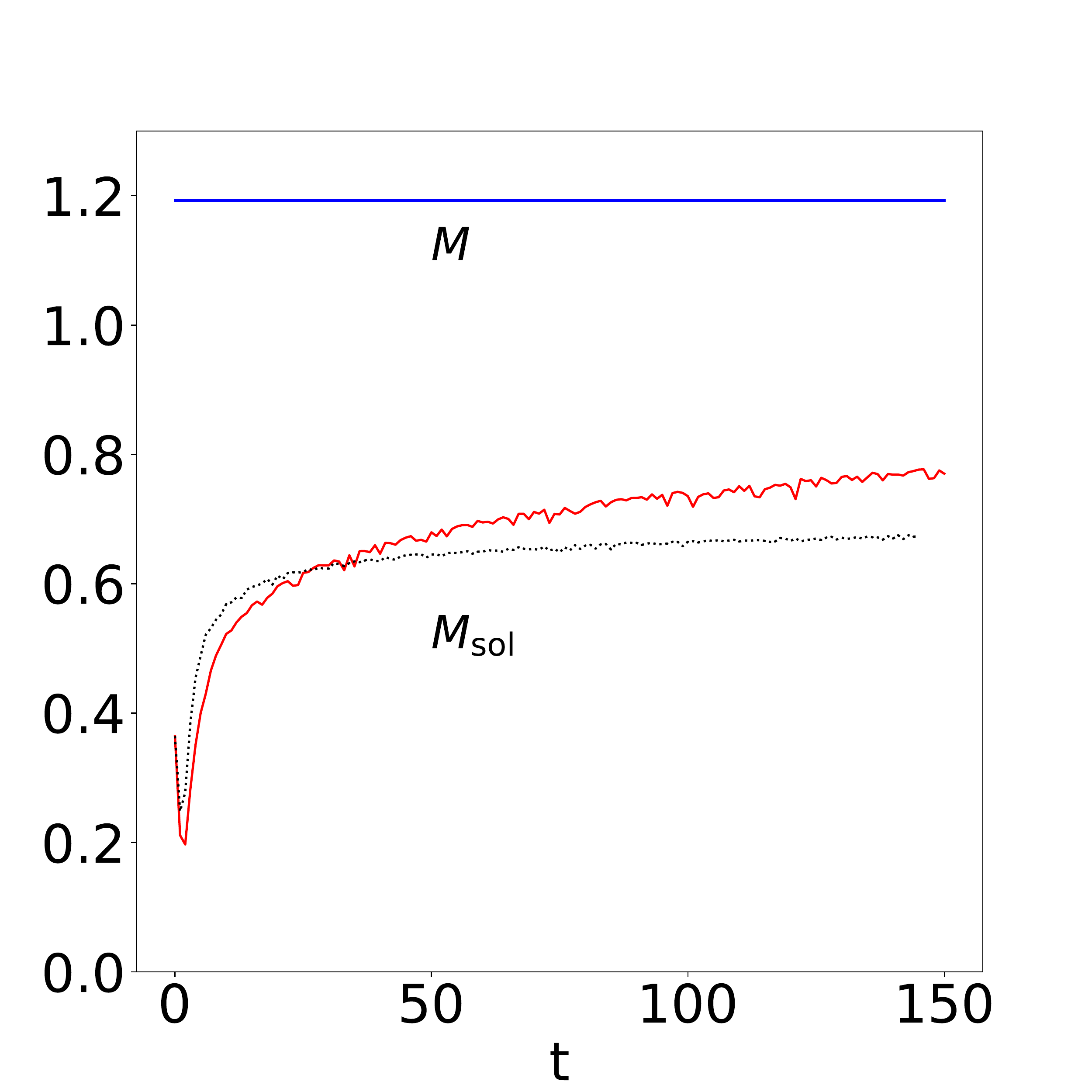}
\includegraphics[height=4.2cm,width=0.28\textwidth]{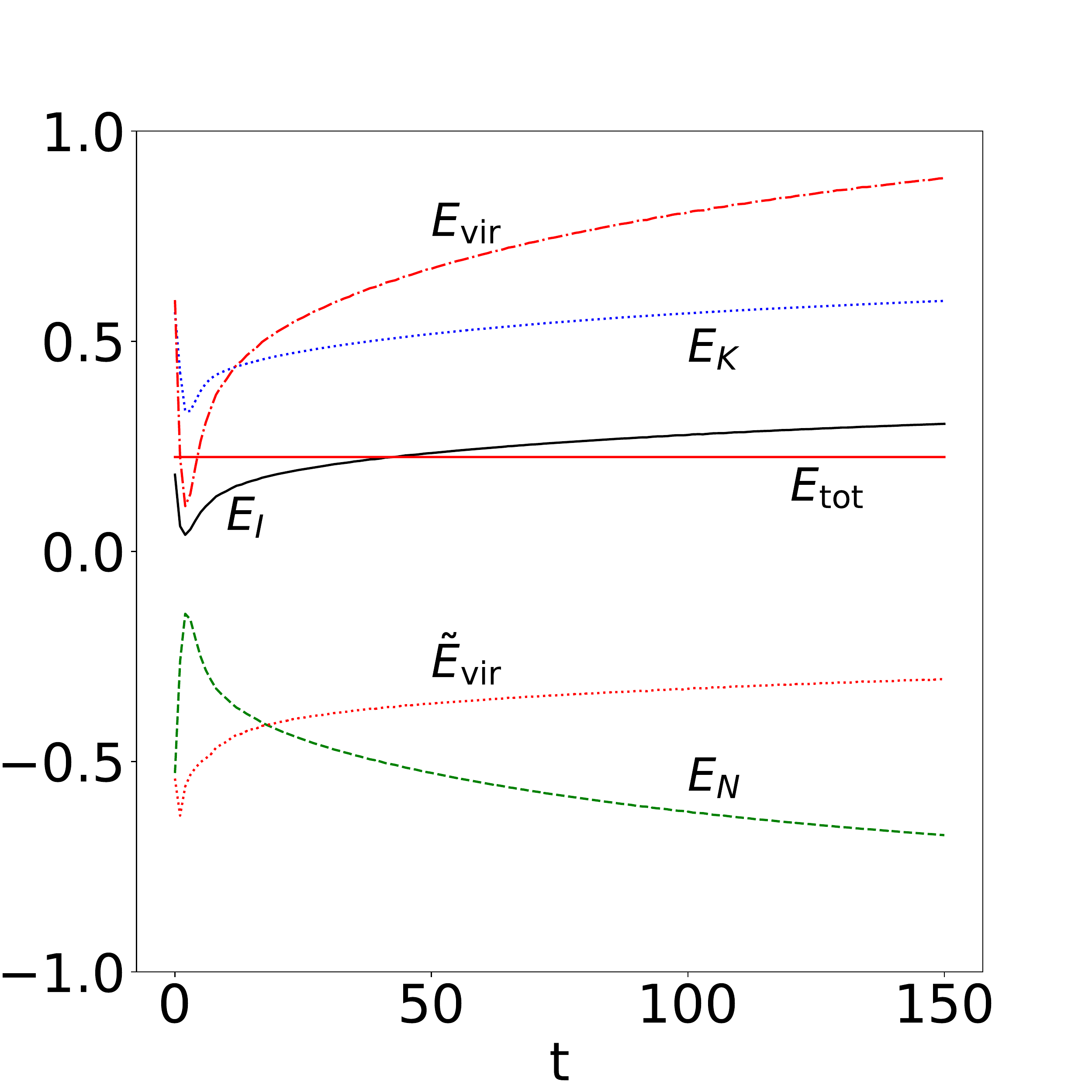}
\caption{
[$R_{\rm sol}=0.5, \;\; \rho_{0{\rm sol}}=0$.]
{\bf (a)} initial density $\rho$ along the $x$ (blue dash-dot line),
$y$ (red dotted line) and $z$ (green dashed line) axis running through the center
of the halo. The smooth brown solid line is the target density profile
(\ref{eq:rho-halo-sinc}) and the black wiggly solid line is
the averaged density $\langle\rho_{\rm halo}\rangle$ of Eq.(\ref{eq:rho-average-nl})
(they can hardly be distinguished in the figure).
There is no central soliton in this initial condition.
{\bf (b)} initial 2D density map on the $(x,y)$ plane at $z=0$.
{\bf (c)} density profile along the $x$, $y$ and $z$ axis
that run through the point $\vec r_{\rm max}$ where the density is maximum,
at time $t=150$. The lower brown solid line is the initial target density profile
as in panel {\bf (a)}, while the upper purple solid line is the density profile of a soliton
(\ref{eq:rho-soliton}) that would contain the mass $M_{\rm sol}(t)$ enclosed within the
radius $R_{\rm sol}$ around $\vec r_{\rm max}$.
{\bf (d)} 2D density map at time $t=150$ on the $(x,y)$ plane centered on $\vec r_{\rm max}$.
{\bf (e)} total potential $\Phi=\Phi_N+\Phi_I$ and gravitational potential $\Phi_N$ at
$t=150$, along the $x$, $y$ and $z$ axis passing through $\vec r_{\rm max}$.
{\bf (f)} evolution with time of the total mass $M$ of the system and of the mass
$M_{\rm sol}$ enclosed within the radius $R_{\rm sol}$ around $\vec r_{\rm max}$.
The black dotted line is the result obtained from a simulation with a twice bigger box.
{\bf (g)} evolution with time of the energy components of the system.
}
\label{fig:rho-plane-sinc-R0p5-rho0}
\end{figure*}

\subsection{Large soliton radius, $R_{\rm sol}=0.5$}

We first consider cases where the radius $r_a$ associated with the self-interactions
is of the order of the halo radius. Thus, in this section we take
$R_{\rm sol}=0.5$. In the cosmological context, this corresponds to the
first overdensities that can collapse just above the Jeans length $\sim r_a$,
as gravity can overcome the pressure associated with the repulsive
$\phi^4$ self-interaction.

\subsubsection{Halo without initial soliton}

We first investigate the dynamics when there is no initial soliton inside the halo,
$\rho_{0{\rm sol}}=0$.
We show in the first two panels (a) and (b) in Fig.~\ref {fig:rho-plane-sinc-R0p5-rho0}
our initial condition for one
realization of the random phases $\Theta_{m\ell m}$ in Eq.(\ref{eq:a-E}).
As seen in the upper left panel, the averaged density $\langle\rho_{\rm halo}\rangle$, defined by
Eq.(\ref{eq:rho-average-nl}) where the interferences between the different modes
$\hat\psi_{n\ell m}$ vanish, provides a good approximation of the target density
(\ref{eq:rho-halo-sinc}).
Moreover, $\langle\rho_{\rm halo}\rangle$ is identical along any axis that
runs through the origin as there is no angular dependence left in
Eq.(\ref{eq:rho-average-nl}), which is consistent with the spherical
symmetry of the target profile (\ref{eq:rho-halo-sinc}).
Thus, for $\epsilon=0.01$ the WKB approximation (\ref{eq:a(E)-f(E)}) for the
amplitude of the coefficients $a_{n\ell m}$ is already rather good. As expected, it
fares somewhat less well at the center of the halo, dominated by low $(n,\ell)$ modes.
On the other hand, the exact random initial density
$\rho_{\rm halo}=|\sum a_{n\ell m} \hat\psi_{n\ell m}|^2$ shows strong fluctuations
around $\langle\rho_{\rm halo}\rangle$ and depends on the chosen axis running
through the center.
In agreement with Eq.(\ref{eq:lambda-dB-epsilon}),
these spikes have a width $\Delta x \sim \epsilon$ that decreases
in the semiclassical regime but their amplitude remains of order unity. Thus,
$\rho_{\rm halo}$ only converges in a weak sense to the target classical density
profile, after coarse-graining over a finite-size window.
Note that for a classical system of discrete particles the density field is also very
noisy, as it is a sum of Dirac peaks in the point-mass limit. Here the width of the
spikes is set by the de Broglie wavelength (\ref{eq:lambda-dB-epsilon}).
More precisely, from Eqs.(\ref{eq:rho-average-nl}) and (\ref{eq:a(E)-f(E)}),
we can see that the number $N$ of eigenmodes $\hat\psi_{n\ell m}$ that contribute
to the density at a given point $\vec x$ grows as $1/\epsilon^3$.
We can also write powers of the exact random halo density, $\rho_{\rm halo} = \psi \psi^\star \geq 0$,
as
\be
\rho_{\rm halo}^p = \sum_{i_1,..,i_p=1}^N \sum_{j_1,..,i_p=1}^N
a_{i_1} .. a_{i_p} a_{j_1}^\star .. a_{j_p}^\star \psi_{i_1} ..\psi_{i_p}
\psi^\star_{j_1} .. \psi^\star_{j_p} ,
\ee
where the indices $i$ or $j$ denote $\{n,\ell,m\}$.
Taking the average over the random phases $\Theta_i$ of Eq.(\ref{eq:a-E}),
the only terms that contribute are those where each $a_{i_k}$ can be paired
with a coefficient $a^\star_{j_{k'}}$ with $j_{k'}=i_k$.
This gives $p!$ possible permutations,
\be
\langle \rho_{\rm halo}^p \rangle = p! \, \langle \rho_{\rm halo} \rangle^p ,
\ee
and we obtain the probability distribution
\be
\rho_{\rm halo} \geq 0 : \;\;\; {\cal P}(\rho_{\rm halo}) = \frac{1}{\langle \rho_{\rm halo} \rangle}
e^{-\rho_{\rm halo}/\langle \rho_{\rm halo} \rangle} ,
\label{eq:Proba-rho-halo}
\ee
which does not depend on $N$ nor $\epsilon$.
In particular, the standard deviation is
$\sqrt{\langle \rho_{\rm halo}^2 \rangle_c}= \langle \rho_{\rm halo} \rangle$,
in agreement with the relative fluctuations of order unity seen in the upper left panel
in Fig.~\ref{fig:rho-plane-sinc-R0p5-rho0}. Thus, the initial density shows strong relative
fluctuations of order unity throughout the halo.

\begin{figure*}[ht]
\centering
\includegraphics[height=4.cm,width=0.24\textwidth]{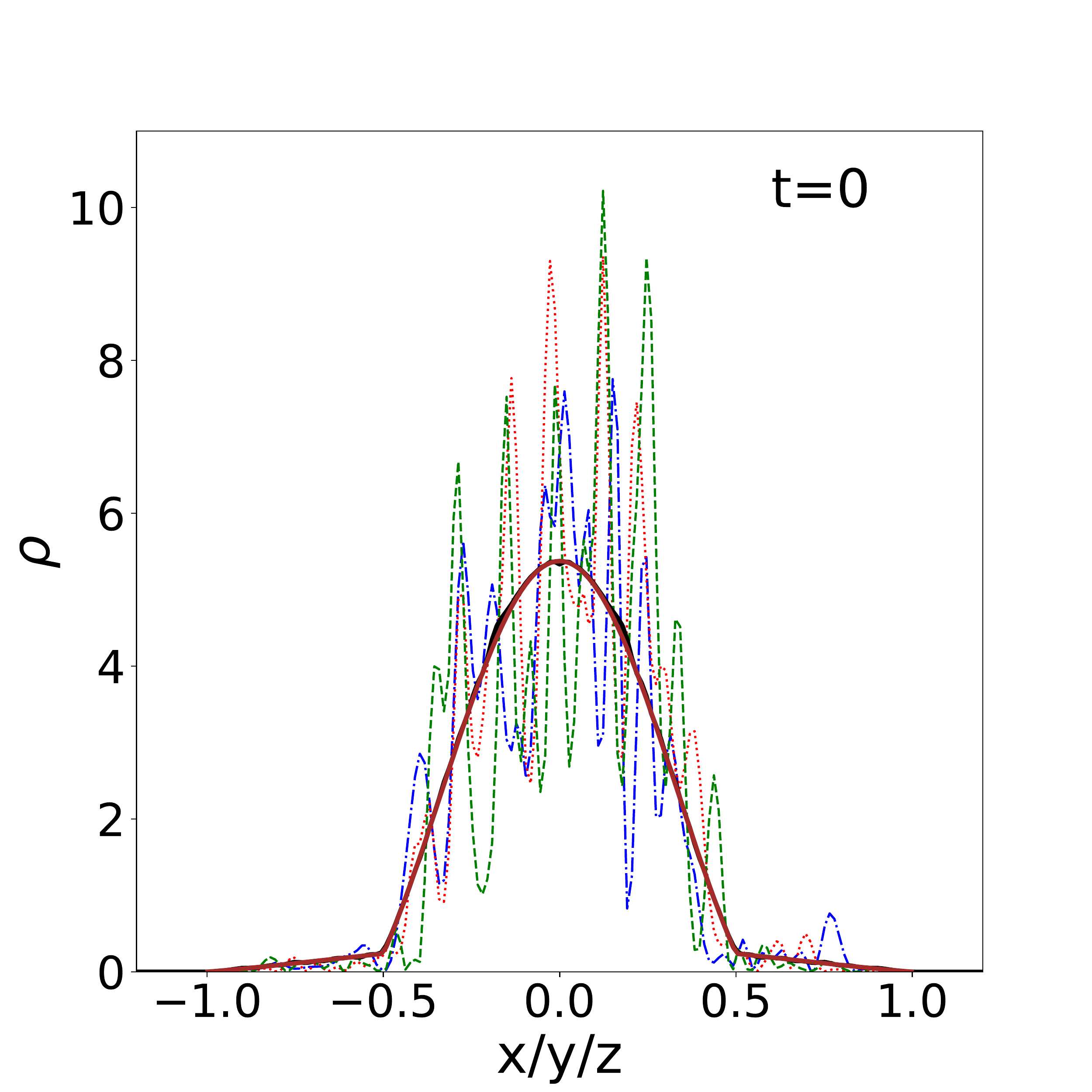}
\includegraphics[height=4.cm,width=0.25\textwidth]{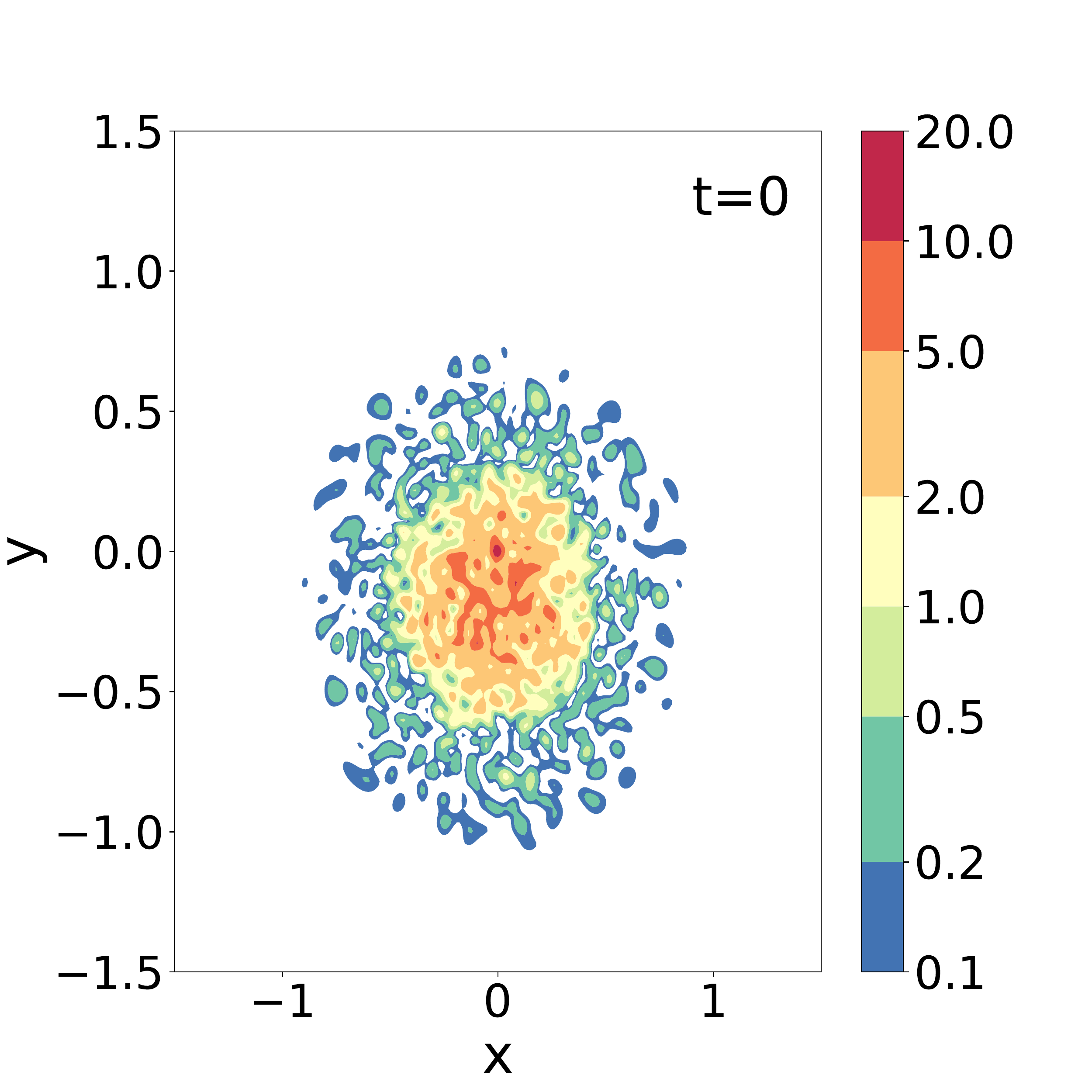}
\includegraphics[height=4.cm,width=0.24\textwidth]{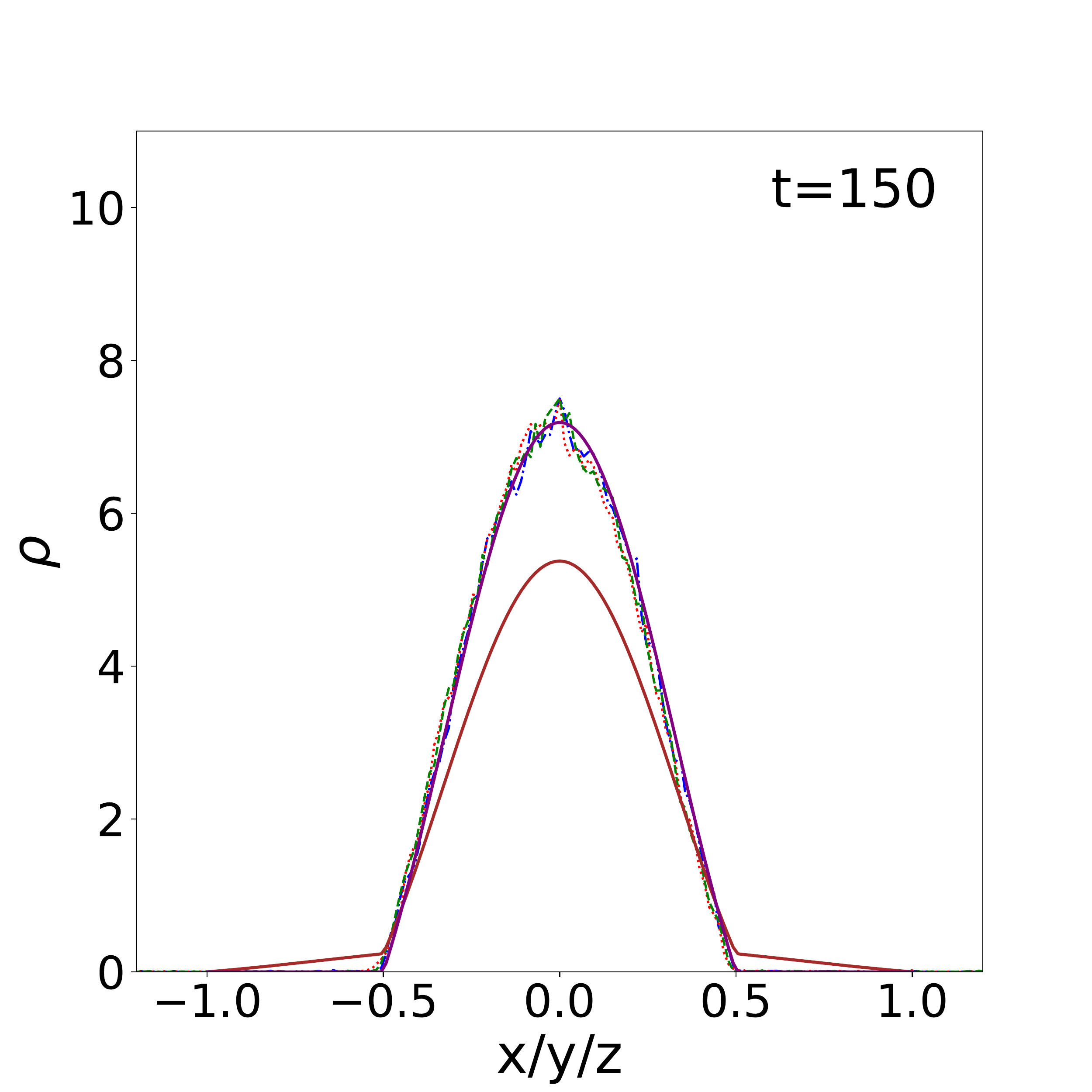}
\includegraphics[height=4.cm,width=0.25\textwidth]{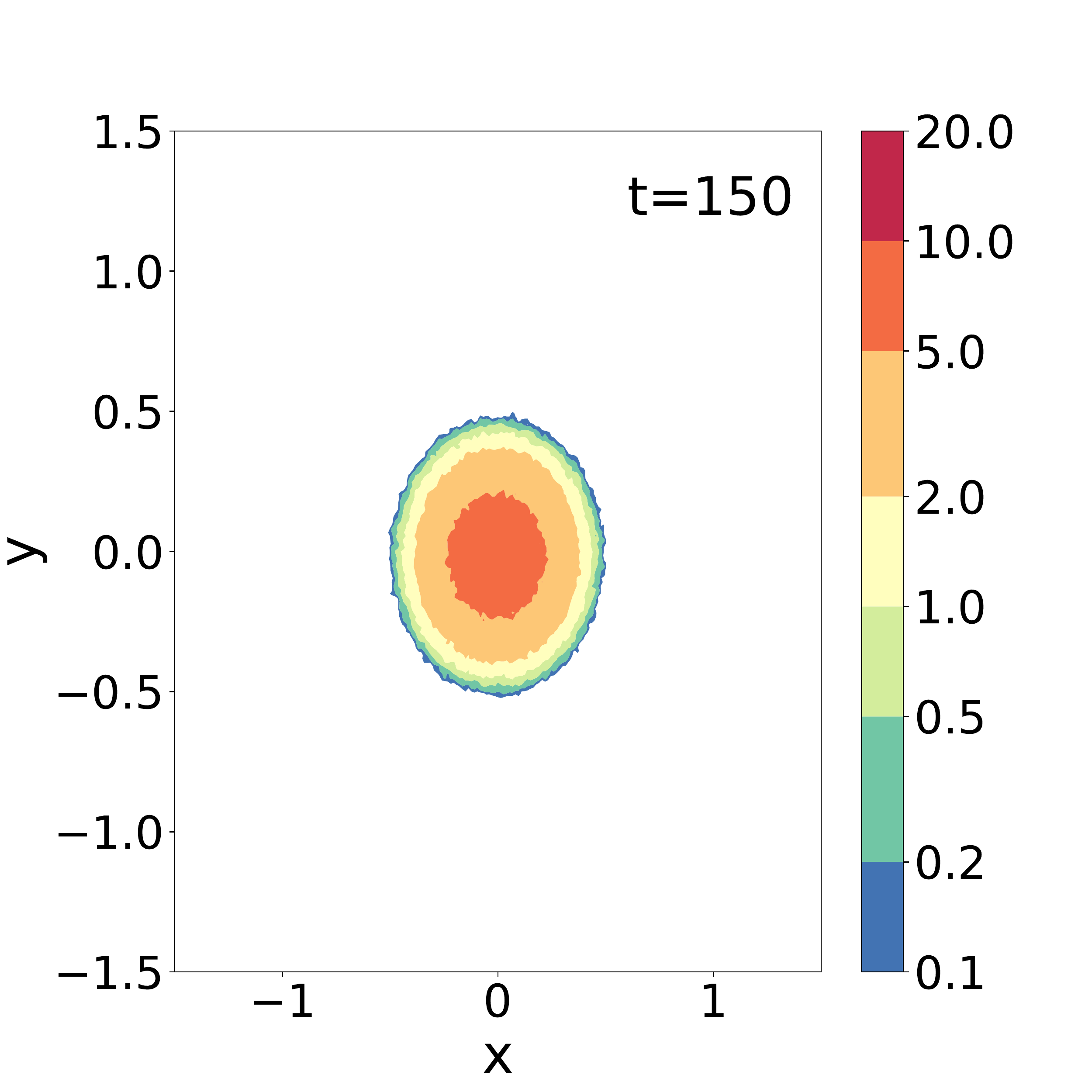}\\
\includegraphics[height=4.2cm,width=0.32\textwidth]{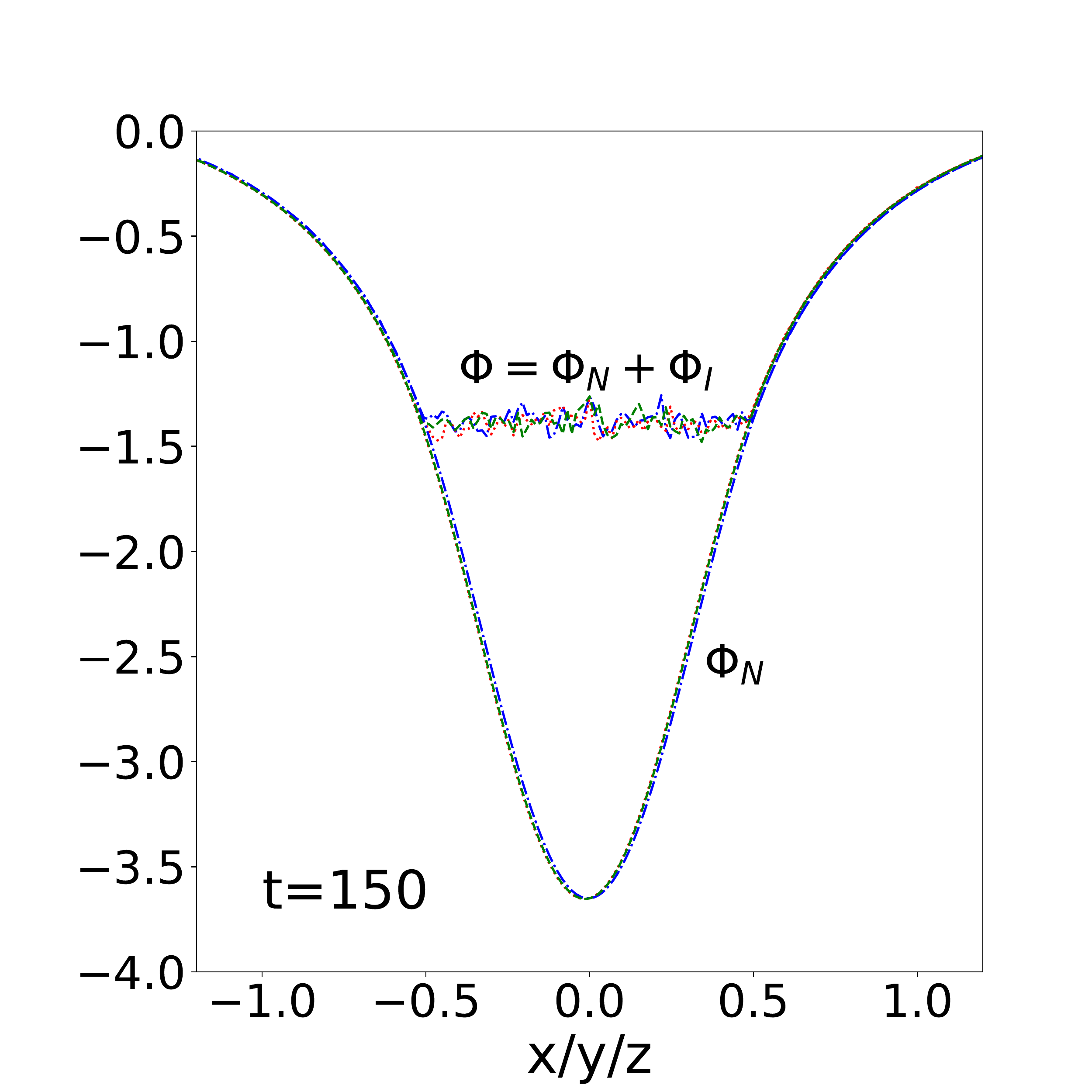}
\includegraphics[height=4.2cm,width=0.28\textwidth]{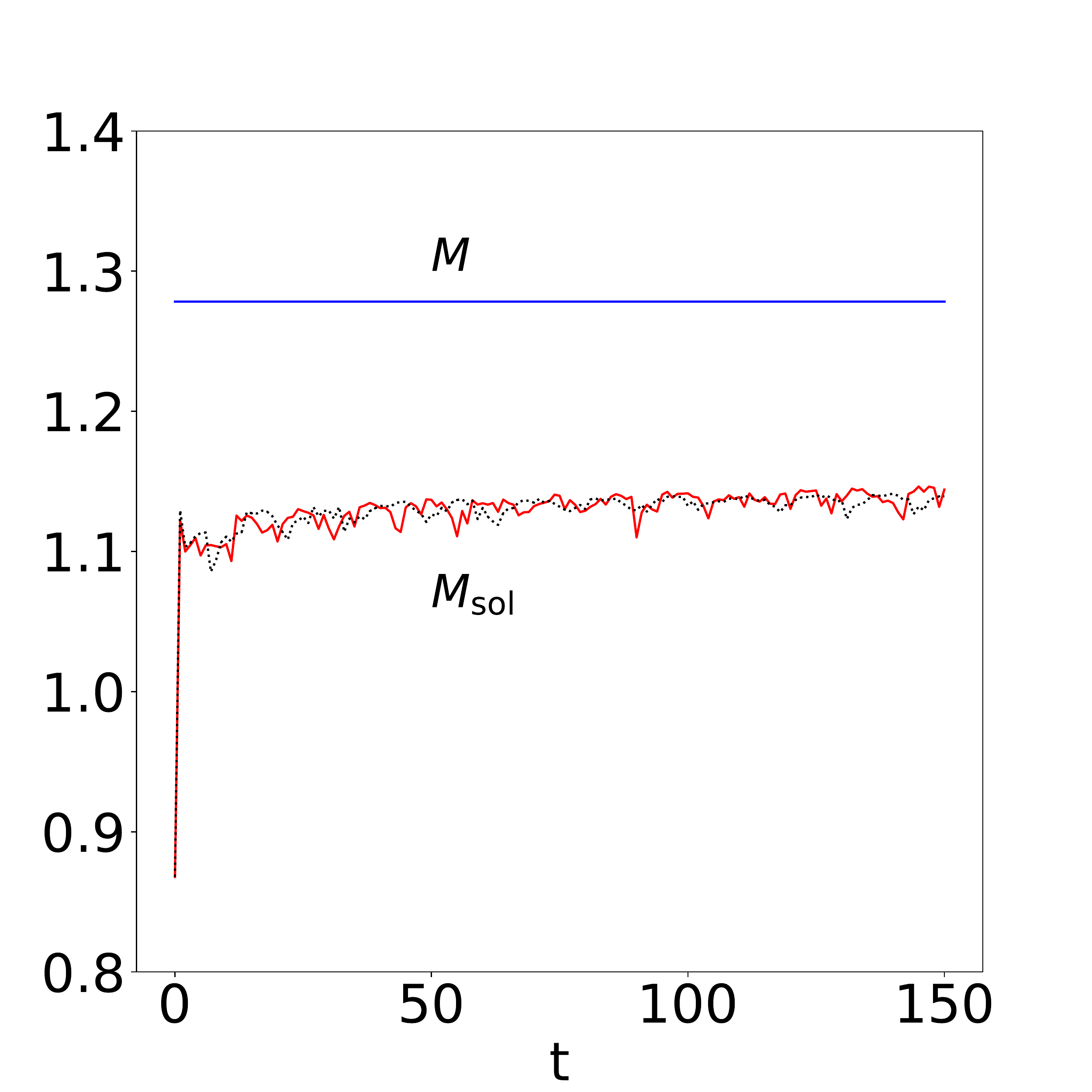}
\includegraphics[height=4.2cm,width=0.28\textwidth]{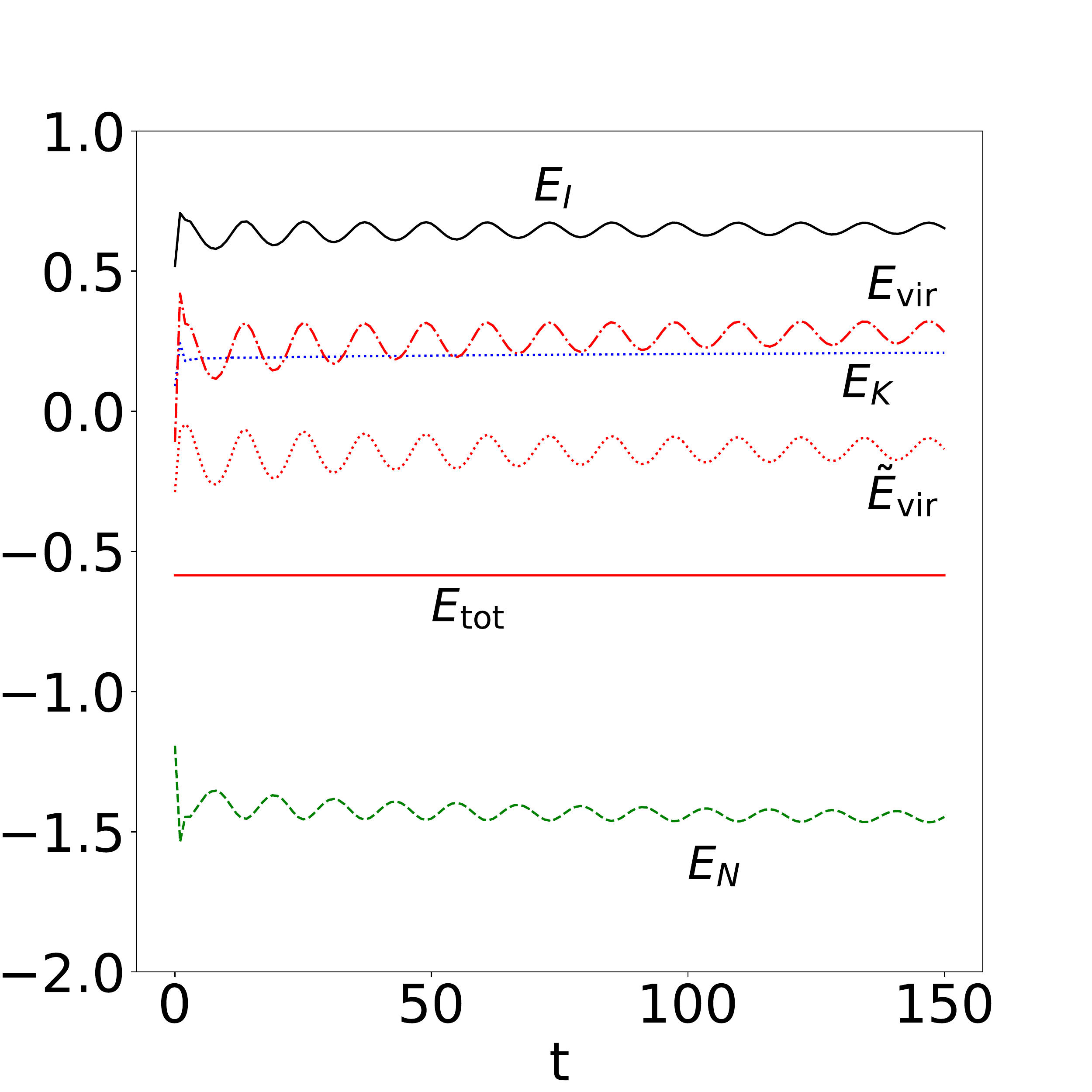}
\caption{
Same as Fig.~\ref{fig:rho-plane-sinc-R0p5-rho0}, with again $R_{\rm sol}=0.5$, but when
there is an initial soliton of density $\rho_{0{\rm sol}}=5$.
}
\label{fig:rho-plane-sinc-R0p5-rho5}
\end{figure*}

We show the evolution with time of the system in the other panels in
Fig.~\ref{fig:rho-plane-sinc-R0p5-rho0}.
Because the soliton moves somewhat around the center of the halo, at the last time
$t=150$ we show the profiles along the $x/y/z$ axis or on the 2D $(x,y)$ plane
that run through the point $\vec r_{\rm max}$ where $\rho$ is maximum
and reaches the value $\rho_{\rm max}$.
We show the final density profiles in panels (c) and (d) and the final potentials
in panel (e).
Panel (f) displays the total mass $M$ of the system and the mass $M_{\rm sol}$ enclosed
within the radius $R_{\rm sol}$ around $\vec r_{\rm max}$, as a function of time.
Panel (g) shows the evolution with time of the energy components of the system.
In our dimensionless units, they are given by
\bea
&& E_K = \frac{\epsilon^2}{2} \int d\vec x \, \nabla\psi \cdot \nabla \psi^* , \;\;\;
E_N = \frac{1}{2}  \int d\vec x \, \rho \Phi_N , \nonumber \\
&& E_I =  \int d\vec x \, {\cal V}_I = \frac{\lambda}{2} \int d\vec x \, \rho^2 .
\label{eq:E-def}
\eea
The total mass $M$ and the total energy $E_{\rm tot}=E_K+E_N+E_I$ are conserved by the
equation of motion and we can check that they are conserved in the numerical simulations
until the final time shown in the figures.
As we consider compact halos $R_{\rm halo}=1$, the density is very low at large radii and
negligible matter amounts can reach the boundaries of the simulation box.
The kinetic energy $E_K$ comes with a prefactor $\epsilon^2$ in Eq.(\ref{eq:E-def}).
This means that in the semiclassical limit, $\epsilon \to 0$, it is negligible for smooth static
configurations such as the equilibrium soliton (\ref{eq:hydrostatic-alpha}), which is thus governed
by the balance between gravity and self-interactions.
However, this is not the case for the halo for two reasons. First, nonzero orbital velocities $\vec v$
of order unity (i.e., of the order of the virial velocity) correspond locally to a phase
$e^{i \vec v \cdot \vec x /\epsilon}$ in the wavefunction (i.e., orbital quantum number $\ell \sim 1/\epsilon$).
This implies that $\psi$ shows large gradients that grow as $1/\epsilon$ and balance
the prefactor $\epsilon^2$ in Eq.(\ref{eq:E-def}).
Second, as seen above the halo also shows strong density fluctuations on a spatial
width $\Delta x \sim \epsilon$, which again lead to large gradients that balance
the prefactor $\epsilon^2$.
Therefore, even for small $\epsilon$ the wavelike nature of the system, governed by the
Schr\"odinger equation rather than by the hydrodynamical Euler equation, remains important.

We can see that within a few dynamical times, $t \lesssim 8$, the system
reaches a quasi-stationnary state where about half of the mass is contained in
a central soliton that follows the profile (\ref{eq:rho-soliton}).
Afterwards, the soliton mass and the energies only show a slow evolution.
The central equilibrium soliton is clearly seen on the density profiles shown
in panels (c) and (d), with its radius $R_{\rm sol}=0.5$.
In panel (c), the purple solid line shows the density profile of a soliton of mass $M_{\rm sol}(t)$
given by the enclosed mass within radius $R_{\rm sol}$ around the maximum density point
$\vec r_{\rm max}$.
We can see that the density profile obtained in the simulation closely follows this analytical shape
(\ref{eq:rho-soliton}) and that $M_{\rm sol}(t)$ is indeed a good approximation of the soliton mass.
Superimposed on this soliton, there remains a depleted halo, with the remaining
half of the initial mass, with again relative fluctuations of order unity
as in in Eq.(\ref{eq:Proba-rho-halo}). The fluctuations are
somewhat lower than in the initial state as the halo mass has been decreased by half.
The hydrostatic equilibrium (\ref{eq:hydrostatic-alpha}) is also clearly
shown in panel (e) by the constant plateau of the total potential
$\Phi=\Phi_N+\Phi_I$ over the extent of the soliton, $r \leq 0.5$
(with small wiggles associated with the excited halo modes that cross the
central region).
Outside of the soliton, the rapid decrease of the density means that
the self-interaction potential $\Phi_I$ becomes small as compared with the
gravitational potential $\Phi_N$ and $\Phi \simeq \Phi_N$. 
This is why we can only distinguish $\Phi_I$ from $\Phi_N$ in the figure in the soliton
domain $r \leq 0.5$.

In the panel (f) we show the growth of the soliton mass $M_{\rm sol}(t)$ for our fiducial run
(red solid line) and for a simulation with a box size that is twice larger (black dotted line).
We can see that both cases recover the growth of the soliton to about half the system mass
in a few dynamical time, although the greater box gives a growth rate that is smaller at late times.
Therefore, our main result on the fast formation of the soliton is robust but the growth rate at late times may be somewhat overestimated because of the small box
size and the periodic boundary conditions (i.e., the oscillations of the other halos in the neighboring
boxes might increase the soliton growth rate).

We can see in the last panel that the energy components (\ref{eq:E-def}) show a fast
variation during the formation stage of the soliton and afterwards only show a slow and
monotonic evolution, in parallel with the slow growth of the soliton.
We also display the virial quantities
\be
E_{\rm vir} = 2 E_K + E_N + 3 E_I , \;\;\;
\tilde{E}_{\rm vir} =  E_N + 3 E_I .
\ee
To mitigate the impact of the periodic boundary conditions, in this sum we renormalize the
gravitational energy by a constant offset, as $E_N \to E_N + \Delta E_{N {\rm max}}$, 
so that $E_N(\vec r_{\rm max})$ is the gravitational energy at the point $\vec r_{\rm max}$
that would be computed for an isolated system (i.e. we compute
$\Phi_N = - \int d\vec r^{\,'} \rho(\vec r^{\, '})/|\vec r^{\, '}-\vec r|$ instead of solving the Poisson
equation by the Fourier transform).
According to the virial theorem, we expect $E_{\rm vir}=0$ for a fully relaxed system.
As the other energy components, $E_{\rm vir}$ shows an early fast variation and next a slow growth.
This slow time dependence is due to the slow growth of the soliton mass.
However, $E_{\rm vir}$ does not seem to converge towards zero until the end of the simulation,
whereas $\tilde{E}_{\rm vir}$ is much closer to zero.
This shows the strong impact of the kinetic energy $E_K$ on $E_{\rm vir}$.
This can be understood as follows. For a fully relaxed soliton, we have $E_{\rm vir}=0$ and
$\tilde{E}_{\rm vir}\simeq 0$, as $E_K \simeq 0$ in the semiclassical regime that we consider
with $\epsilon \ll 1$.
At the final time $t=150$ shown in Fig.~\ref{fig:rho-plane-sinc-R0p5-rho0}, about half the mass
of the system is contained in the central soliton. It also contains most of the self-interaction energy
and of the gravitational energy, as seen in panel (e). This means that $\tilde{E}_{\rm vir}\simeq 0$
as it is then approximately given by the soliton energies. On the other hand, as seen in panel
(g) the kinetic energy is not negligible and $E_{\rm vir} \simeq 2 E_K$. Therefore, the violation
of the virial equilibrium condition $E_{\rm vir}=0$ is mostly due to the fluctuations of the density field
around the mean soliton profile. Because of their small finite width, $\Delta x \sim 1/\epsilon$,
they still give rise to large density gradients and a significant kinetic energy.
If the soliton finally manages to eat all the mass of the system, then we will eventually have
$E_{\rm vir}=0$ and $E_K \simeq 0$. However, the finite time of our simulations and the increasingly
slow growth of the soliton do not permit us to conclude whether all the mass will eventually be absorbed
by the soliton or a small amount of fluctuations will remain.

\subsubsection{Initial soliton $\rho_{0\rm sol}=5$}

We now consider the case where there is an initial soliton of density
$\rho_{0{\rm sol}}=5$
on top of the halo profile.
This initial condition is shown in panels (a) and (b) in
Fig.~\ref{fig:rho-plane-sinc-R0p5-rho5}.
We can see that very quickly, in a few dynamical times $t \lesssim 2$,
the mass of the soliton grows to about $85\%$ of the total mass and seems
to remain stable thereafter. This decreases the halo density and the amplitude
of the density fluctuations, as compared with the initial state.
Again, this process is clearly apparent in the shape of the potential $\Phi$
and the density maps shown in panels (c) and (d).
They clearly display the smoothing of the density field over the soliton extent and the
damping of the initial fluctuations.

We can check in panels (f) and (g) that the total mass and energy of the systems are
conserved by the numerical simulation. We now find that the soliton mass is the same
for the larger box simulation. Therefore, this regime is robust with
respect to finite box size effects. The energy components show small decaying collective oscillations,
left by the early sudden growth of the soliton. Again, $\tilde{E}_{\rm vir}$ is closer to zero than
$E_{\rm vir}$ but their difference is smaller than for the case shown in 
Fig.~\ref{fig:rho-plane-sinc-R0p5-rho0} because of the lower kinetic energy $E_K$, as the 
fluctuations are much smaller.

Together with the results of Fig.~\ref{fig:rho-plane-sinc-R0p5-rho0}, this shows
that the soliton is to some degree an attractor of the dynamics, when
$R_{\rm sol}$ is not much below the size of the system.
It appears from a random initial state to make $50\%$ of the total mass,
as in Fig.~\ref{fig:rho-plane-sinc-R0p5-rho0}, or can grow to larger values
if it is already present with a significant mass, as in
Fig.~\ref{fig:rho-plane-sinc-R0p5-rho5}.
The latter results also suggest that the soliton does not grow to capture all the
mass of the system. However, this numerical simulation cannot rule out secular
effects that would become manifest on timescales that are much greater than the
dynamical time and the time of our simulations.

\begin{figure*}[ht]
\centering
\includegraphics[height=5.cm,width=0.325\textwidth]{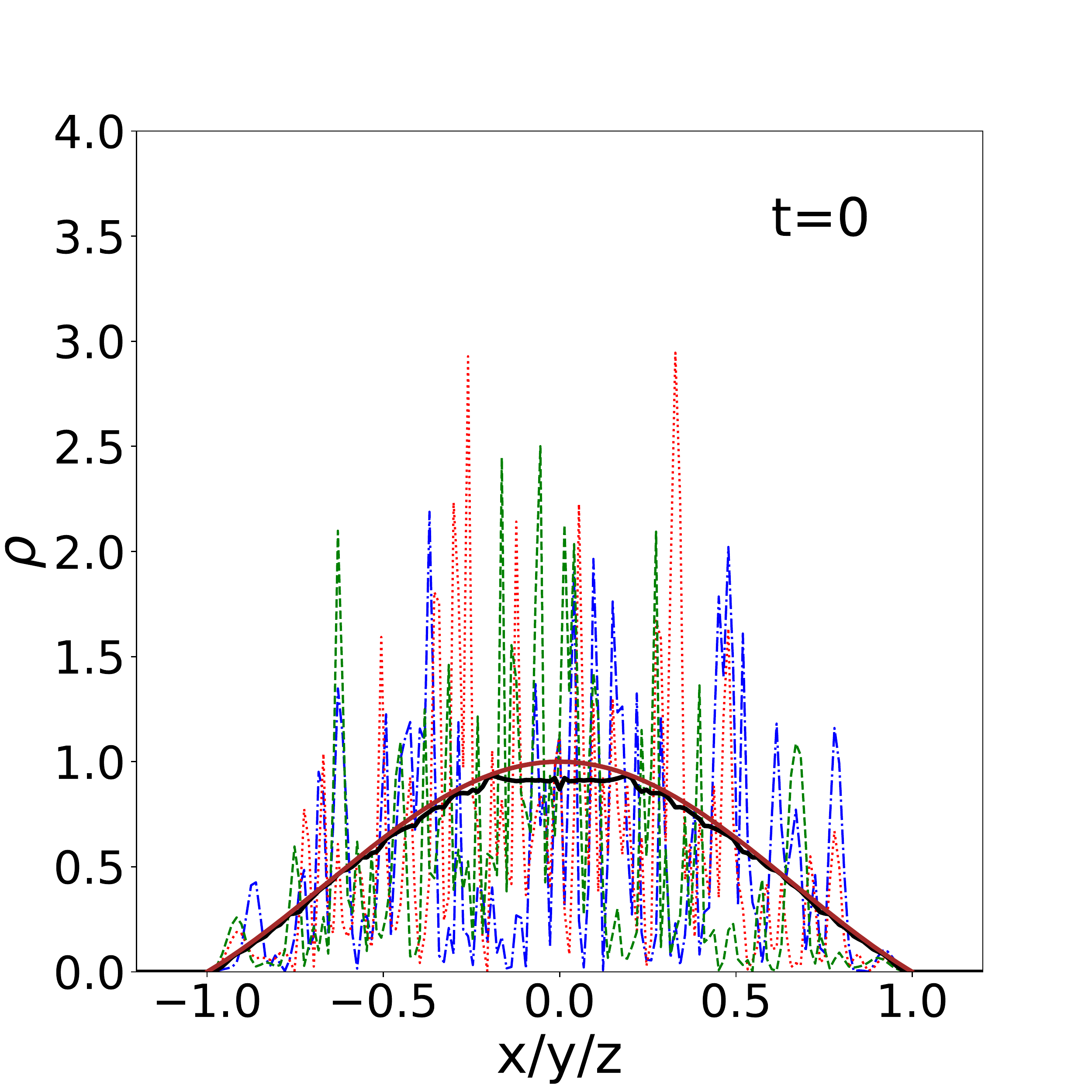}
\includegraphics[height=5.cm,width=0.325\textwidth]{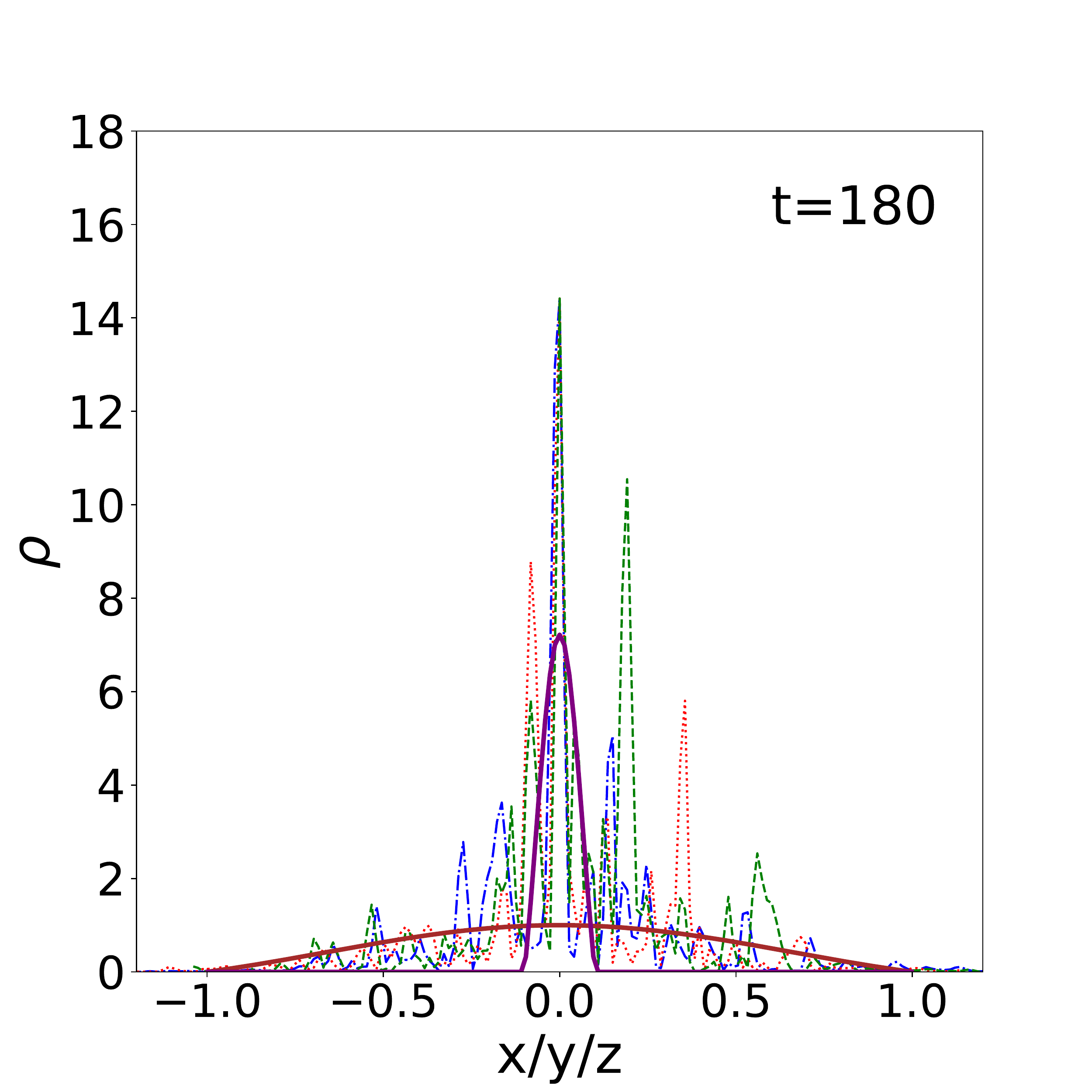}
\includegraphics[height=5.cm,width=0.325\textwidth]{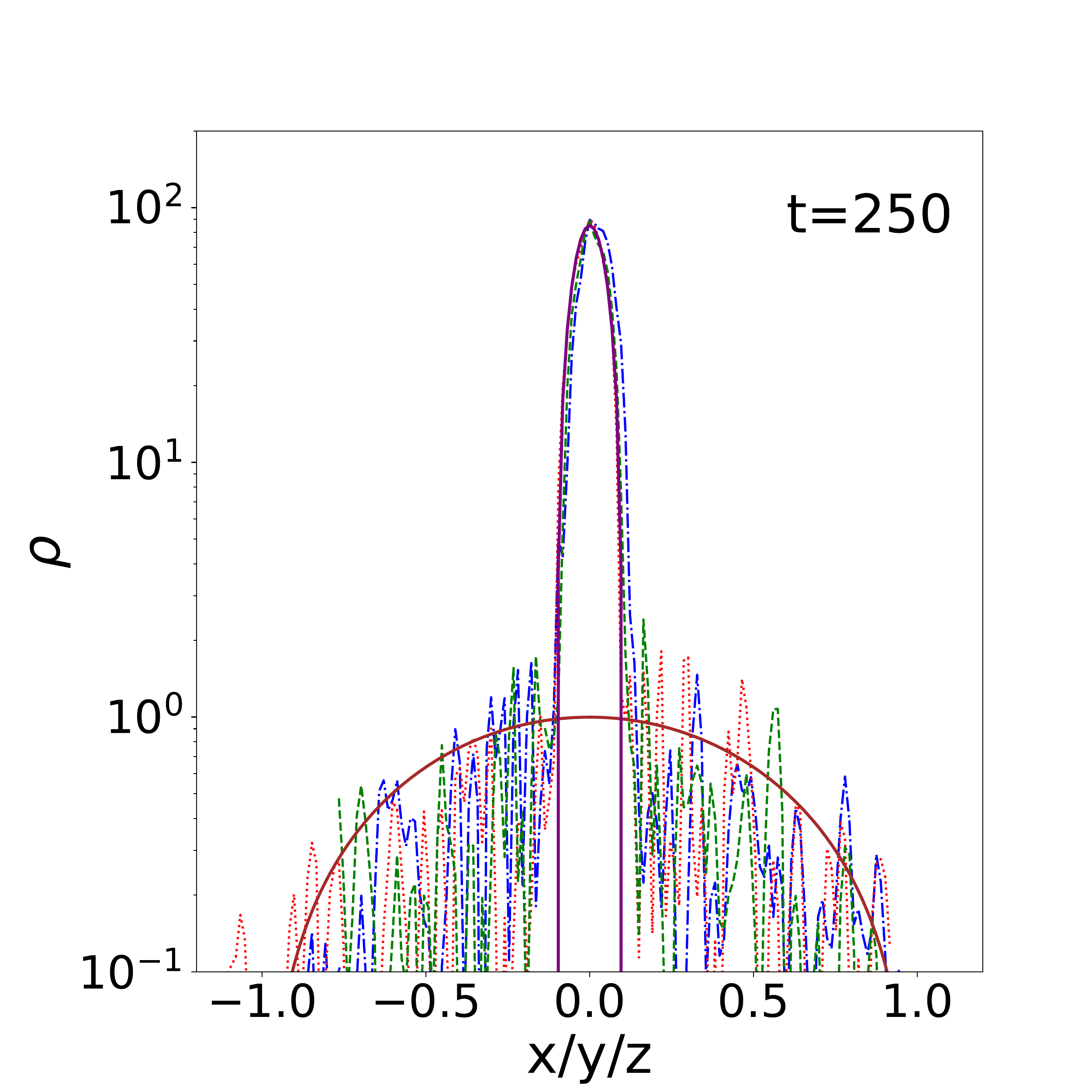}\\
\includegraphics[height=5.cm,width=0.325\textwidth]{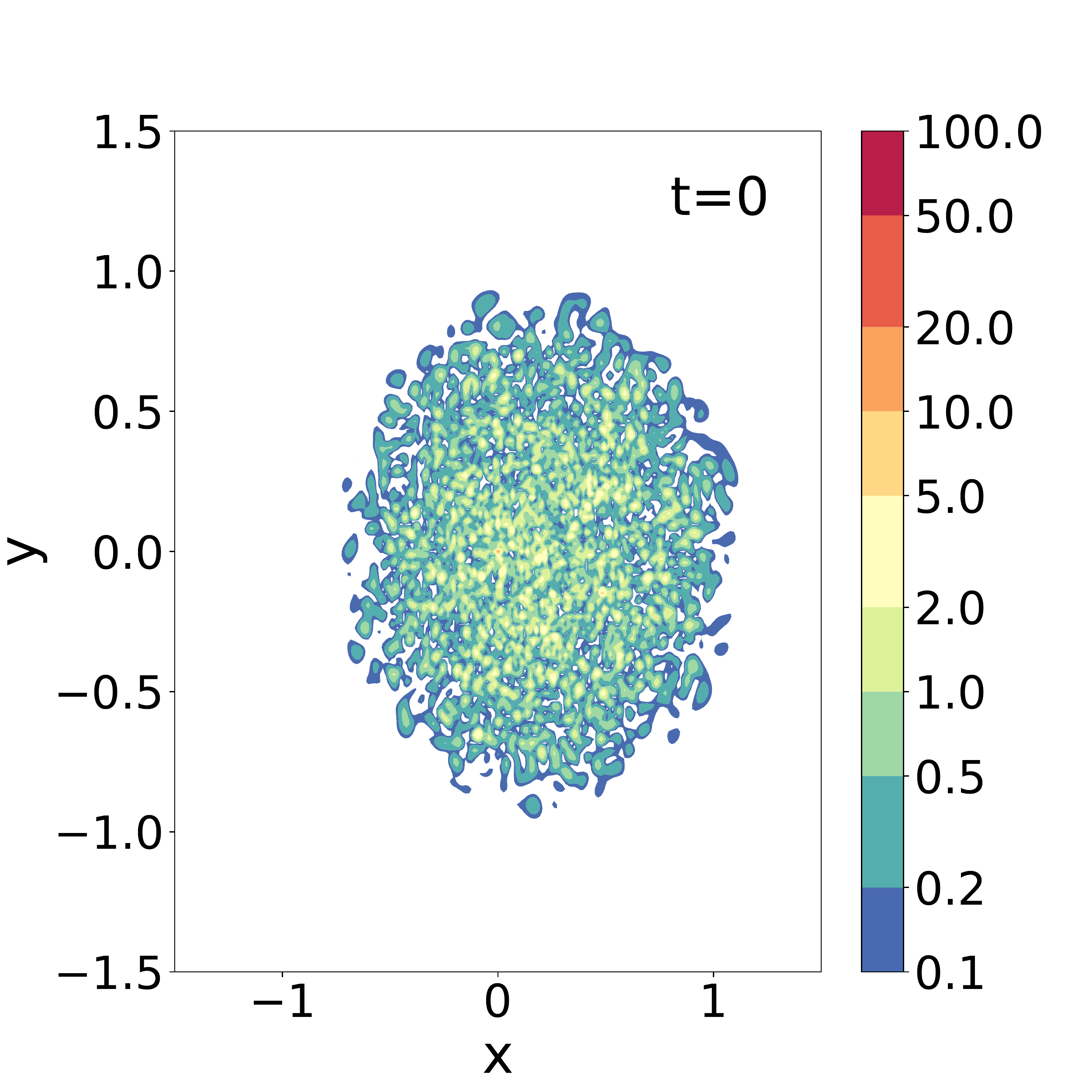}
\includegraphics[height=5.cm,width=0.325\textwidth]{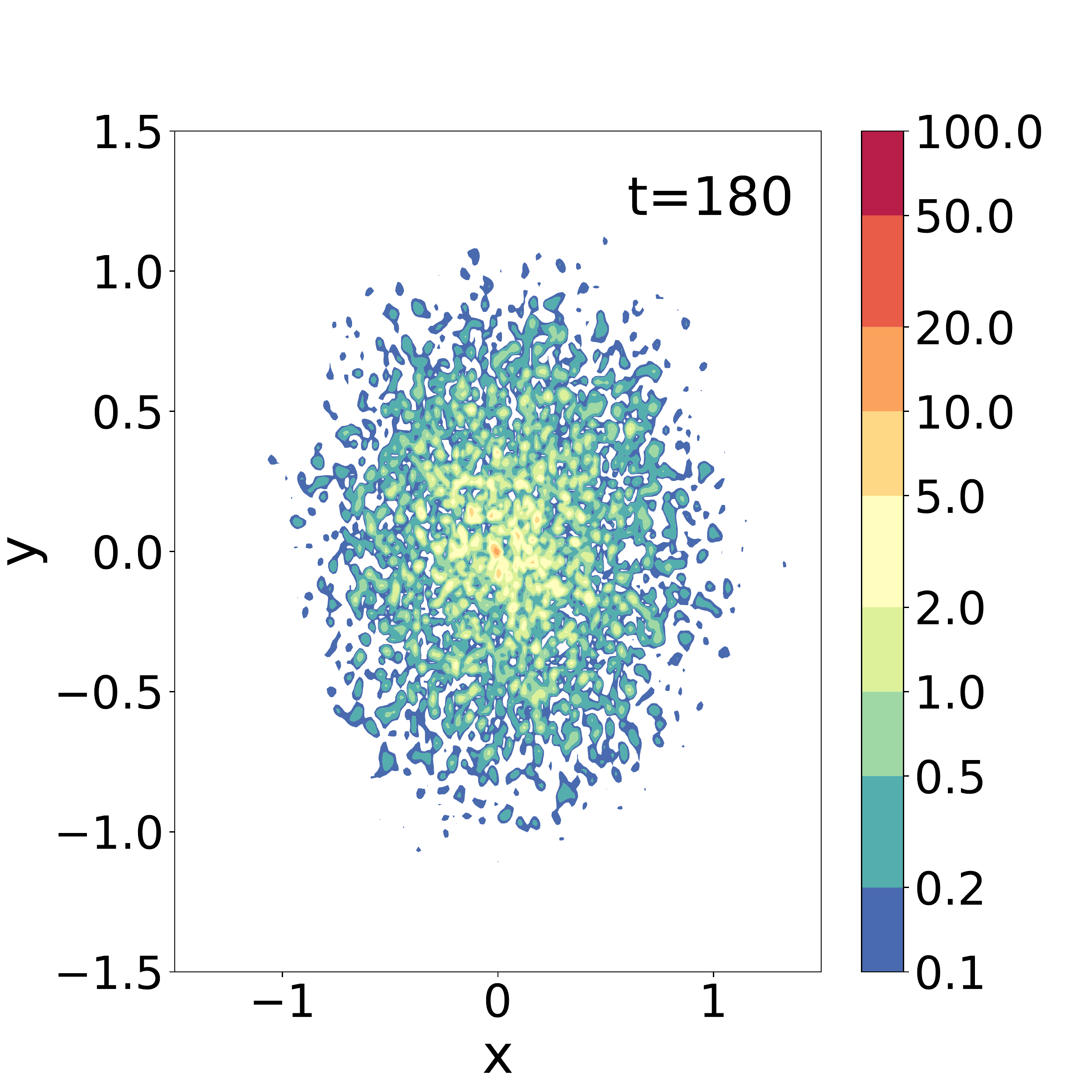}
\includegraphics[height=5.cm,width=0.325\textwidth]{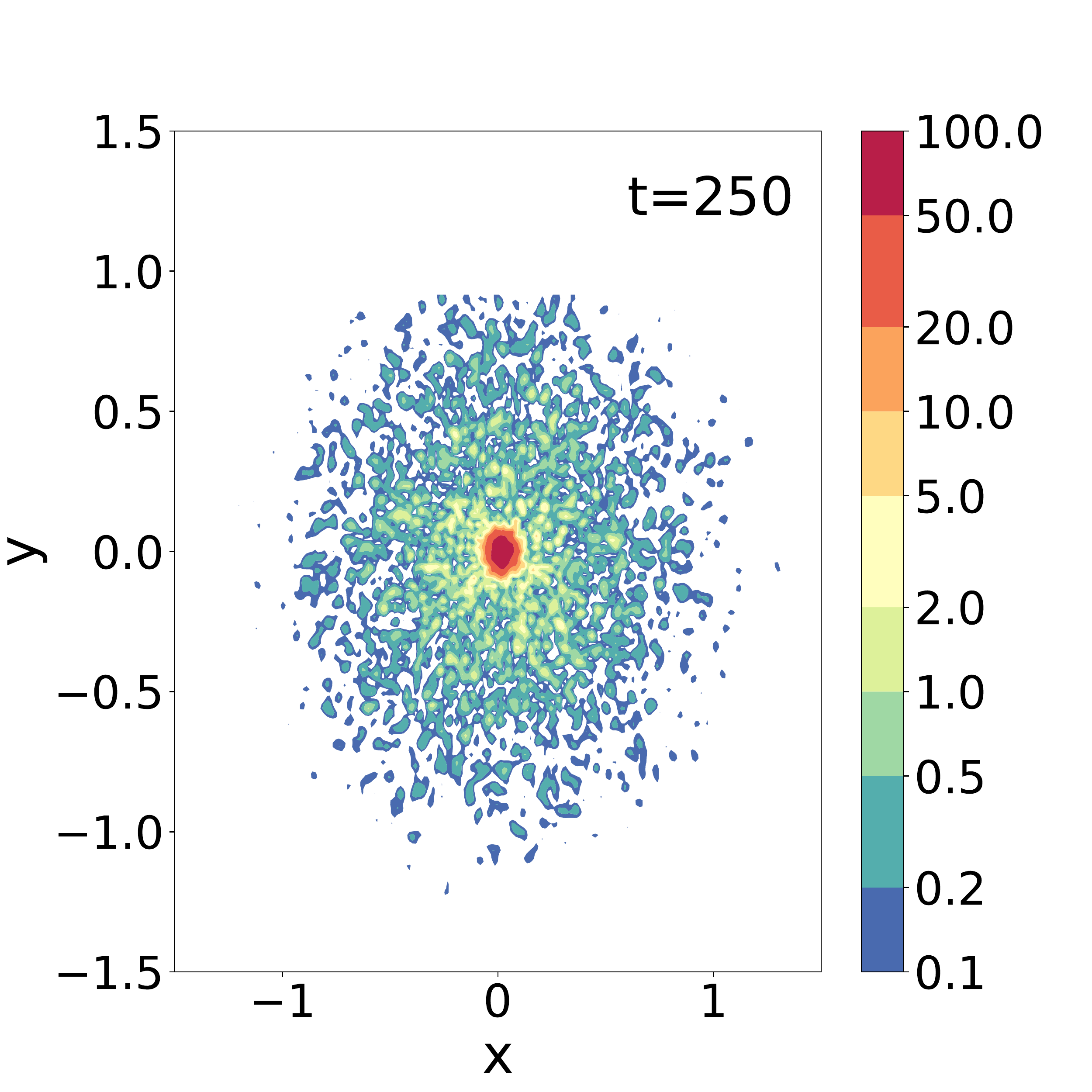}\\
\includegraphics[height=4.2cm,width=0.25\textwidth]{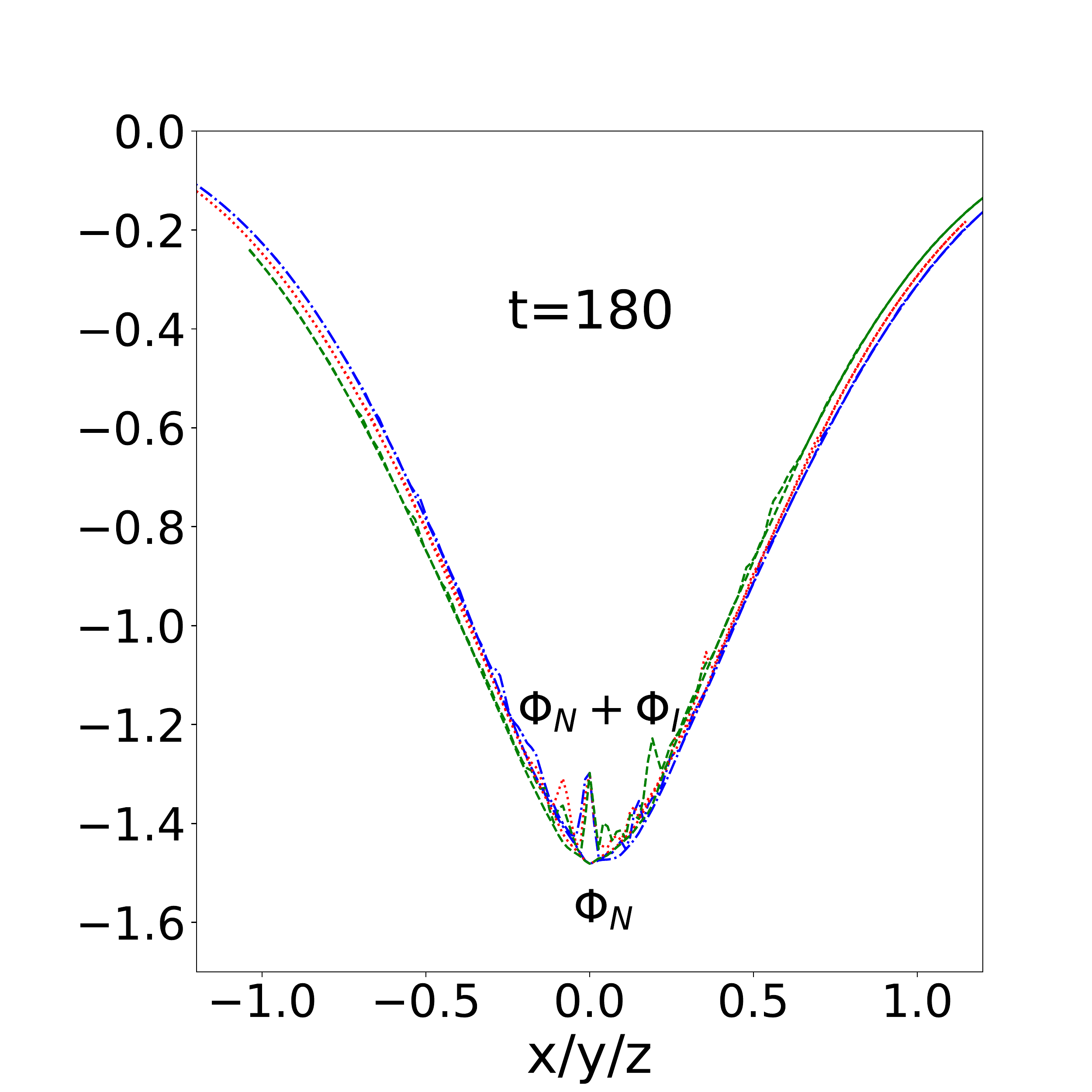}
\includegraphics[height=4.2cm,width=0.25\textwidth]{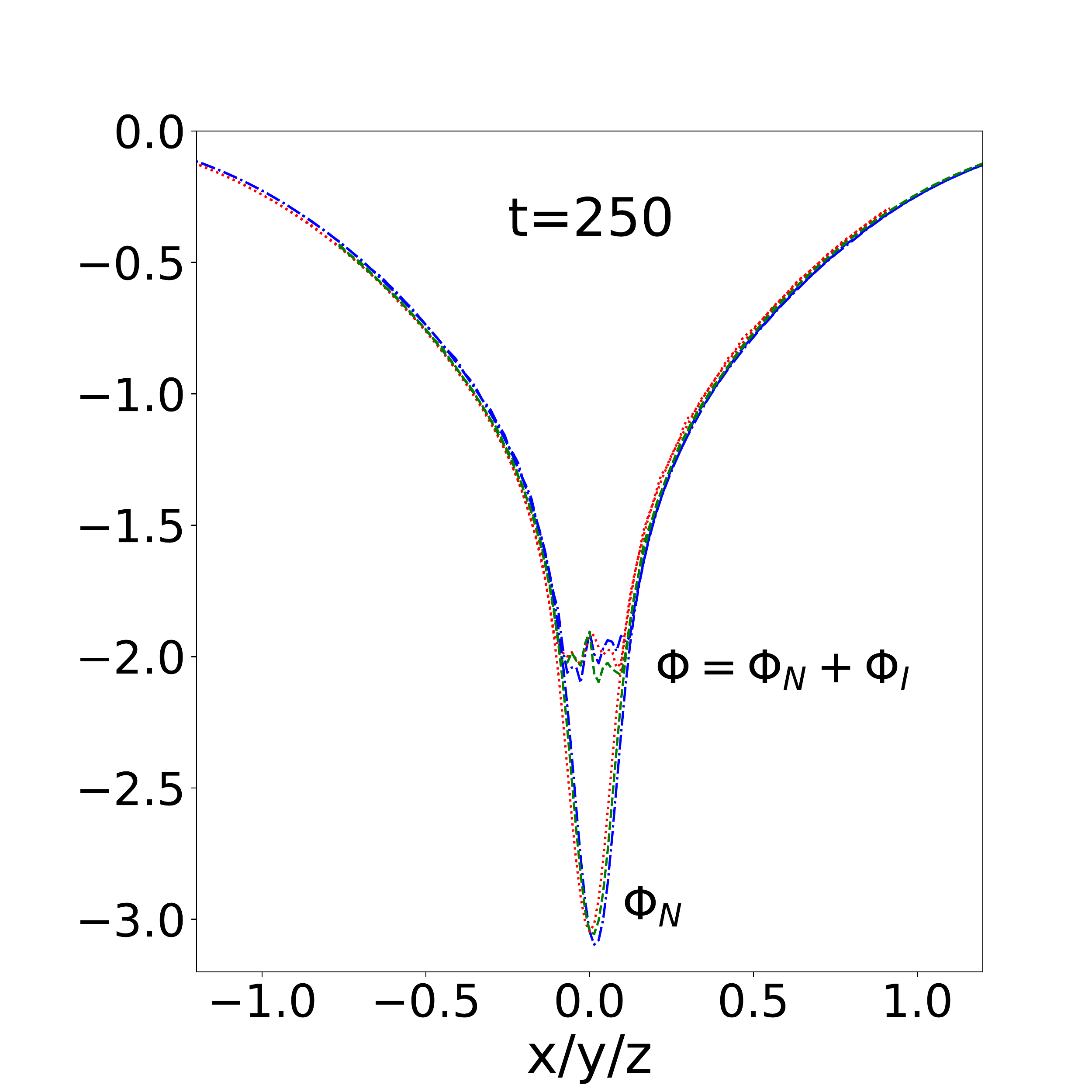}
\includegraphics[height=4.2cm,width=0.24\textwidth]{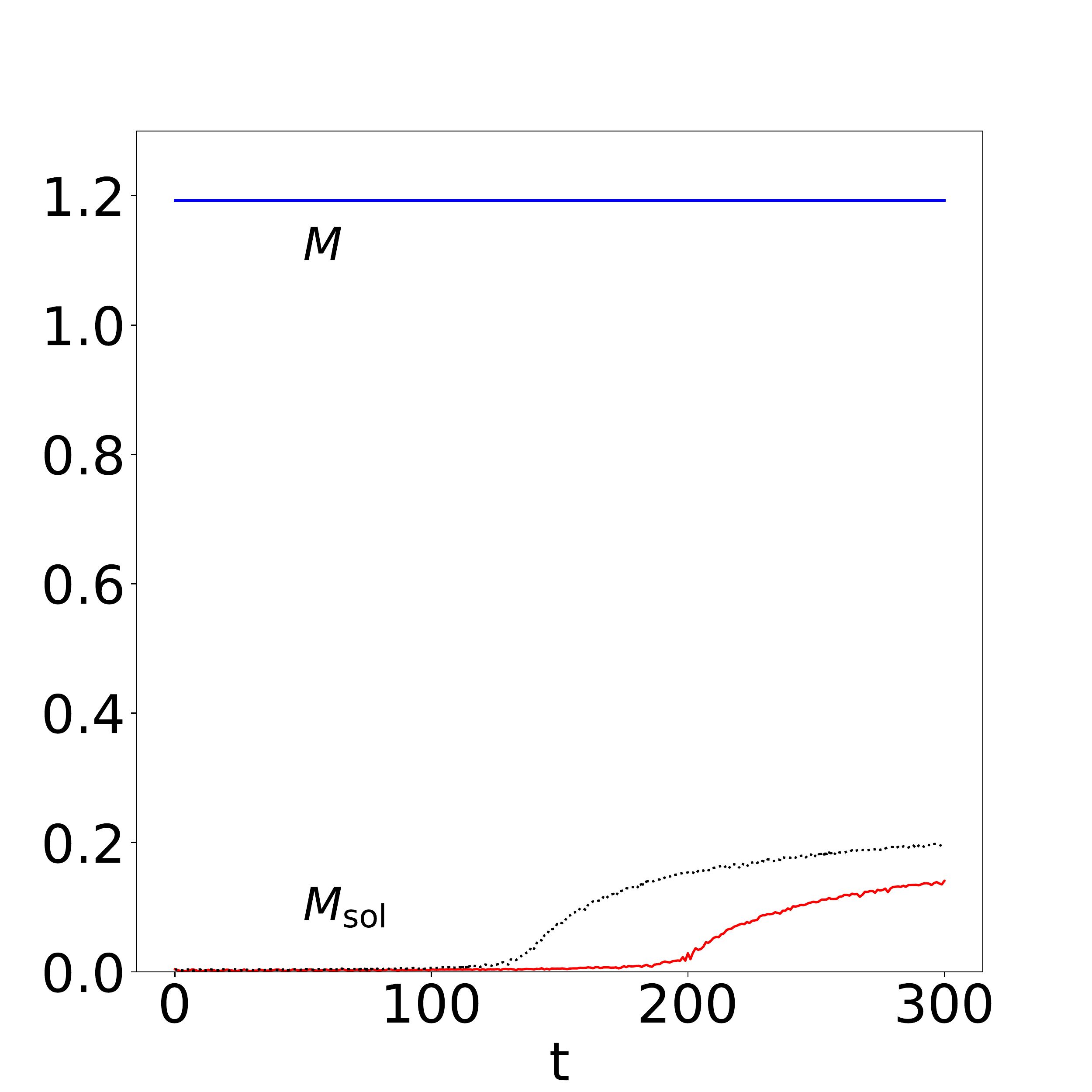}
\includegraphics[height=4.2cm,width=0.24\textwidth]{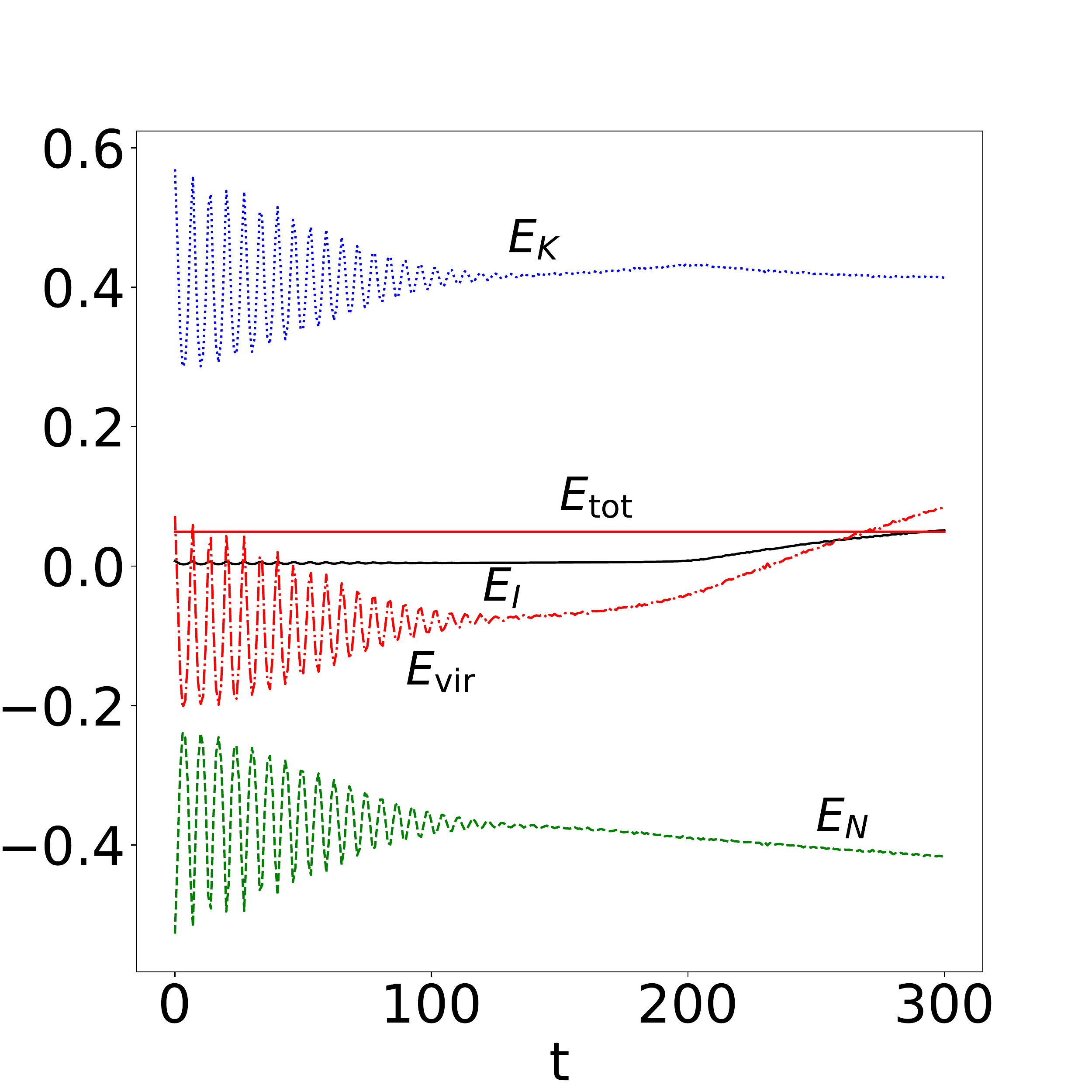}
\caption{
Evolution of a flat system with $R_{\rm sol}=0.1, \;\; \rho_{0\rm sol}=0$.
}
\label{fig:rho-plane-sinc-R0p1-rho0}
\end{figure*}

\begin{figure*}[ht]
\centering
\includegraphics[height=5.cm,width=0.325\textwidth]{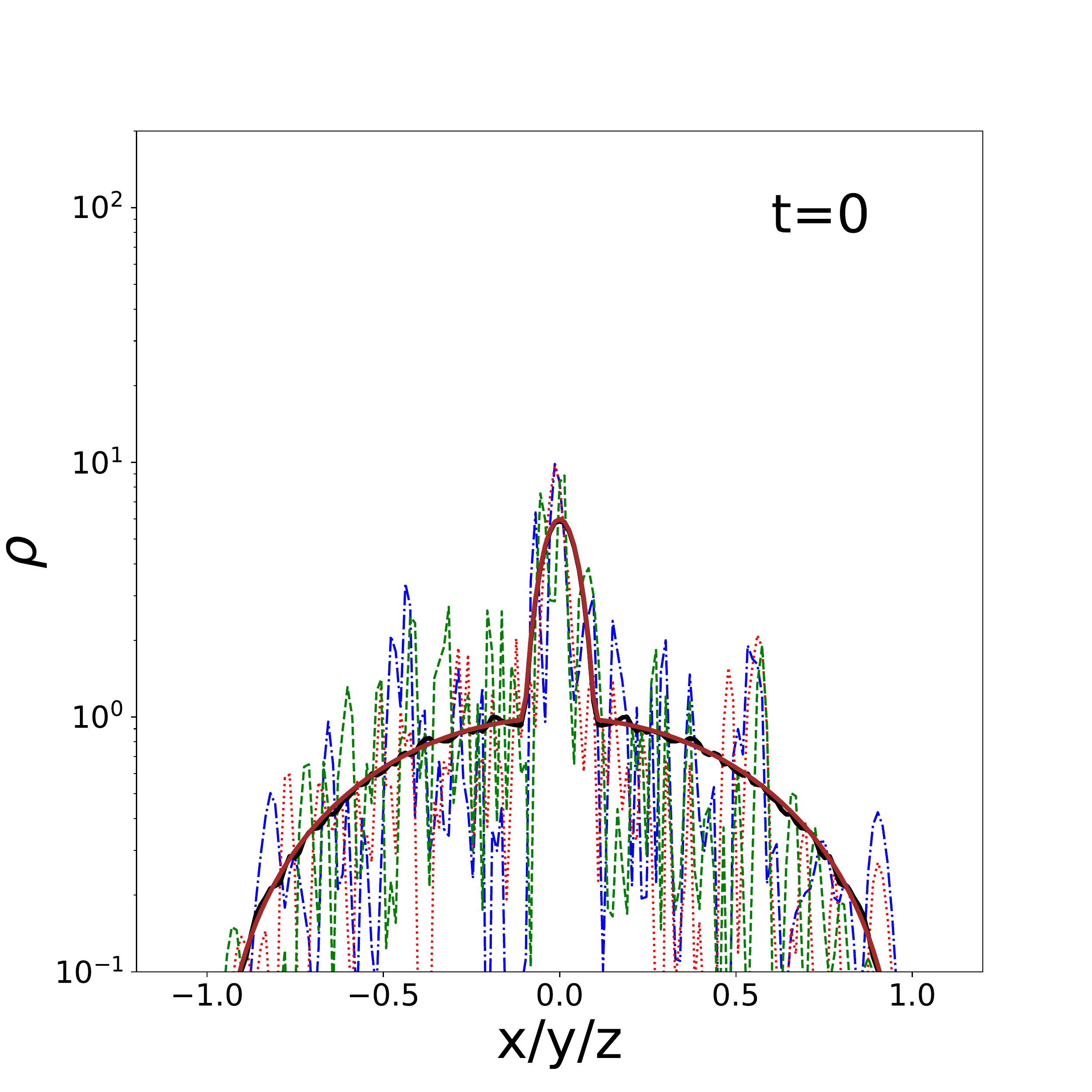}
\includegraphics[height=5.cm,width=0.325\textwidth]{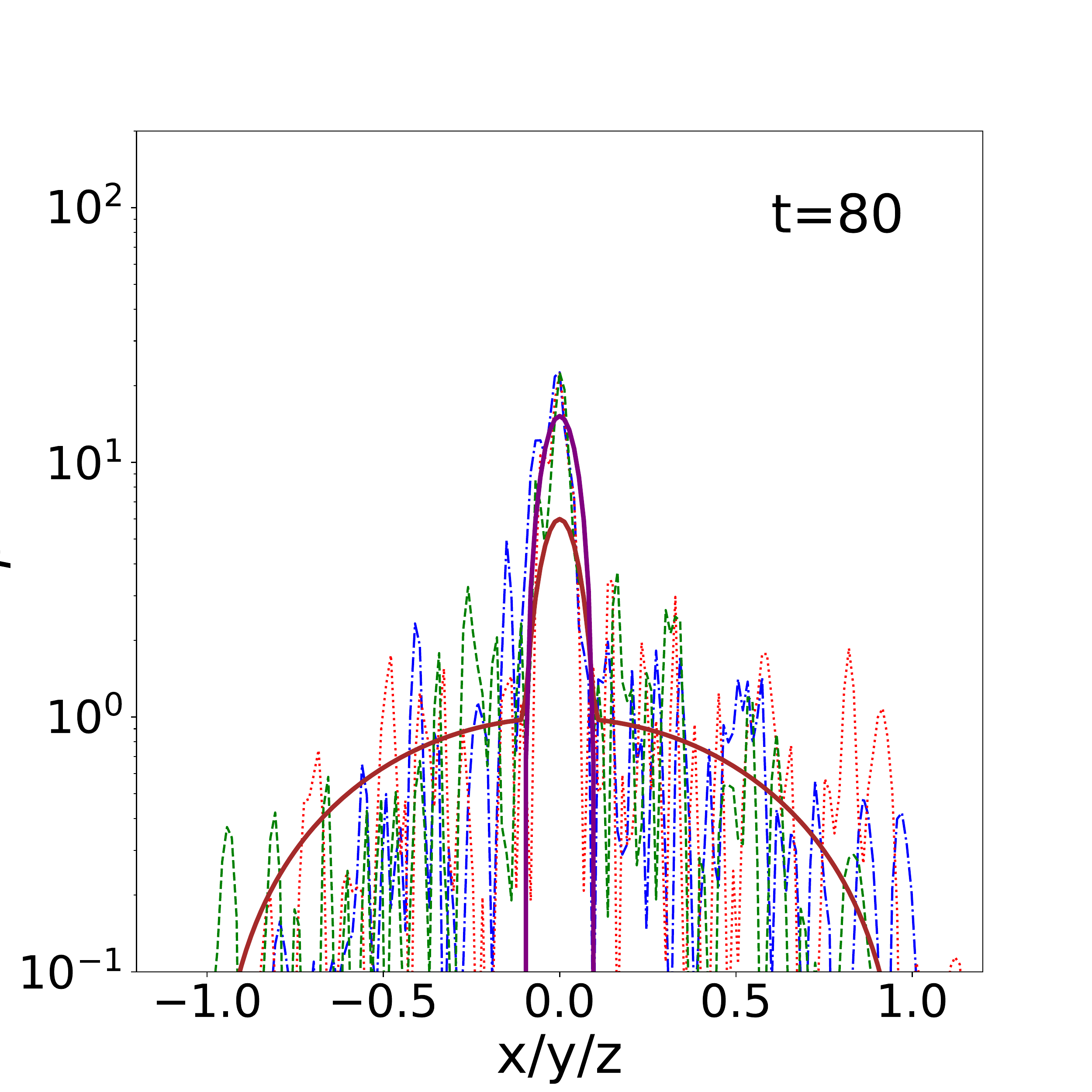}
\includegraphics[height=5.cm,width=0.325\textwidth]{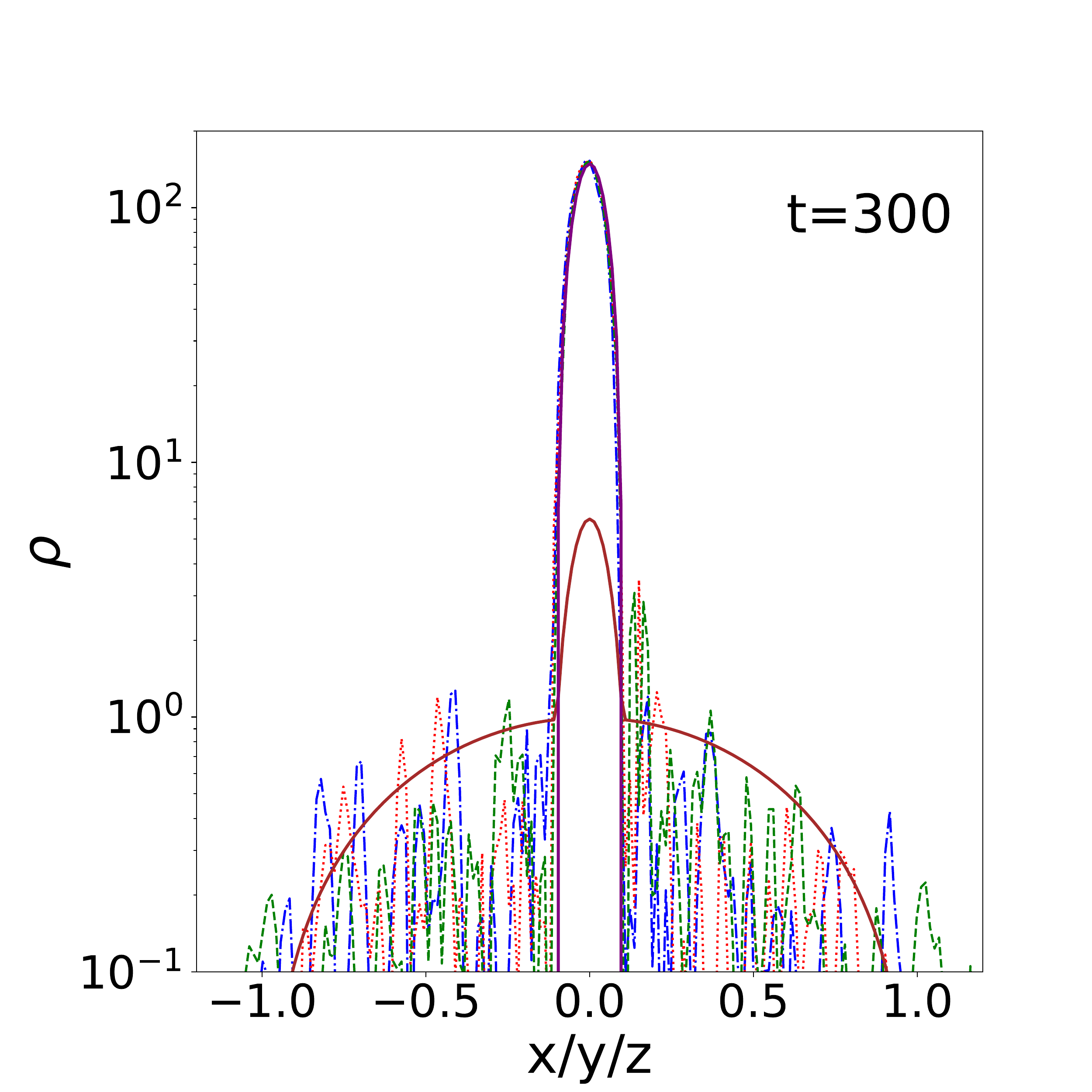}\\
\includegraphics[height=5.cm,width=0.325\textwidth]{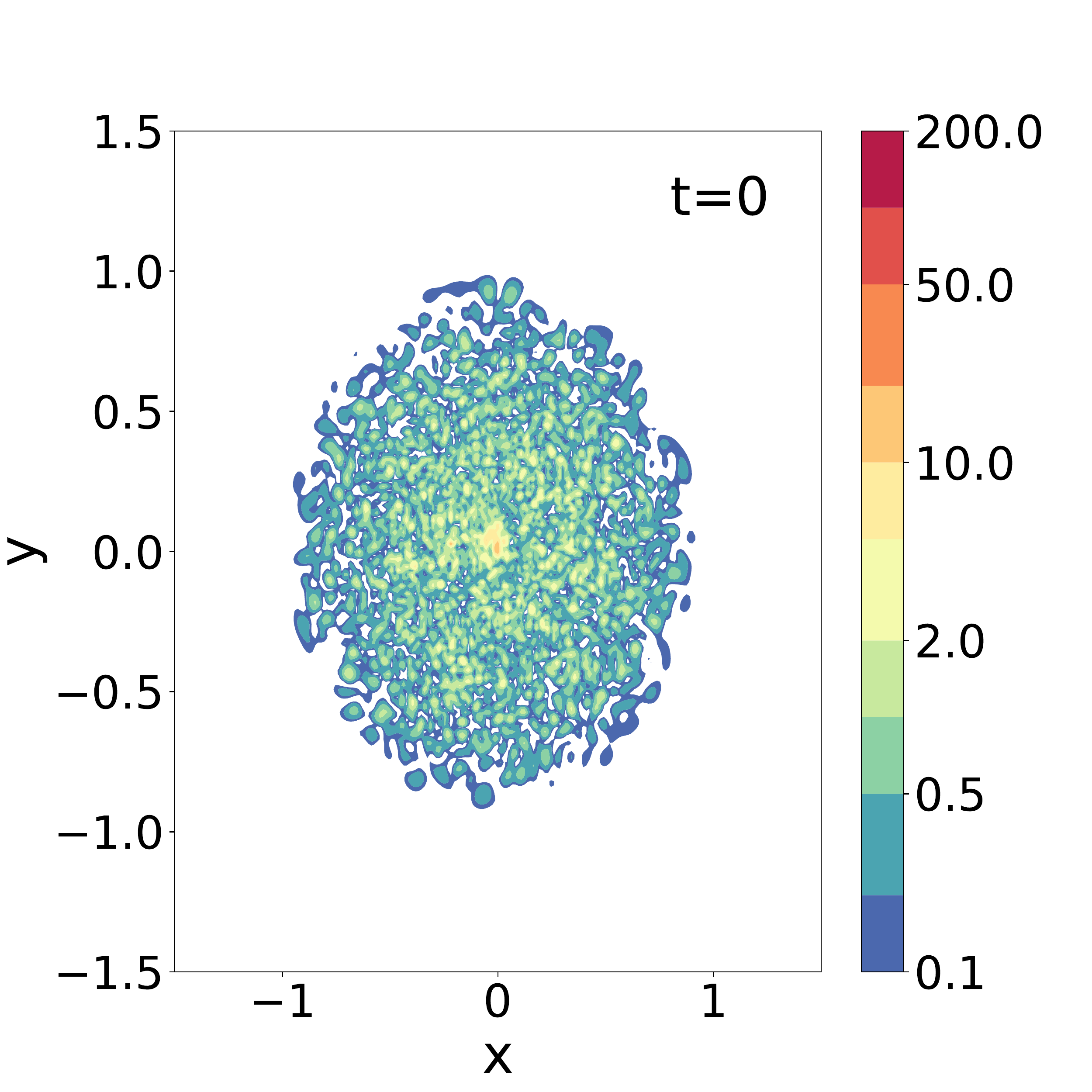}
\includegraphics[height=5.cm,width=0.325\textwidth]{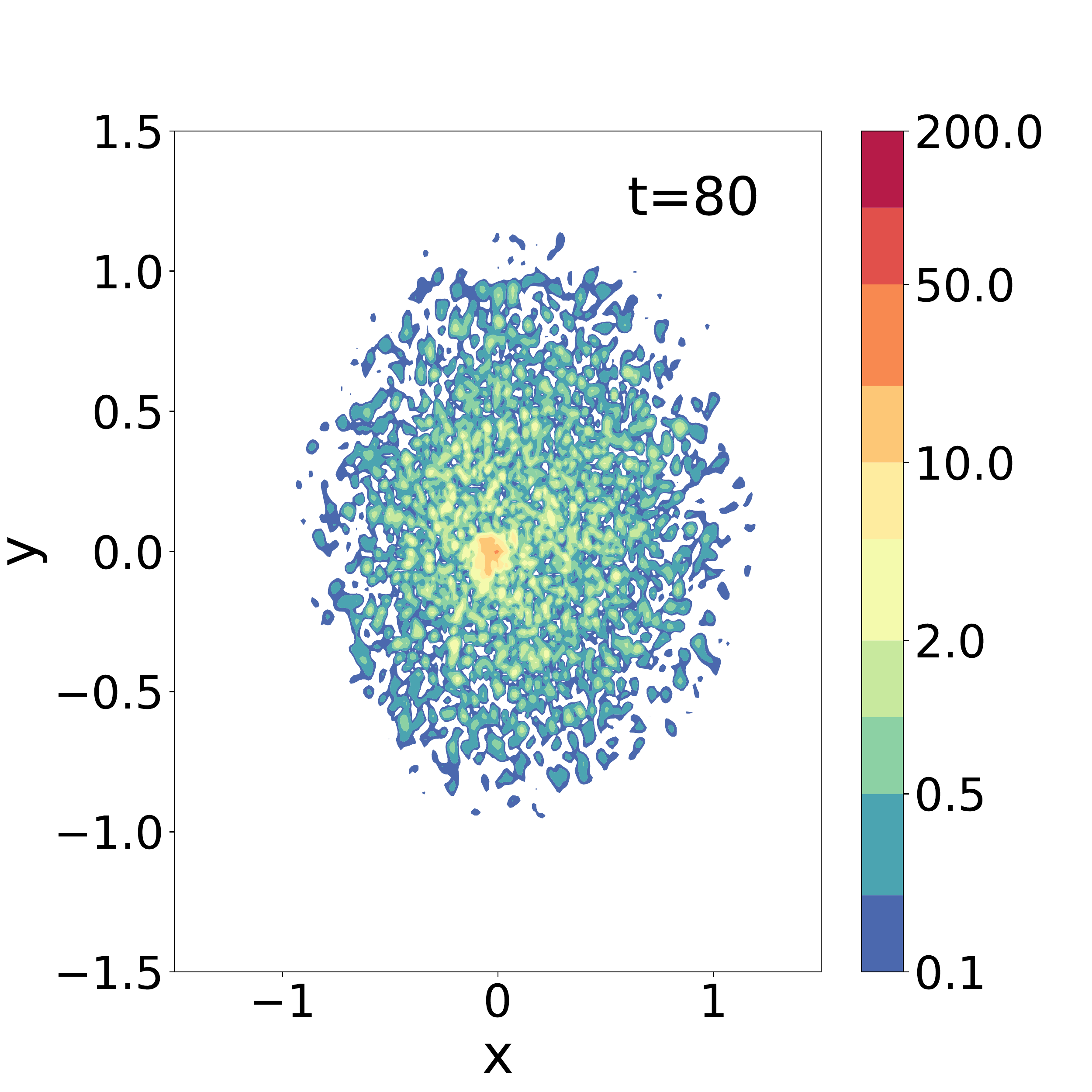}
\includegraphics[height=5.cm,width=0.325\textwidth]{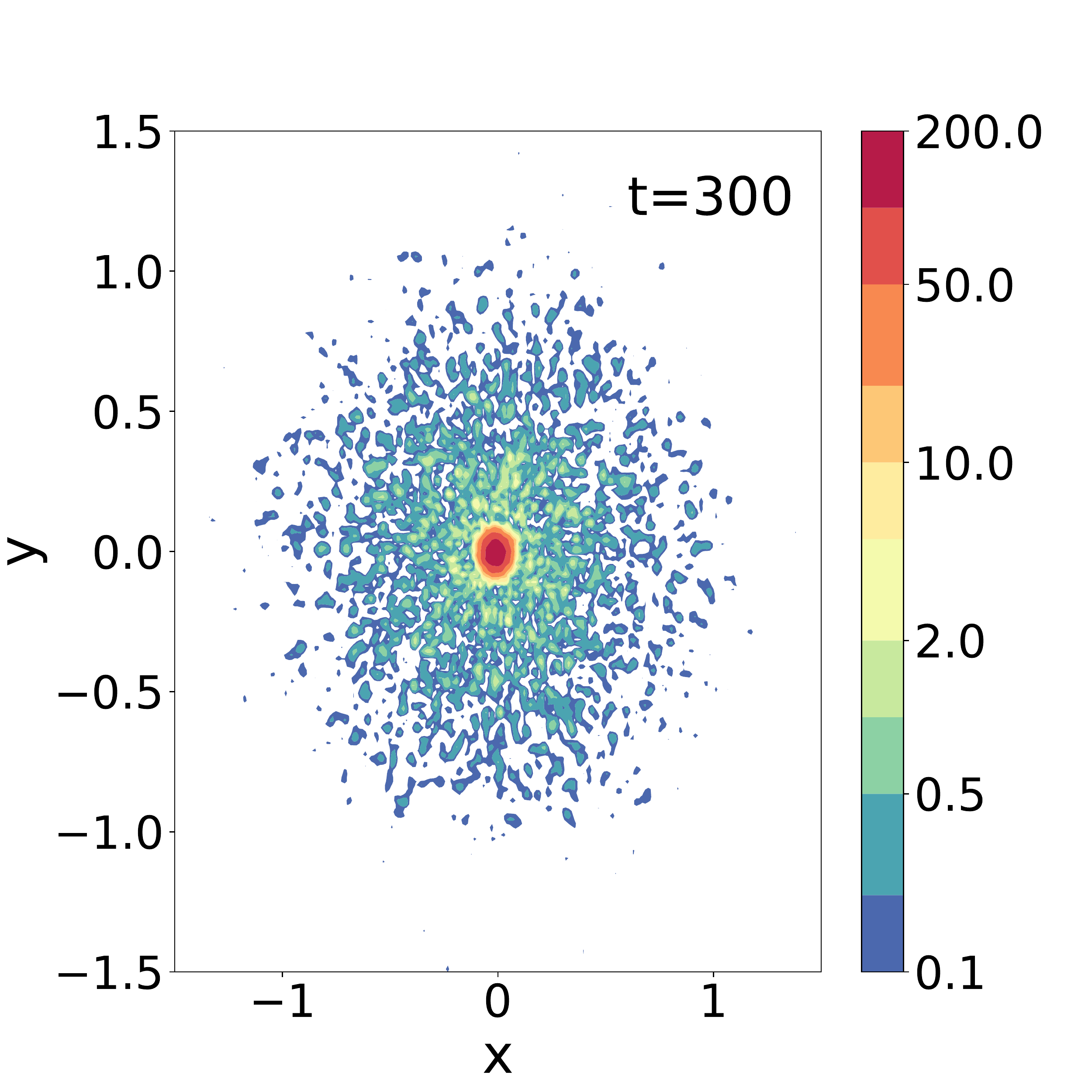}\\
\includegraphics[height=4.2cm,width=0.25\textwidth]{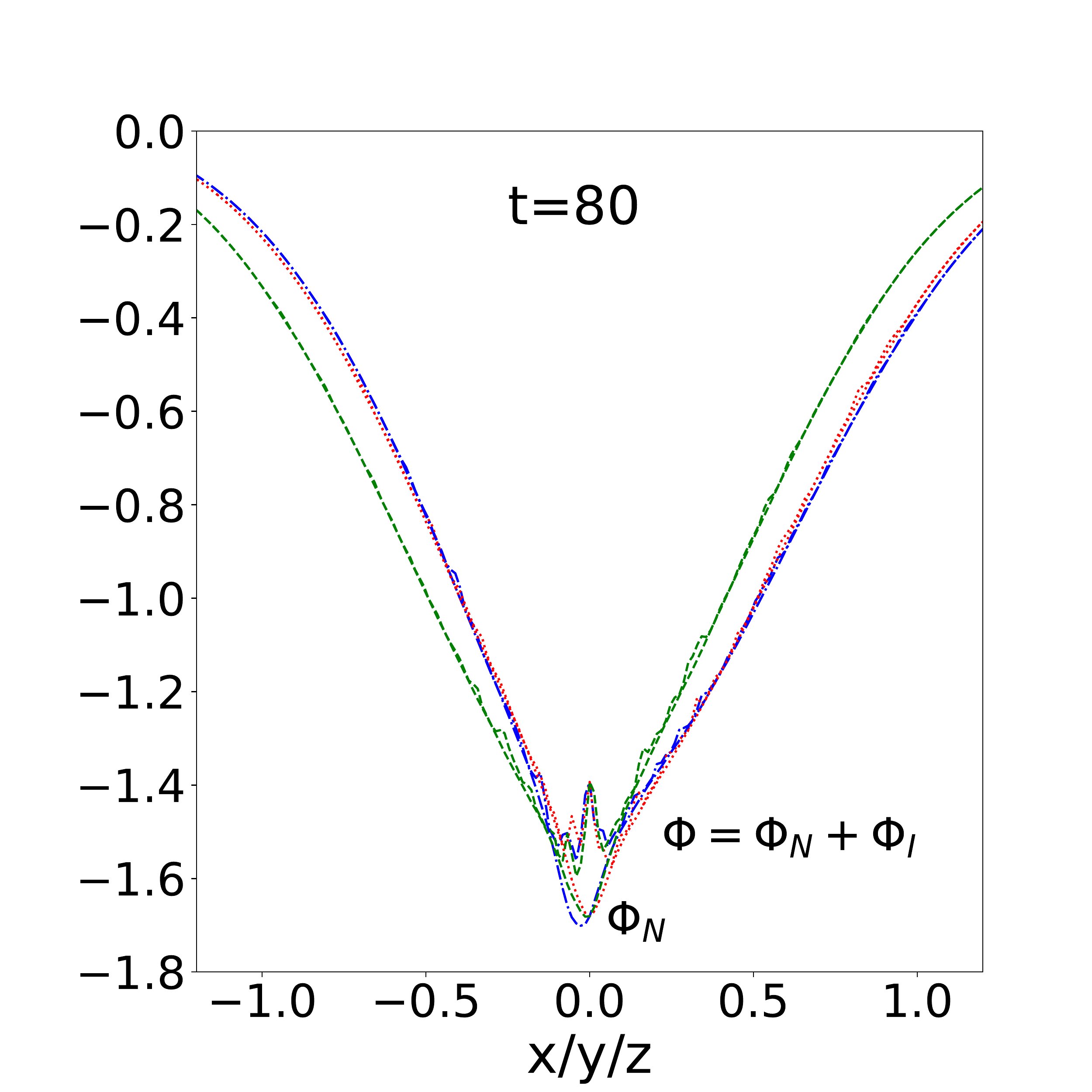}\includegraphics[height=4.2cm,width=0.25\textwidth]{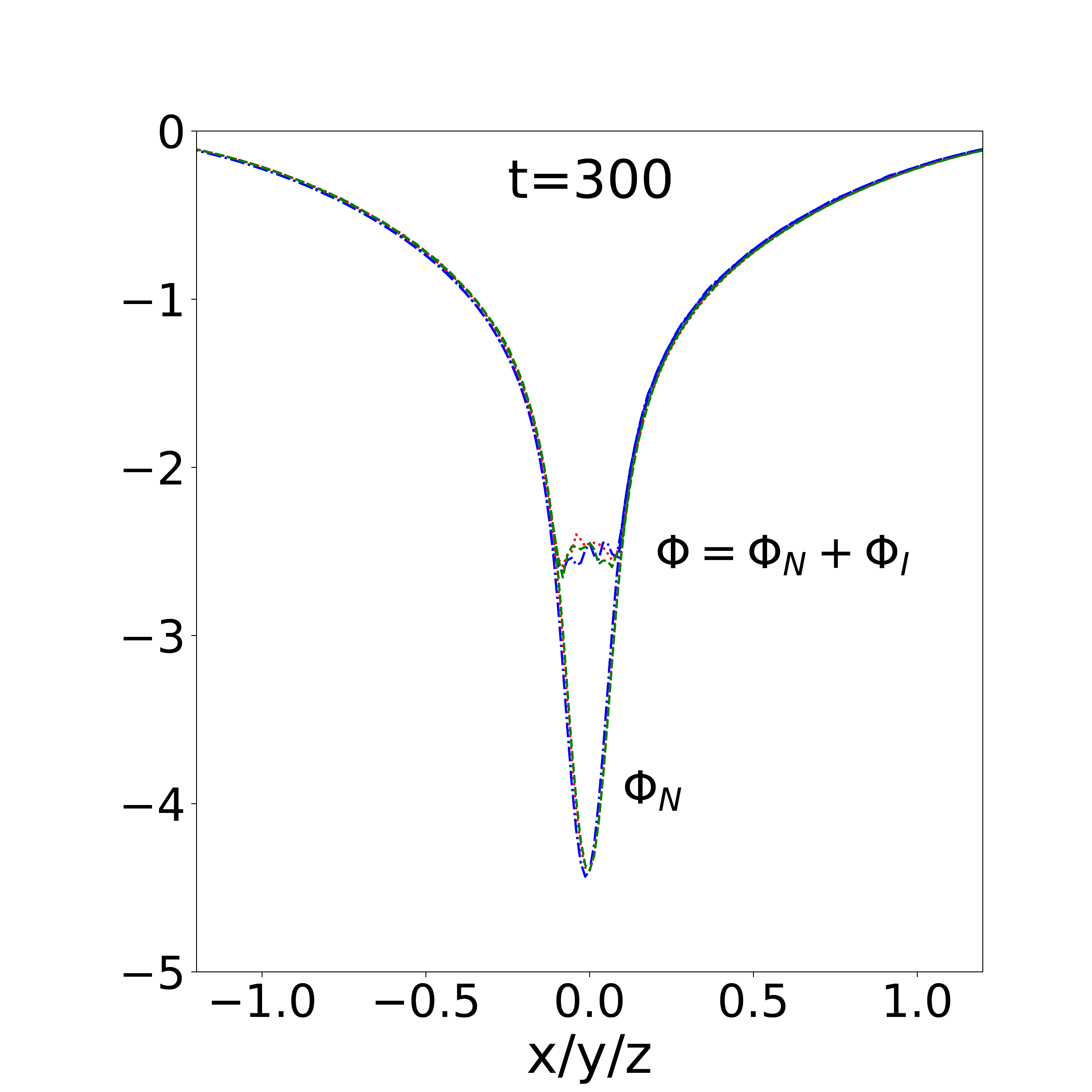}
\includegraphics[height=4.2cm,width=0.24\textwidth]{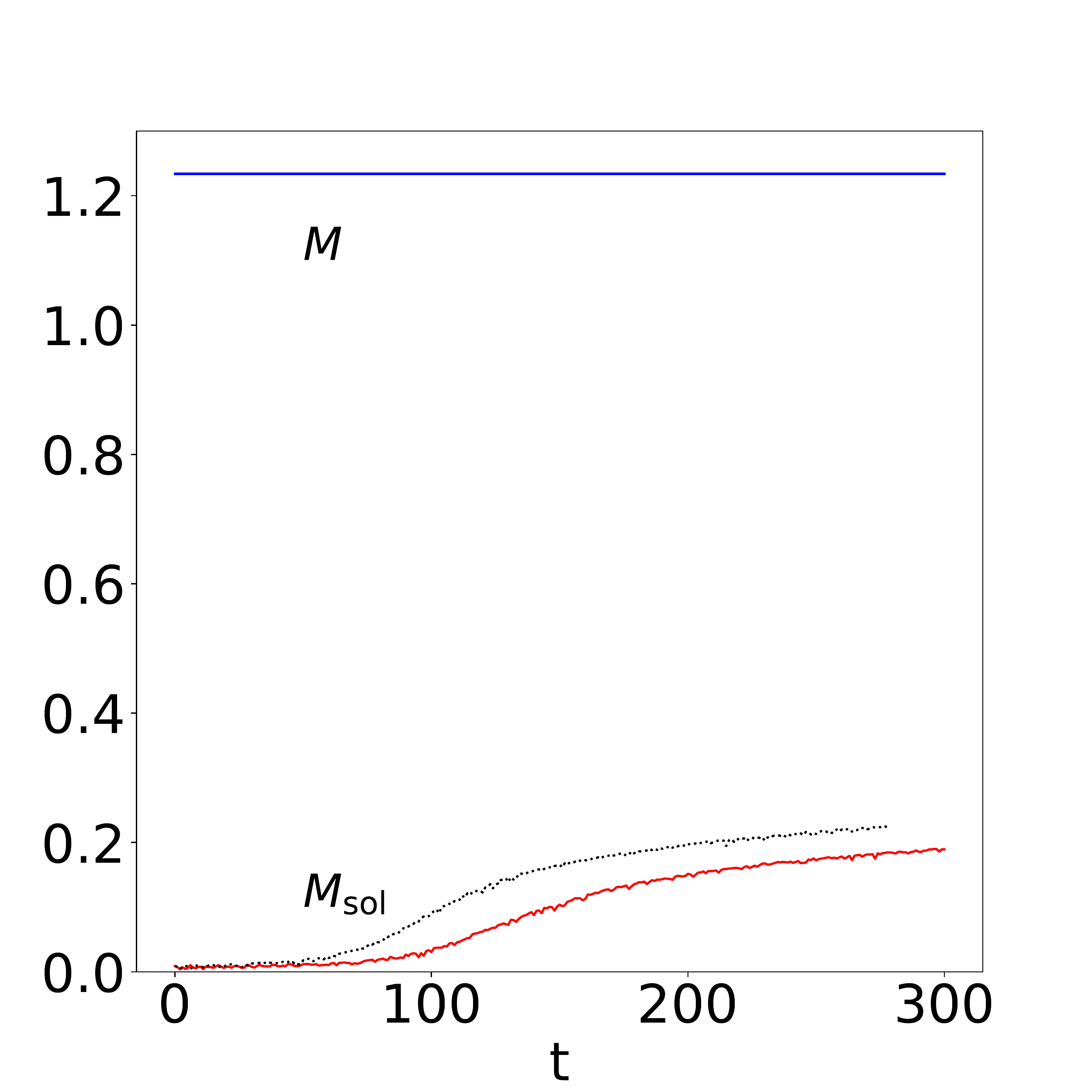}
\includegraphics[height=4.2cm,width=0.24\textwidth]{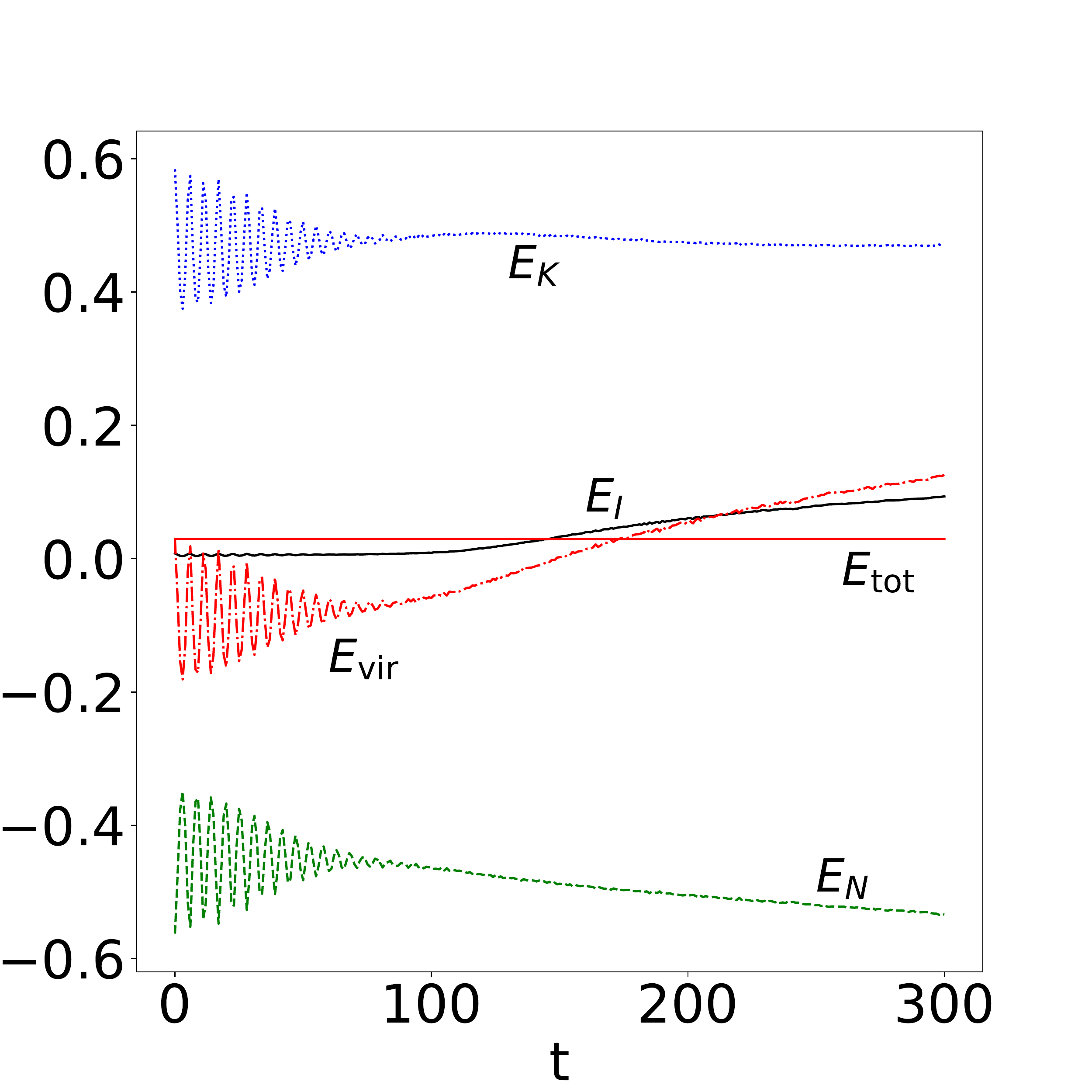}
\caption{
Evolution of a flat system with $R_{\rm sol}=0.1$, $\rho_{0{\rm sol}}=5$.
}
\label{fig:rho-plane-sinc-R0p1-rho5}
\end{figure*}

\subsection{Small soliton radius, $R_{\rm sol}=0.1$}

We now consider cases where the radius $r_a$ associated with the self-interactions
is much smaller than the halo radius. Thus, in this section we take
$R_{\rm sol}=0.1$. In the cosmological context, this would correspond
to late-time structures that collapse on a scale that is much greater than the
characteristic length $r_a$ associated with the self-interactions.
We also take the mass of the system to be constant,
$M_{\rm sol}+M_{\rm halo}=4/\pi$, so that all simulations have about the same mass
(up to the random fluctuations associated with the stochastic initial conditions).

\subsubsection{Halo without initial soliton}
\label{sec:flat-R01-rho0}

We first study in Fig.~\ref{fig:rho-plane-sinc-R0p1-rho0} the dynamics of a halo
without initial soliton, $\rho_{0{\rm sol}}=0$.

We can see that no soliton dominated by the self-interactions
appears until $t \sim 200$. As seen in panels (b) and (e), at $t=180$ the halo is still dominated by
strong fluctuations, associated with the superposition of incoherent modes, and a few rare high-density
spikes that appear randomly. Their spatial width is not set by the radius
$R_{\rm sol}=0.1$ associated with hydrostatic equilibria governed by the balance between
the self-interaction and gravity. Instead, it is of the order of $\Delta x \sim \epsilon=0.01$
and as in FDM scenarios it is governed by the quantum pressure,
that is, by wave effects that appear on the de Broglie scale.
This is clearly seen in panel (b), where the highest density peak is much higher and narrower than
the soliton profile (shown by the purple solid line) that would contain the mass enclosed within radius
$R_{\rm sol}=0.1$ around this point.
This can also be seen in panel (g), where the total potential $\Phi$ does not show a flat plateau of radius
$R_{\rm sol}$. Instead, it closely follows the gravitational potential and only shows small departures at the
bottom of the potential, associated with the various narrow density peaks that are close to the center of the
system.

Eventually, at time $t \sim 200$, one of these high density peaks grows sufficiently to dominate over 
all other peaks and to form a stable soliton governed by the quantum pressure rather than the self-interactions.
This leads to a sharp rise of the central density and of the mass $M_{\rm sol}$ enclosed within radius
$R_{\rm sol}$ around $\vec r_{\rm max}$.
This corresponds to a change of the physics of the system, with the formation of a new soliton that is 
no longer of the FDM type (balance between gravity and quantum pressure) but of the self-interaction 
type (balance between gravity and self-interactions), as given by Eq.(\ref{eq:hydrostatic-alpha}).
This is clearly seen in panels (c) and (f) at $t=250$, where we can see the characteristic radius
$R_{\rm sol}=0.1$ of such hydrostatic equilibria and the plateau in the total potential $\Phi$ over
the soliton extent.
This also shows as a plateau for $\Phi$ in panel (h). 
In agreement with the condition (\ref{eq:Thomas-Fermi-2}), it is only after
one of the narrow high density peaks has grown sufficiently to reach a density threshold
$\rho \gtrsim 15$ that this transition takes place and a soliton supported by the
self-interactions can appear.

As seen in panel (i), the total mass of the system is conserved in the numerical simulation.
However, the larger box simulation gives an earlier soliton formation time at $t \sim 150$.
This can be understood from the nature of the transition, which is due to the growth of one of the
stochastic density peaks until a density threshold is reached. The exact transition time thus appears
to be strongly dependent on the details of the dynamics as it is due to the interactions of the
random initial fluctuations, which may be perturbed by the small gravitational attraction from the
halos in the neighboring boxes (in the case of periodic boundary conditions).

The oscillations seen in panel (j) until $t \lesssim 100$ are due to global modes, associated with 
a pulsation of the halo radius, whence of its characteristic densities and energies. 
These modes appear to be damped after $t \gtrsim 100$.
Thus, by $t \sim 100$, the halo relaxes to a quasi-stationary state close to the initial conditions obtained
from the WKB approximation. The halo radius is not significantly modified but small-scale
density fluctuations have grown in the central region, with the appearance of high-density
peaks governed by the quantum pressure.
As expected for such a quasi-stationary state, the virial quantity $E_{\rm vir}$ is close to zero.
At late times, $t > 200$, when the soliton has formed and slowly grows, $E_{\rm vir}$ slowly grows
as the system is no longer in equilibrium and is made of two distinct components with a time-dependent
mass ratio.
This is reminiscent of the growth of $E_{\rm vir}$ found in Fig.~\ref{fig:rho-plane-sinc-R0p5-rho0}
for the large soliton case $R_{\rm sol}=0.5$. However, in the case shown in 
Fig.~\ref{fig:rho-plane-sinc-R0p1-rho0} all energy components remain dominated by the halo rather than
by the soliton until the end of the simulation. 
This is why $E_{\rm vir}$ remains close to zero and we do not display $\tilde{E}_{\rm vir}$.

Thus, in this case we find that while the system remains dominated by FDM spikes for a long time
and seems almost stationary, the secular evolution eventually makes one spike grow until the
self-interactions come into play and lead to the formation of a broad soliton supported by these
self-interactions. There is thus a transition in the system from a FDM phase to a self-interacting phase,
embedded in the FDM halo. This transition may only happen after a long time, much greater than the
dynamical time of the system, as the growth of the central density peaks is very slow until one of them
reaches this threshold and suddenly builds a unique massive soliton. This also means that the transition
time strongly depends on the initial conditions and the details of the dynamics.

\subsubsection{Small initial soliton $\rho_{0{\rm sol}}=5$}

We study in Fig.~\ref{fig:rho-plane-sinc-R0p1-rho5} the dynamics of a halo
with a small initial soliton, $\rho_{0{\rm sol}}=5$. 
In agreement with Eq.(\ref{eq:Thomas-Fermi-2}), it is initially strongly perturbed by the 
wave packets from the halo as it is close to this density threshold, but we can see that 
its density slowly grows with time.
Until $t \sim 80$, the soliton cannot be clearly seen as it wanders inside the half-radius 
of the halo and is somewhat masked by the large overlying fluctuations associated with 
higher-energy modes.
This also means that at a given time the location of the highest density peak is not perfectly
centered at the bottom of the gravitational potential. This leads to the anisotropy of the
gravitational potential shown in panel (g), as well as in panel (g) of 
Fig.~\ref{fig:rho-plane-sinc-R0p1-rho0}.

However, as the soliton density slowly grows it becomes less affected by these small-scale 
perturbations and by $t=100$ we can clearly see the characteristic size
$R_{\rm sol}=0.1$ of the central overdensity, much greater than the size $\sim \epsilon=0.01$
of the incoherent fluctuations. This is also apparent in the potential $\Phi$, which shows a flat
plateau at the center perturbed by the wiggles due to the higher-energy modes.
The main behavior is similar to that found in Fig.~\ref {fig:rho-plane-sinc-R0p1-rho0}, where the
initial condition had no soliton but one formed at $t \sim 200$.
Here, thanks to the initial soliton seed, a distinct soliton growth appears earlier at $t \sim 100$.
Again, the bigger box simulation gives a somewhat earlier soliton formation time.
However, because the initial condition already contains a significant soliton seed
the dependence on the details of the dynamics is weaker and the difference between
the fiducial and bigger box simulations is smaller than in 
Fig.~\ref{fig:rho-plane-sinc-R0p1-rho0}.
The soliton is still growing at $t=300$.

Therefore, even reasonably small solitons, with a density a few times greater than the halo
background, survive and grow with time.
This is despite their energy and potential $\Phi$ are much smaller than the halo counterparts.
This is of course consistent with the fact that initial conditions without a soliton eventually form one,
as found in the previous section and in Fig.~\ref {fig:rho-plane-sinc-R0p1-rho0}.
Therefore, solitons governed by the self-interactions appear to be robust attractors.
We checked with numerical simulations that initial solitons with a higher density follow the same
pattern, they are not destroyed and slowly grow with time.

\section{Halo with a cuspy density profile}
\label{sec:cuspy}

\begin{figure*}[ht]
\centering
\includegraphics[height=6.5cm,width=0.39\textwidth]{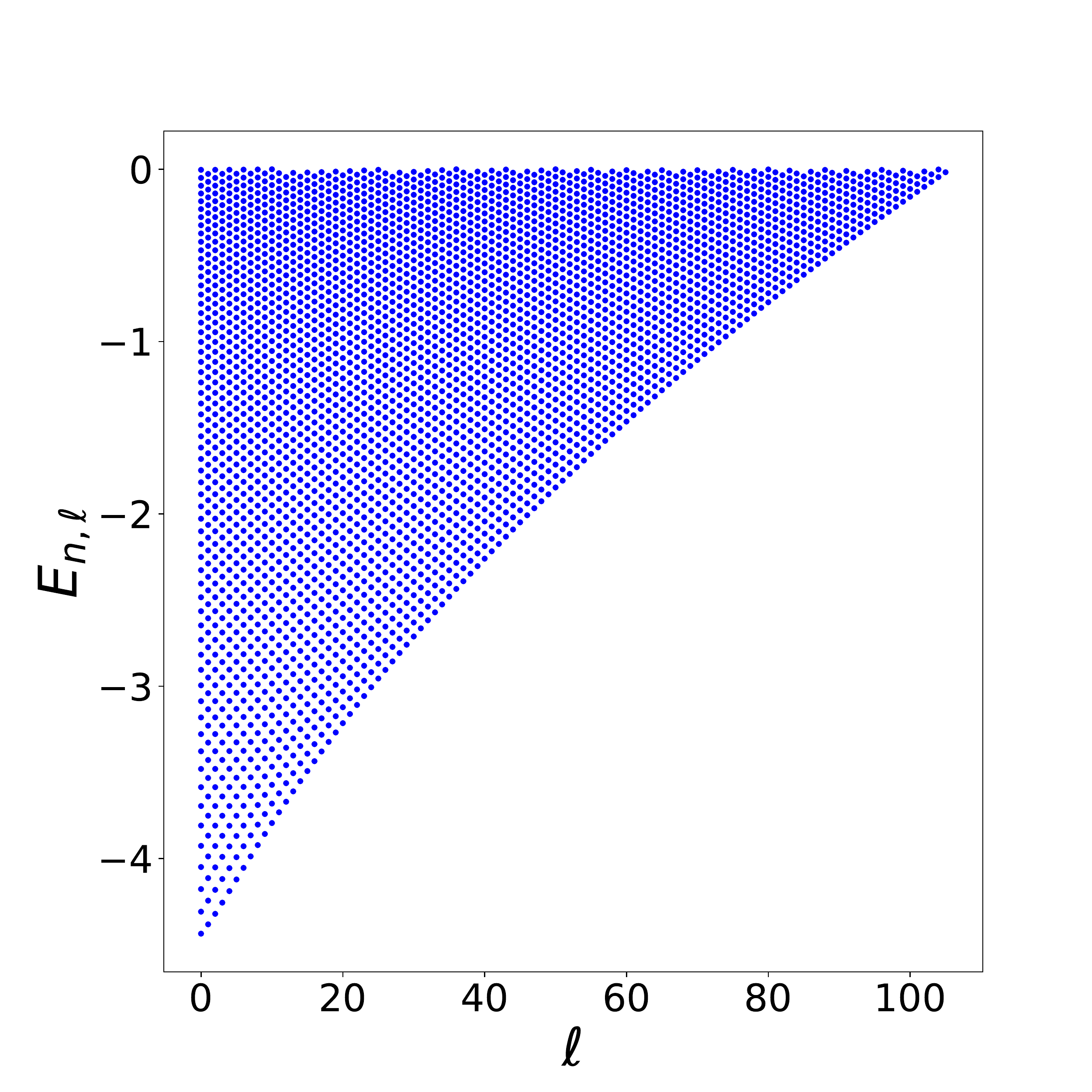}
\includegraphics[height=6.5cm,width=0.39\textwidth]{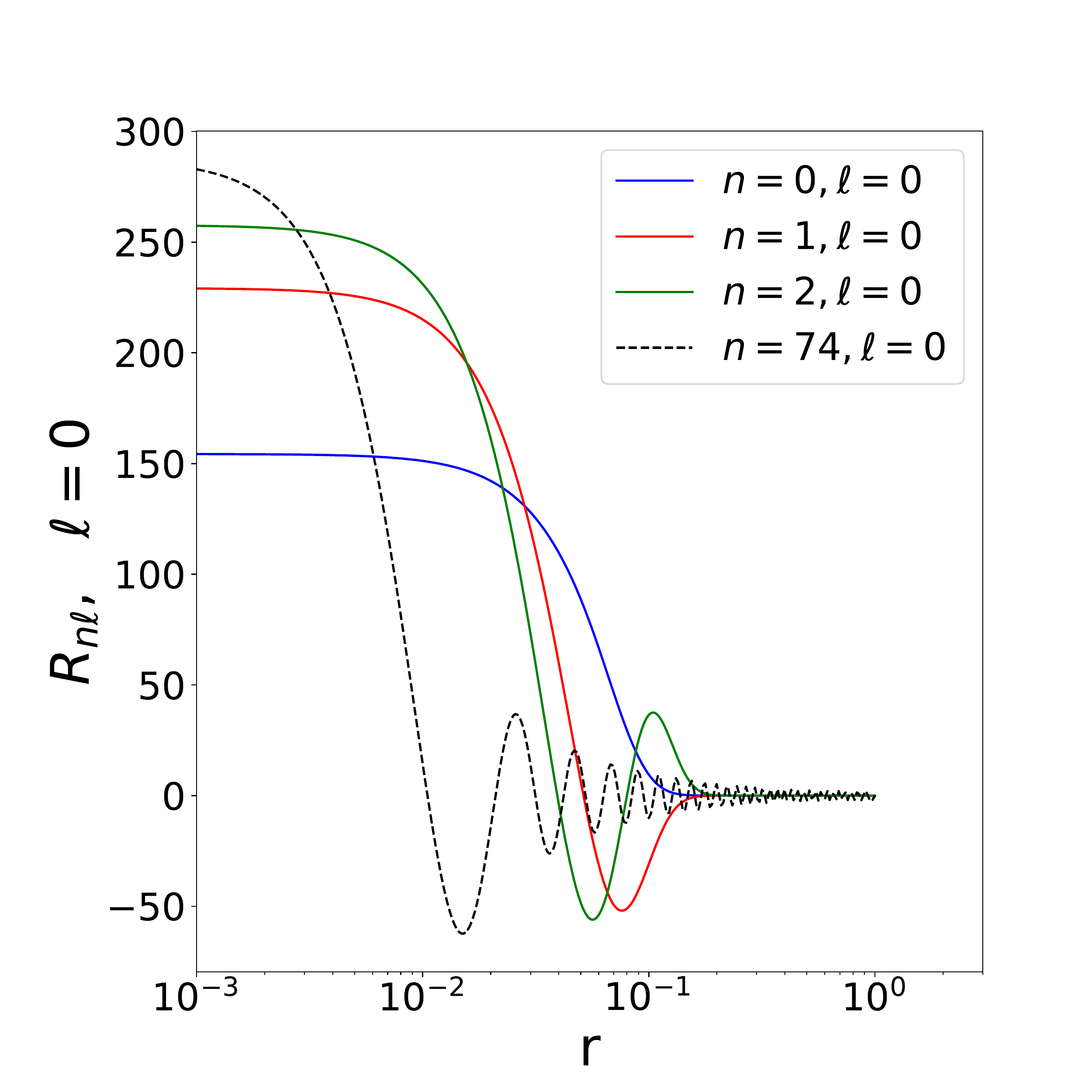}\hfill
\includegraphics[height=6.5cm,width=0.39\textwidth]{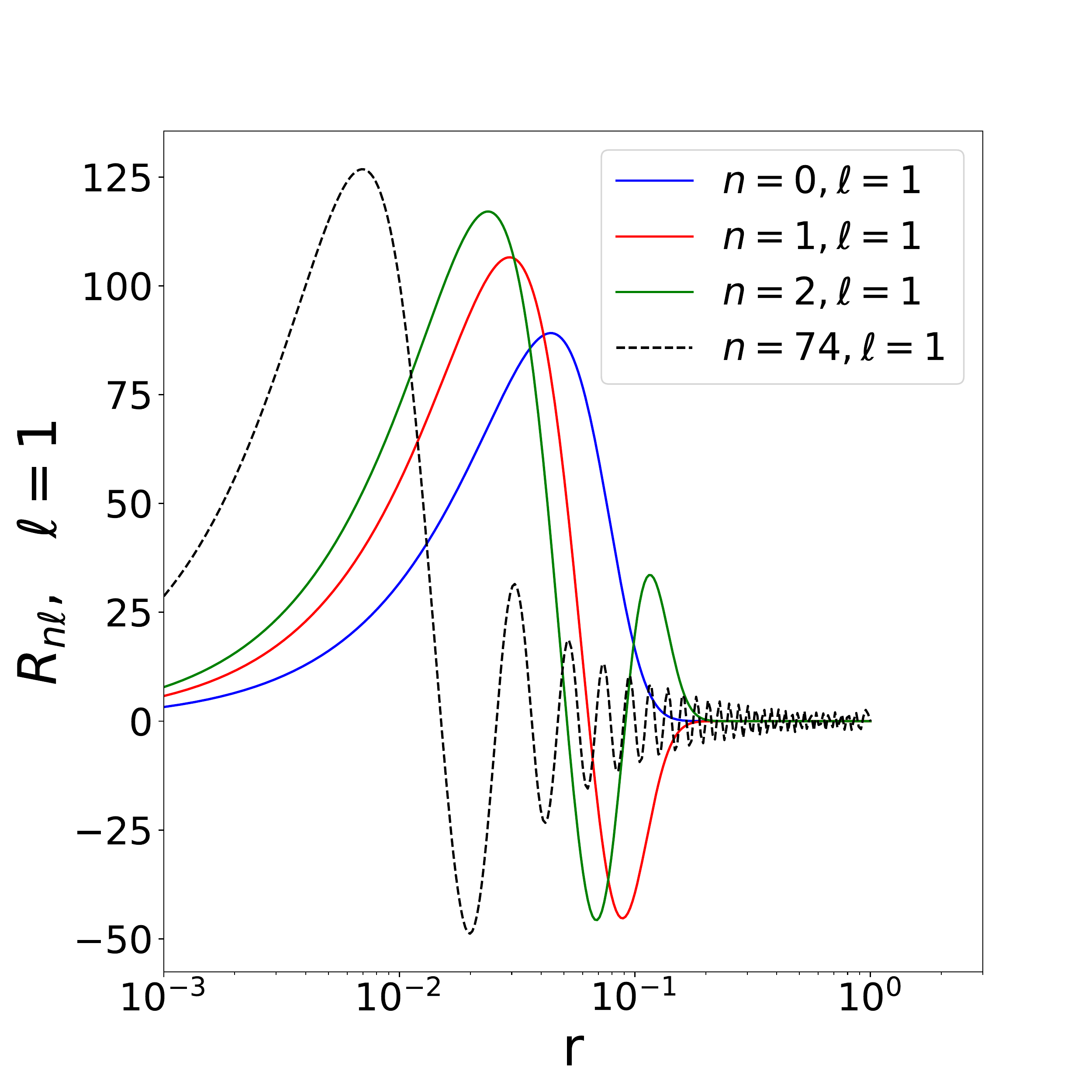}
\includegraphics[height=6.5cm,width=0.39\textwidth]{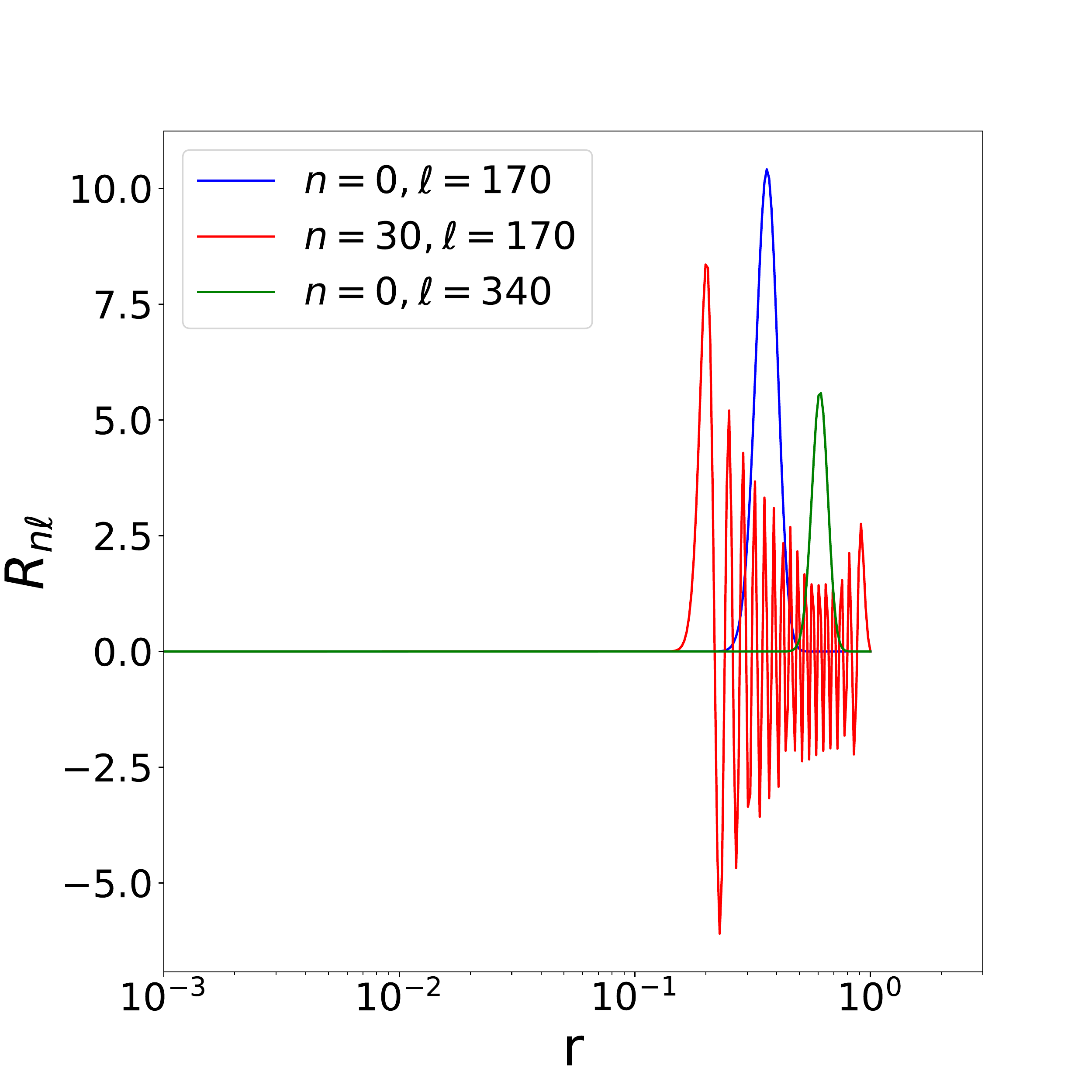}
\caption{
Energy levels $(\ell,E_{n\ell})$ and eigenmodes ${\cal R}_{n\ell}$ for the halo profile (\ref{eq:rho-cos}).
These eigenmodes are normalized to unity as in Eq.(\ref{eq:Rn1-Rn2-orthonormal}).}
\label{fig:E-levels-cos}
\end{figure*}

\begin{figure*}[ht]
\centering
\includegraphics[height=4.cm,width=0.24\textwidth]{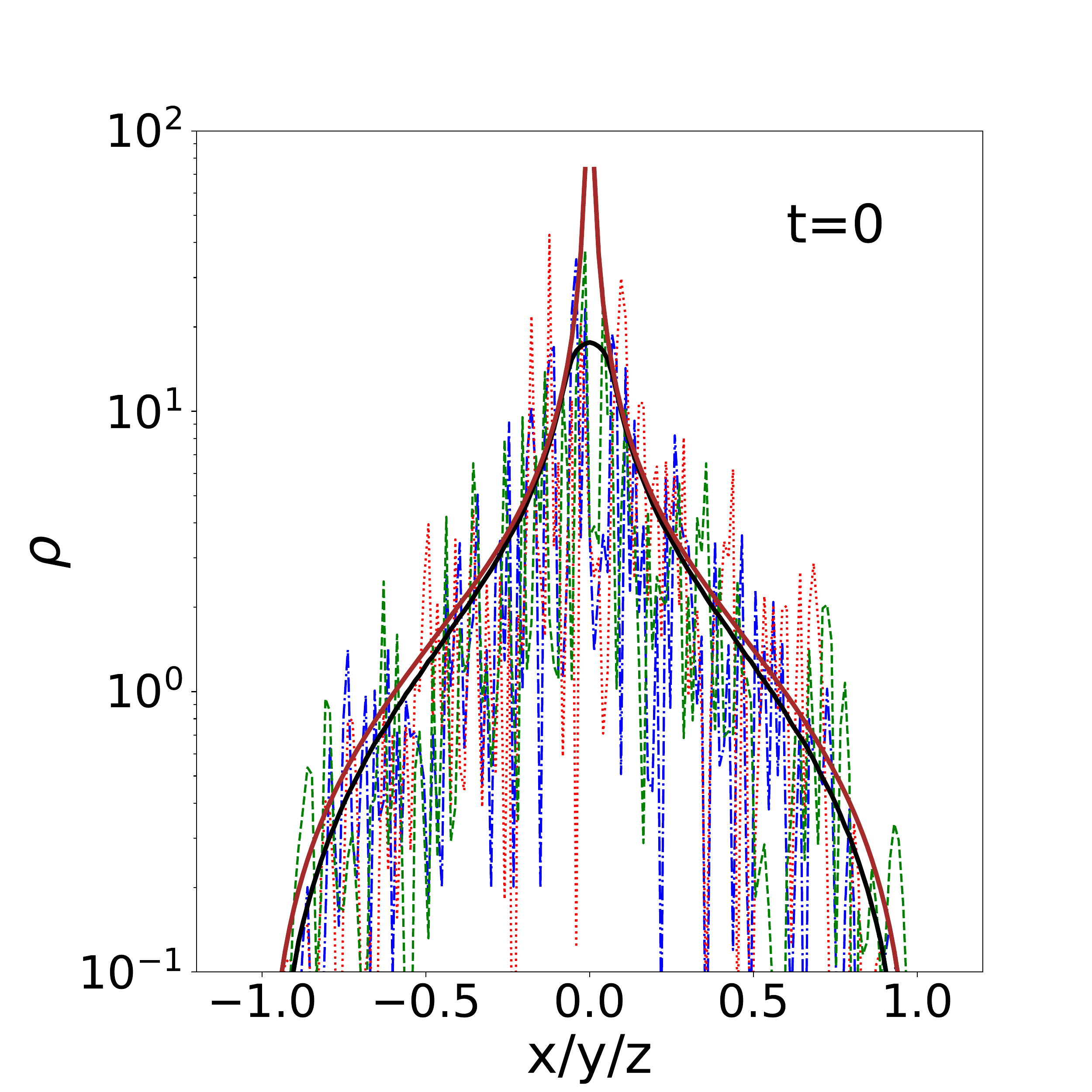}
\includegraphics[height=4.cm,width=0.25\textwidth]{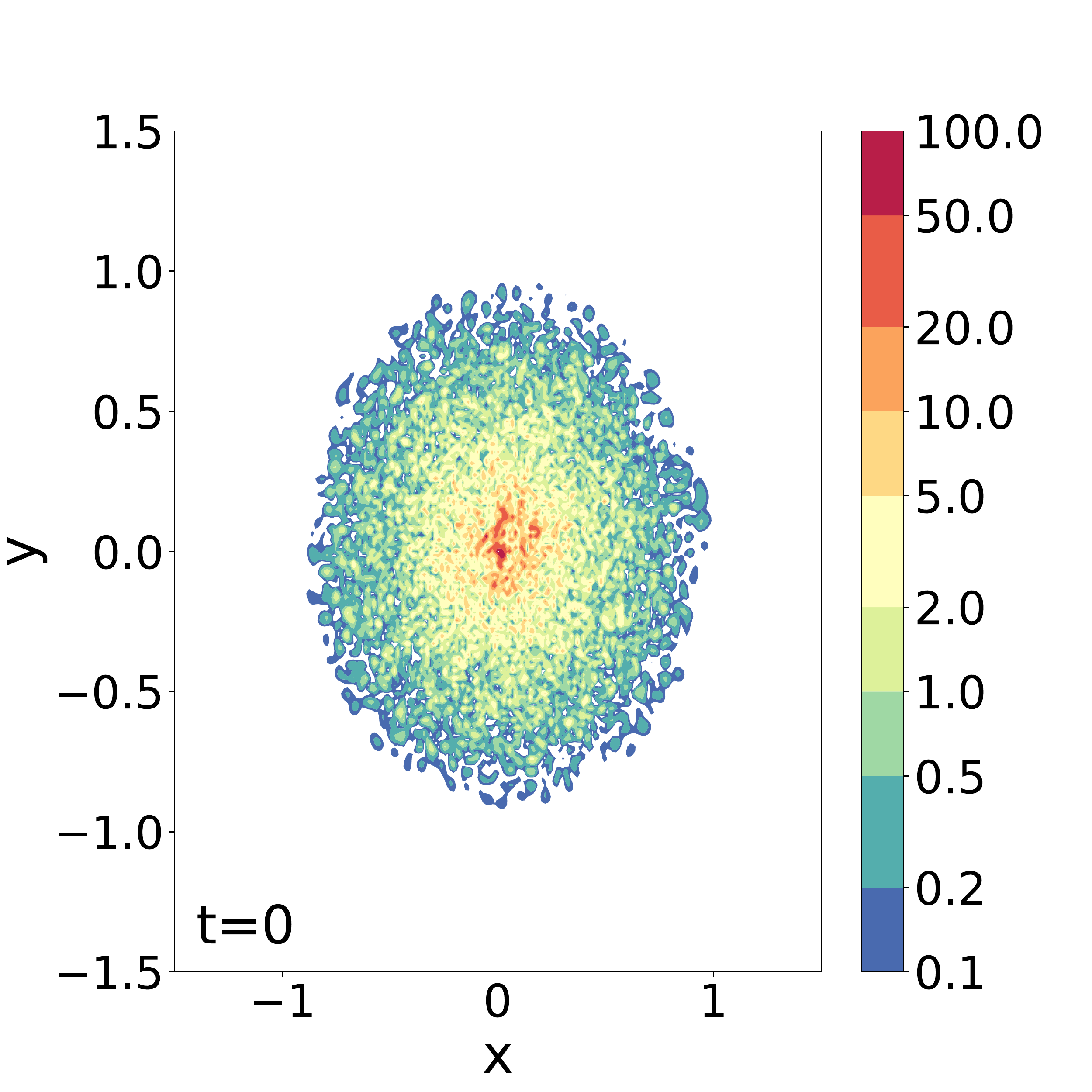}
\includegraphics[height=4.cm,width=0.24\textwidth]{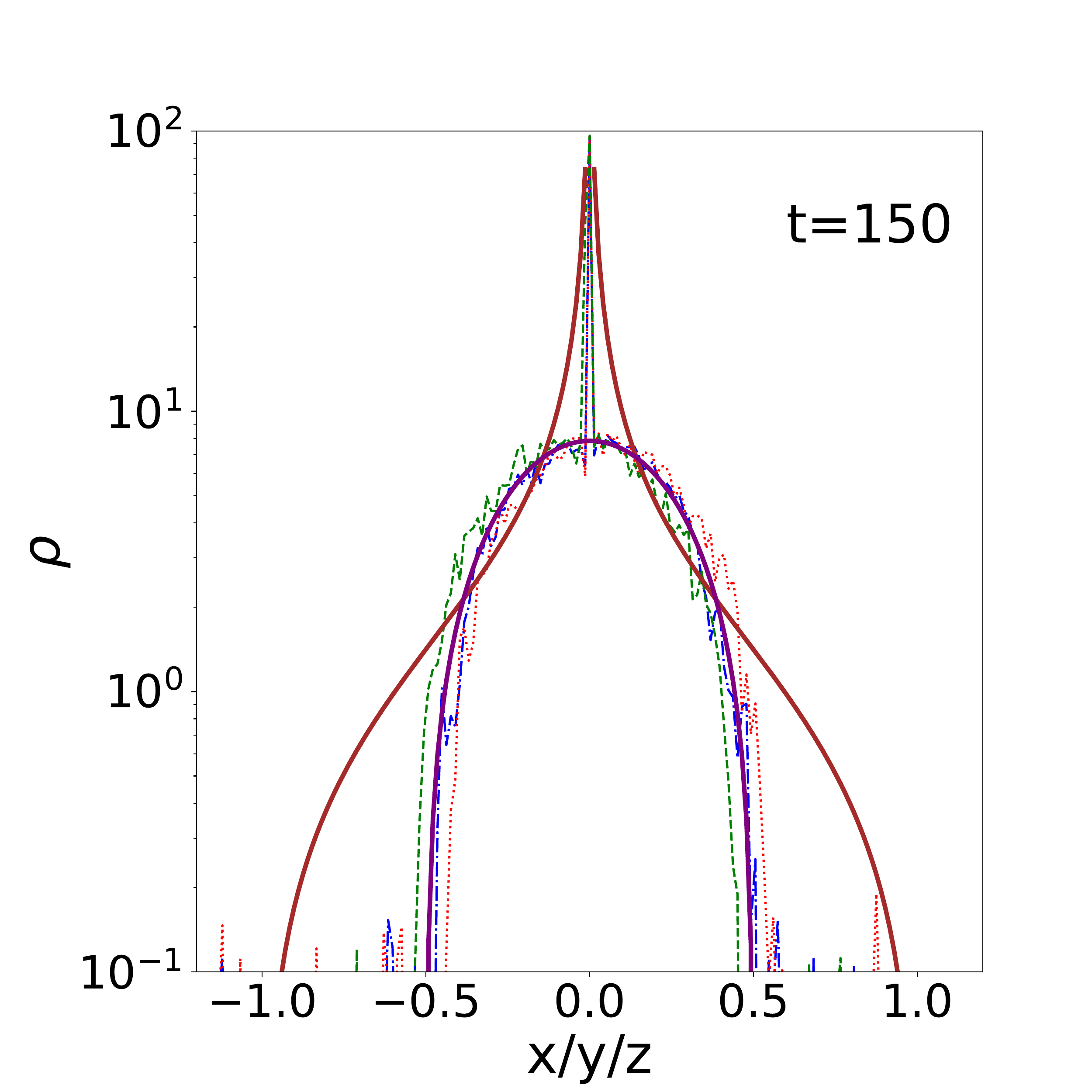}
\includegraphics[height=4.cm,width=0.25\textwidth]{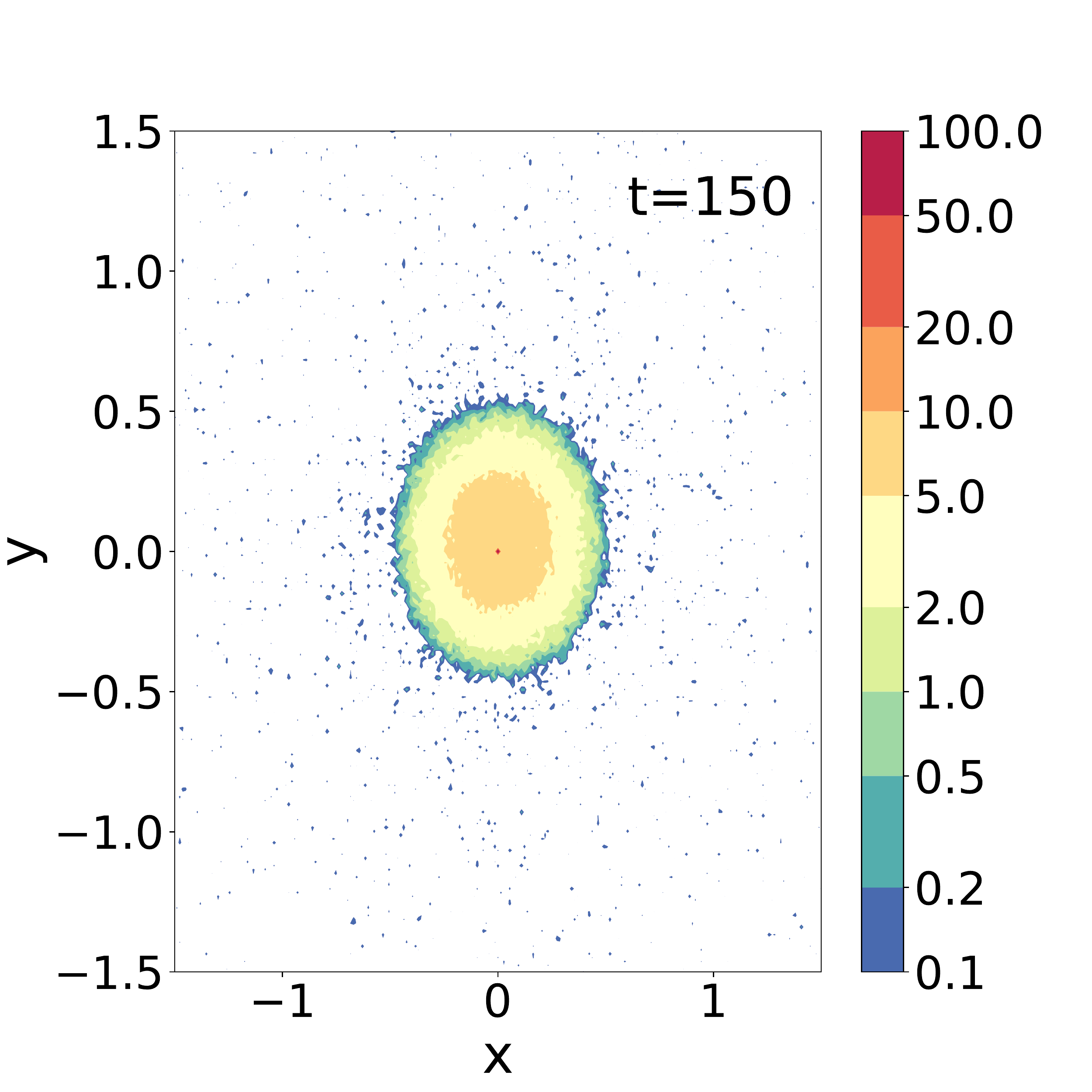}\\
\includegraphics[height=4.2cm,width=0.32\textwidth]{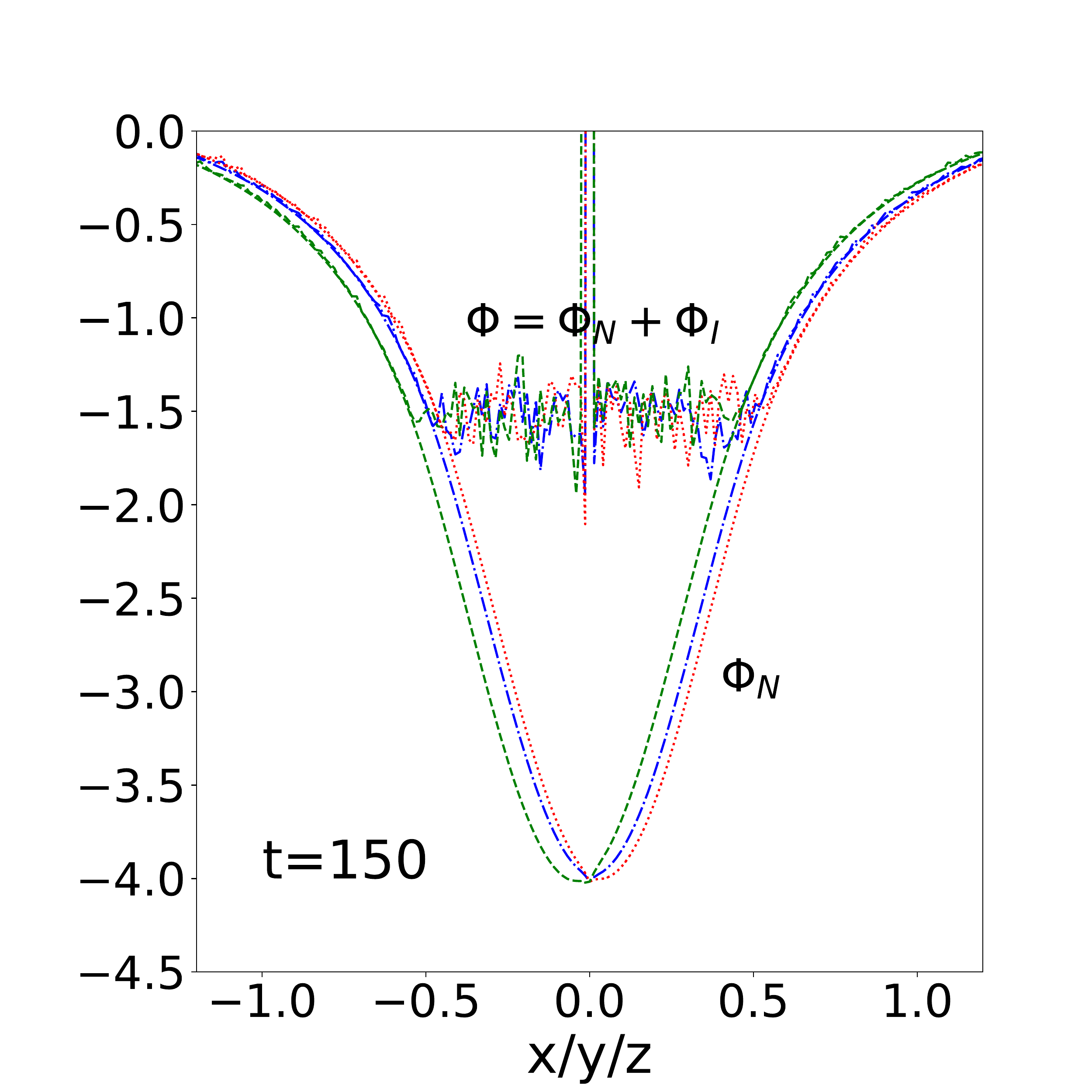}
\includegraphics[height=4.2cm,width=0.28\textwidth]{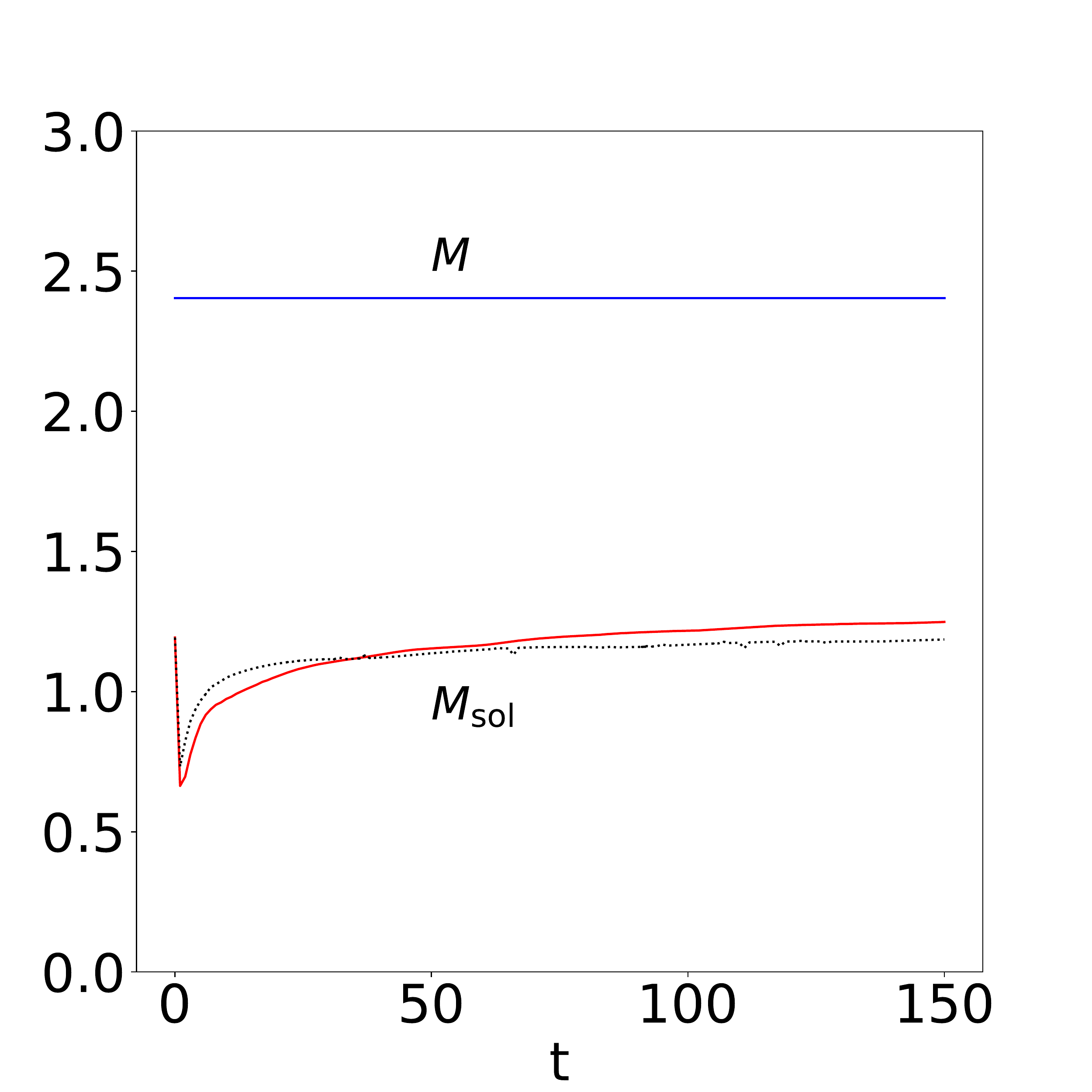}
\includegraphics[height=4.2cm,width=0.28\textwidth]{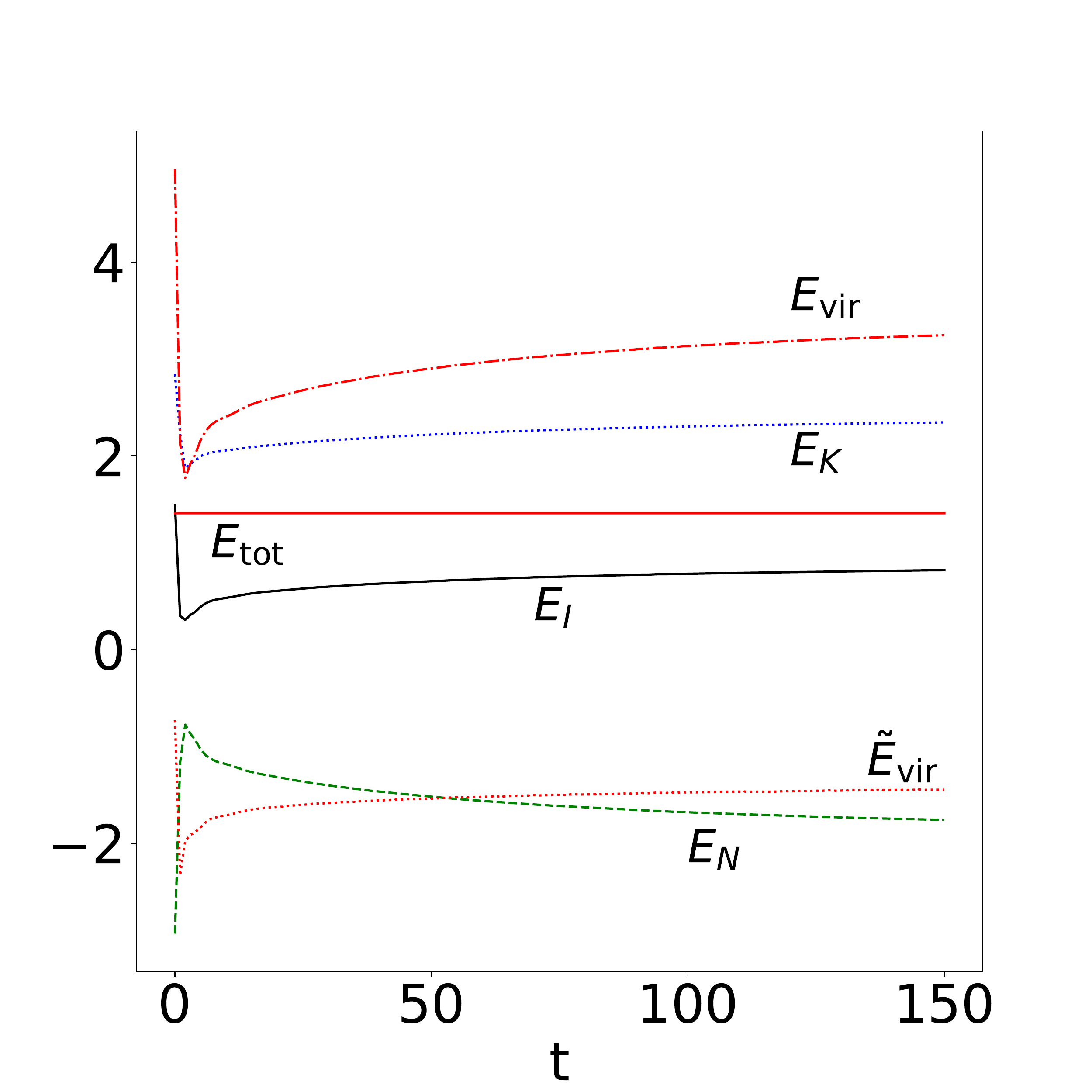}
\caption{
Evolution of a cuspy system with $R_{\rm sol}=0.5$, $\rho_{0{\rm sol}}=0$.
}
\label{fig:rho-plane-cos-R0p5-rho0}
\end{figure*}

\begin{figure*}[ht]
\centering
\includegraphics[height=4.cm,width=0.24\textwidth]{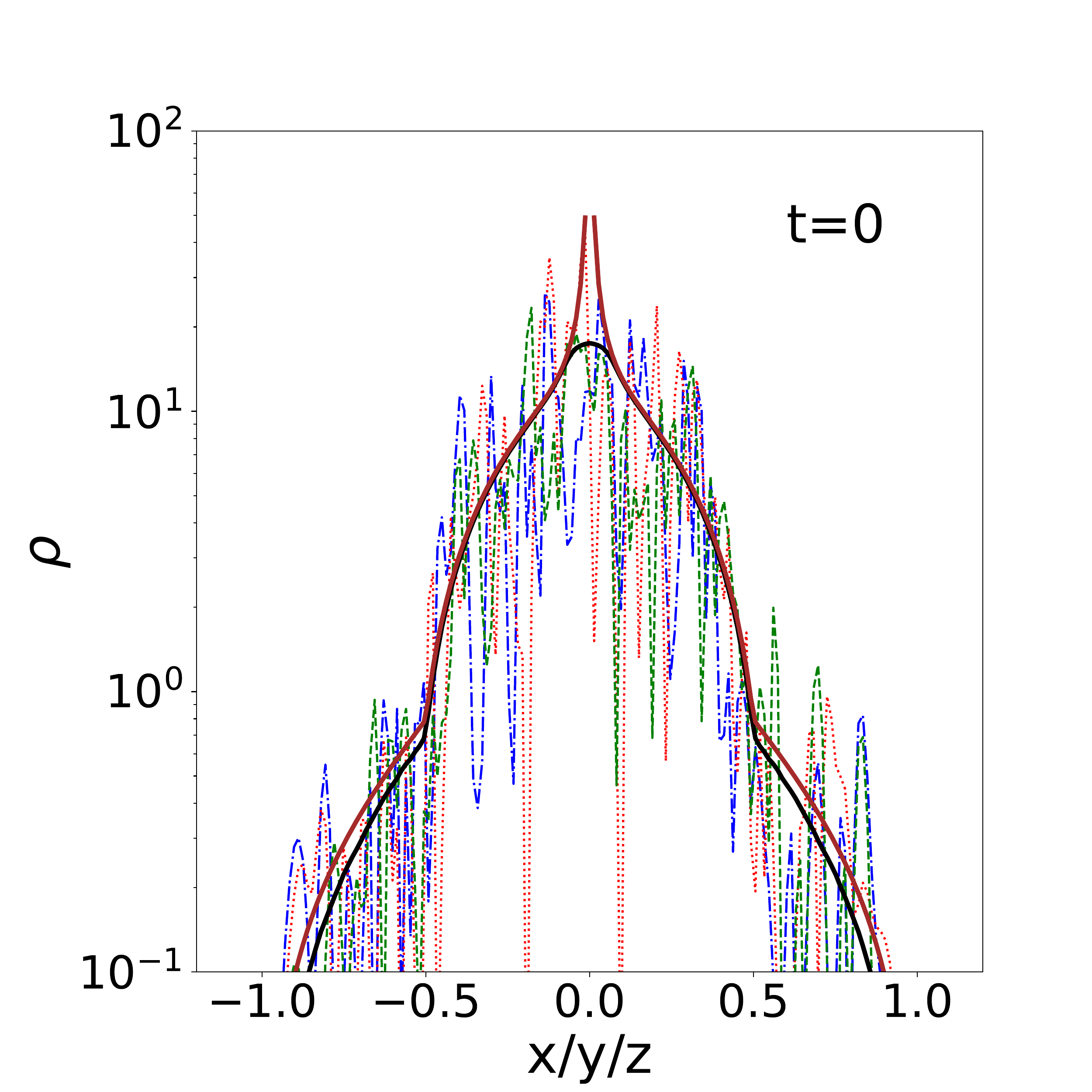}
\includegraphics[height=4.cm,width=0.24\textwidth]{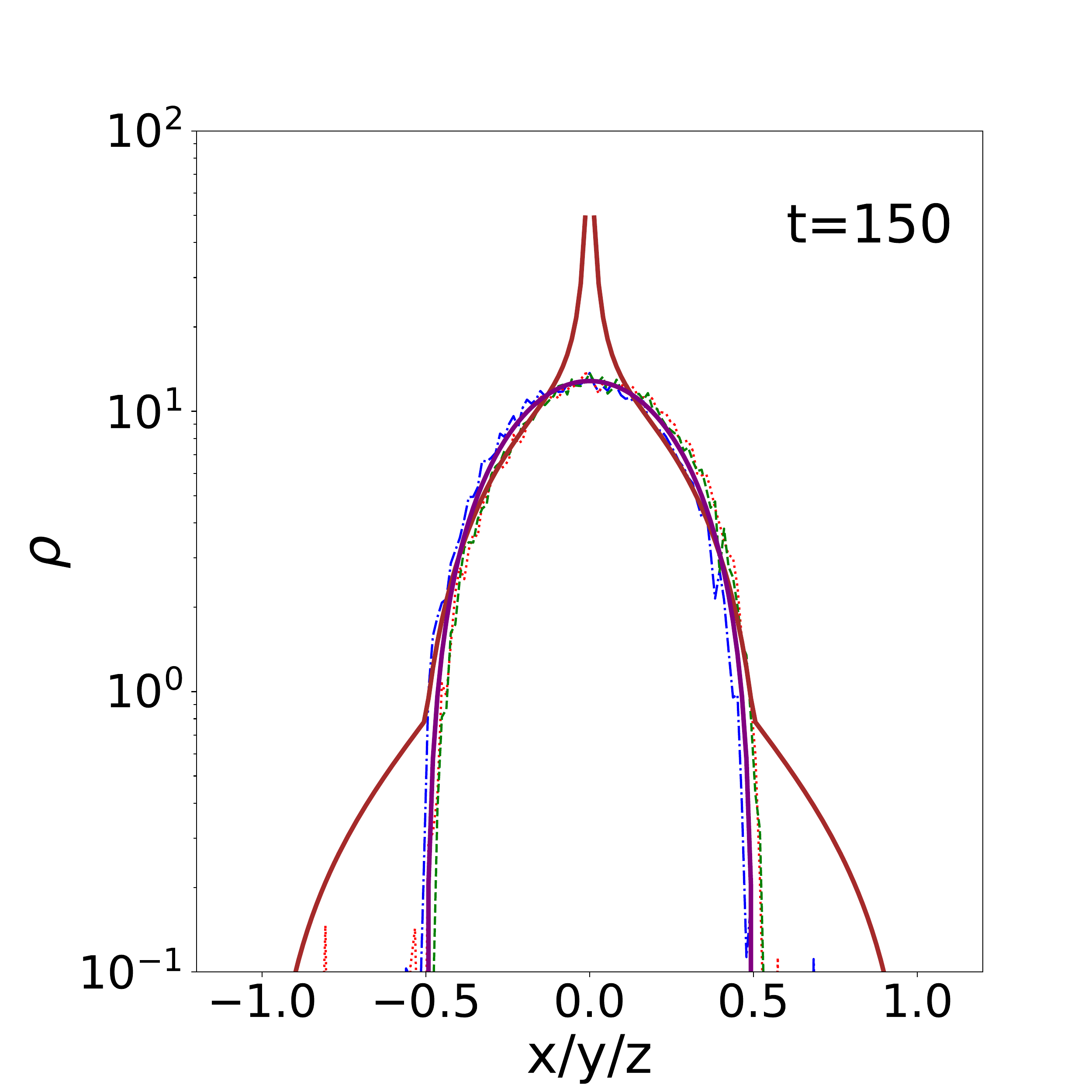}
\includegraphics[height=4.cm,width=0.25\textwidth]{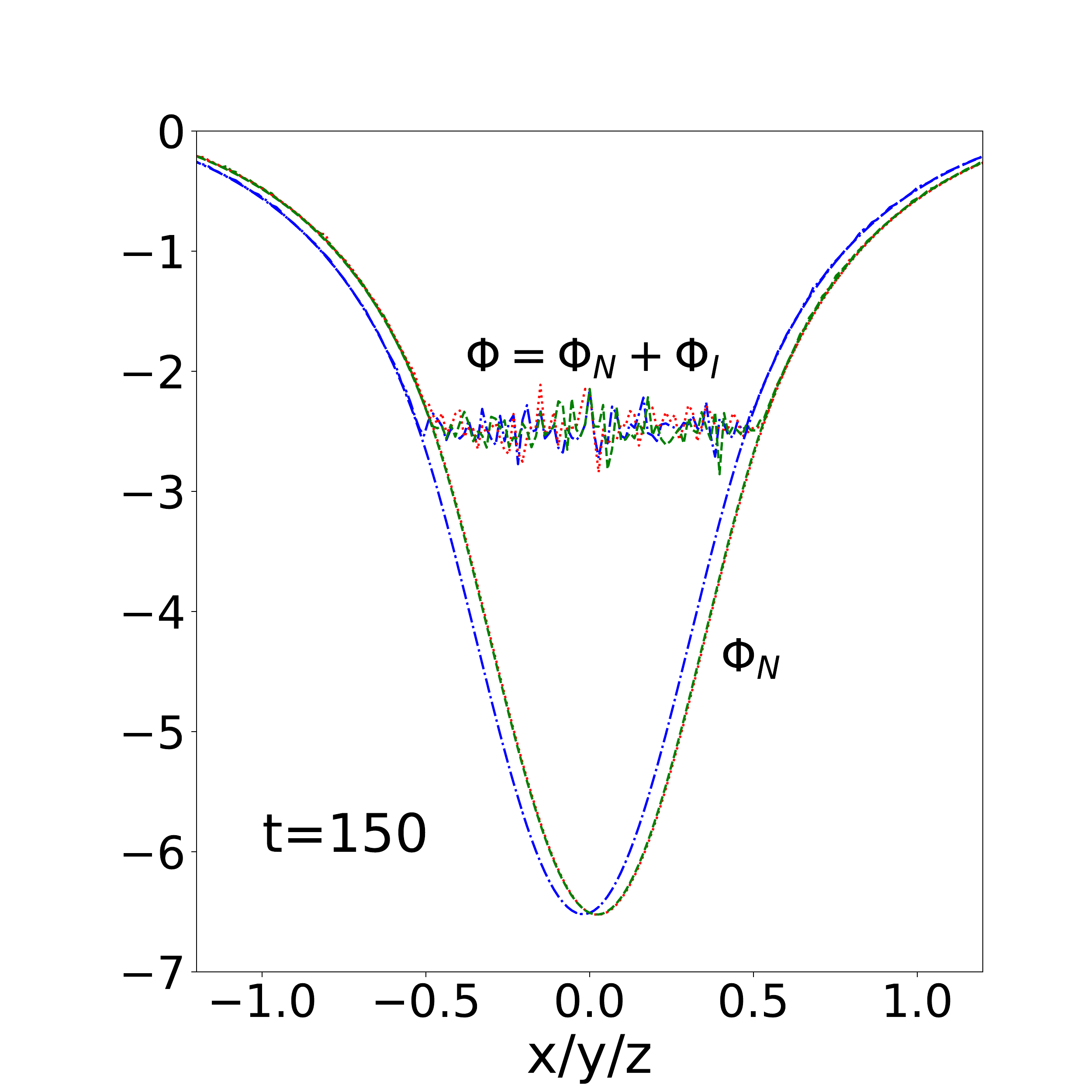}
\includegraphics[height=4.cm,width=0.24\textwidth]{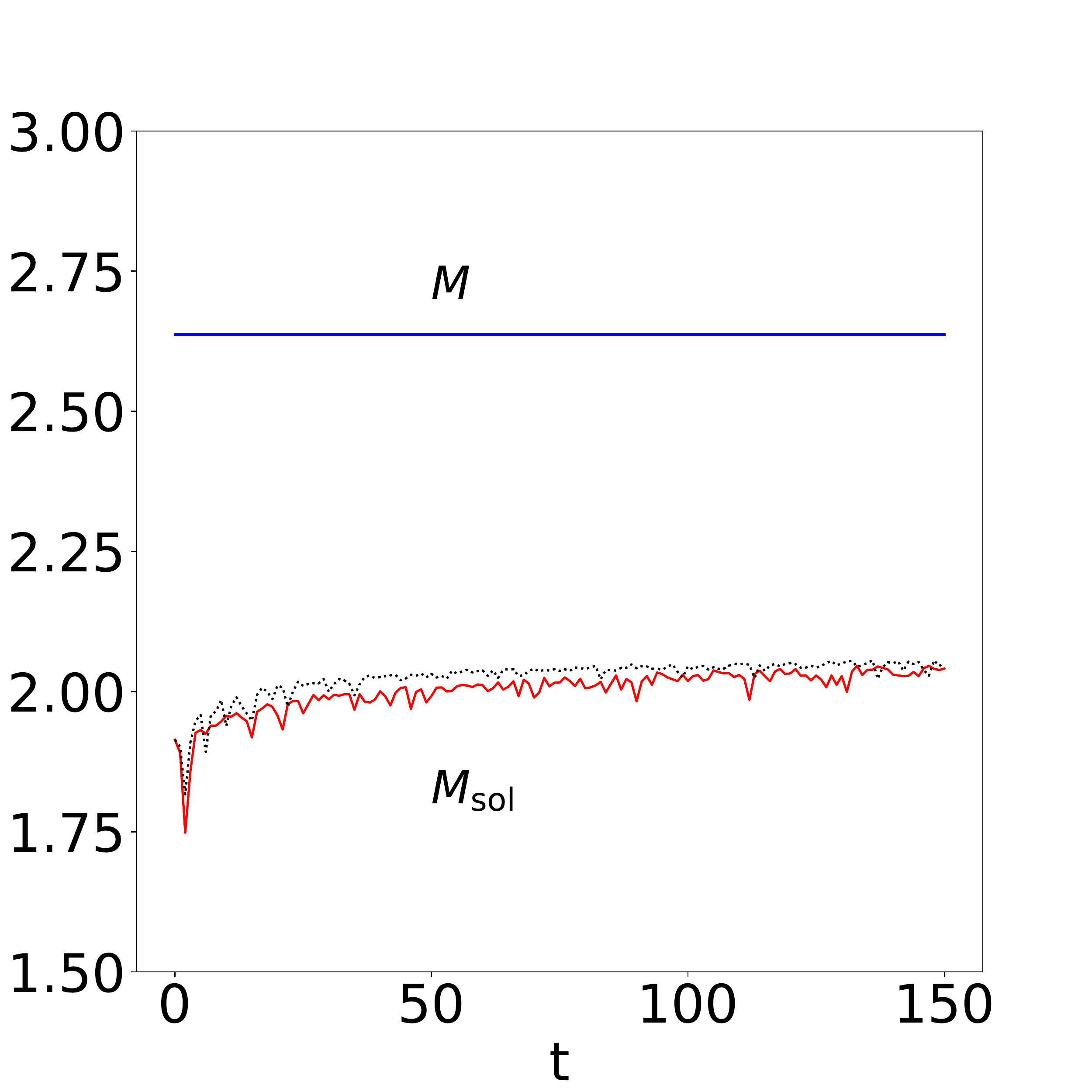}
\caption{
Evolution of a cuspy system with $R_{\rm sol}=0.5$, $\rho_{0{\rm sol}}=8$.
}
\label{fig:rho-plane-cos-R0p5-rho8}
\end{figure*}

\begin{figure*}[ht]
\centering
\includegraphics[height=4.cm,width=0.24\textwidth]{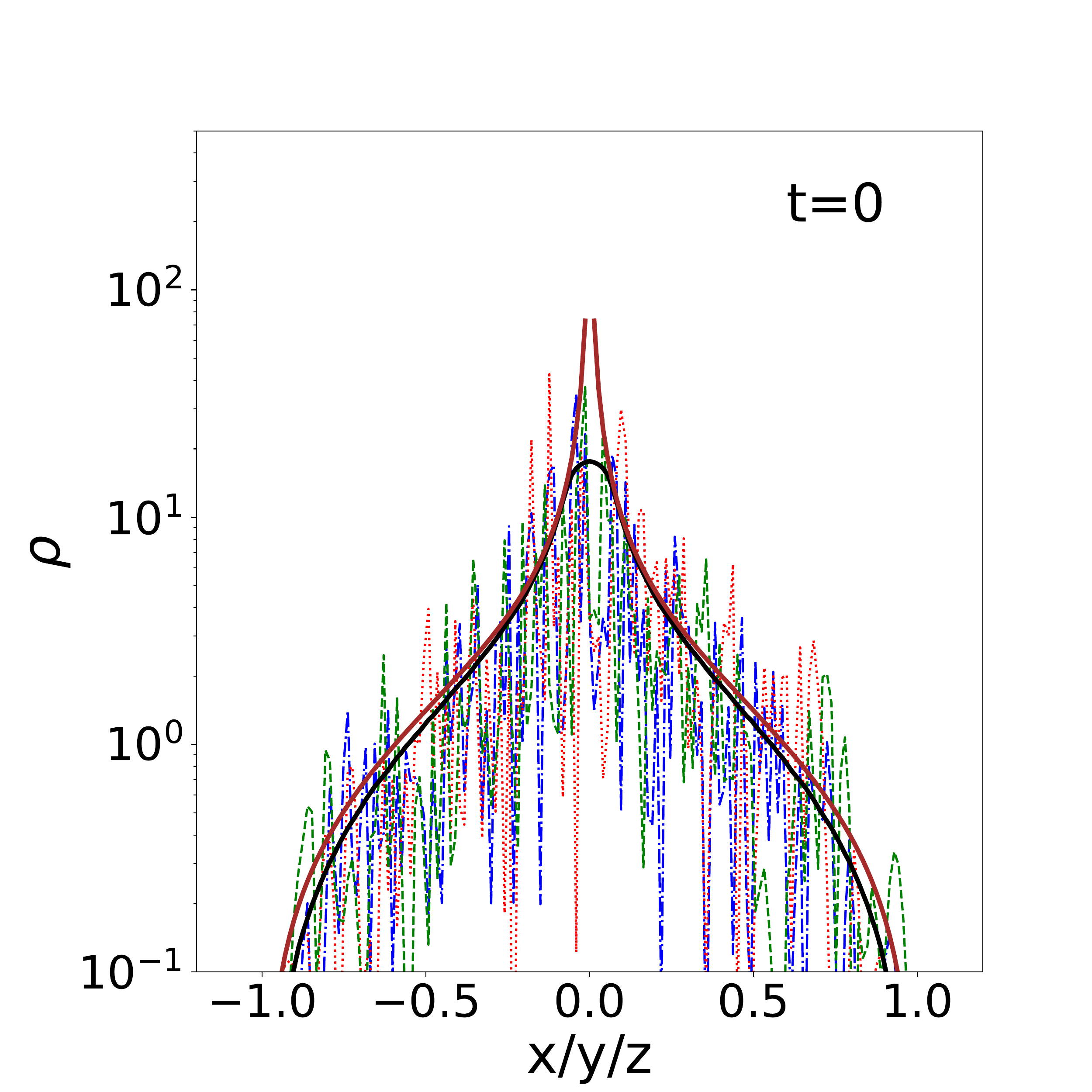}
\includegraphics[height=4.cm,width=0.25\textwidth]{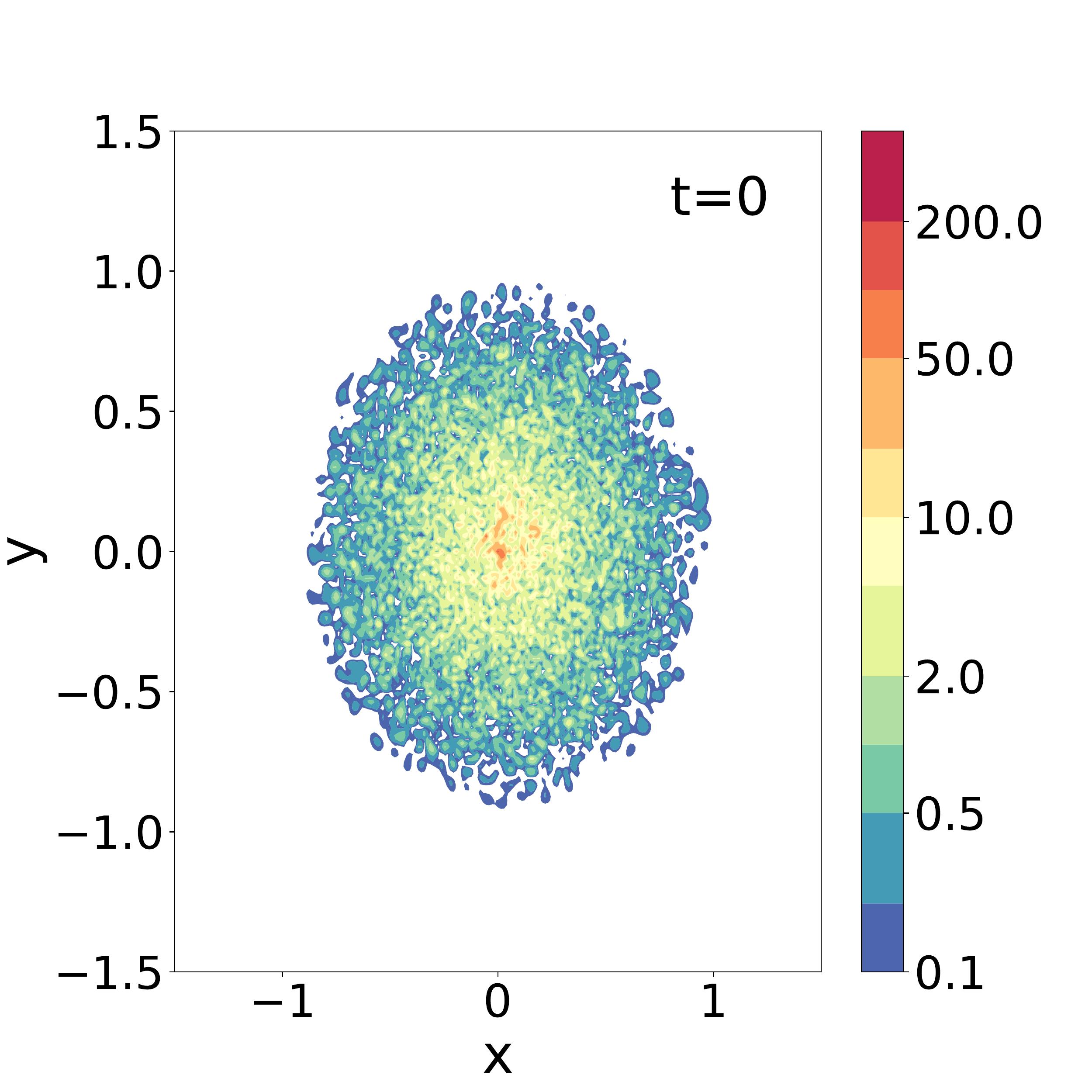}
\includegraphics[height=4.cm,width=0.24\textwidth]{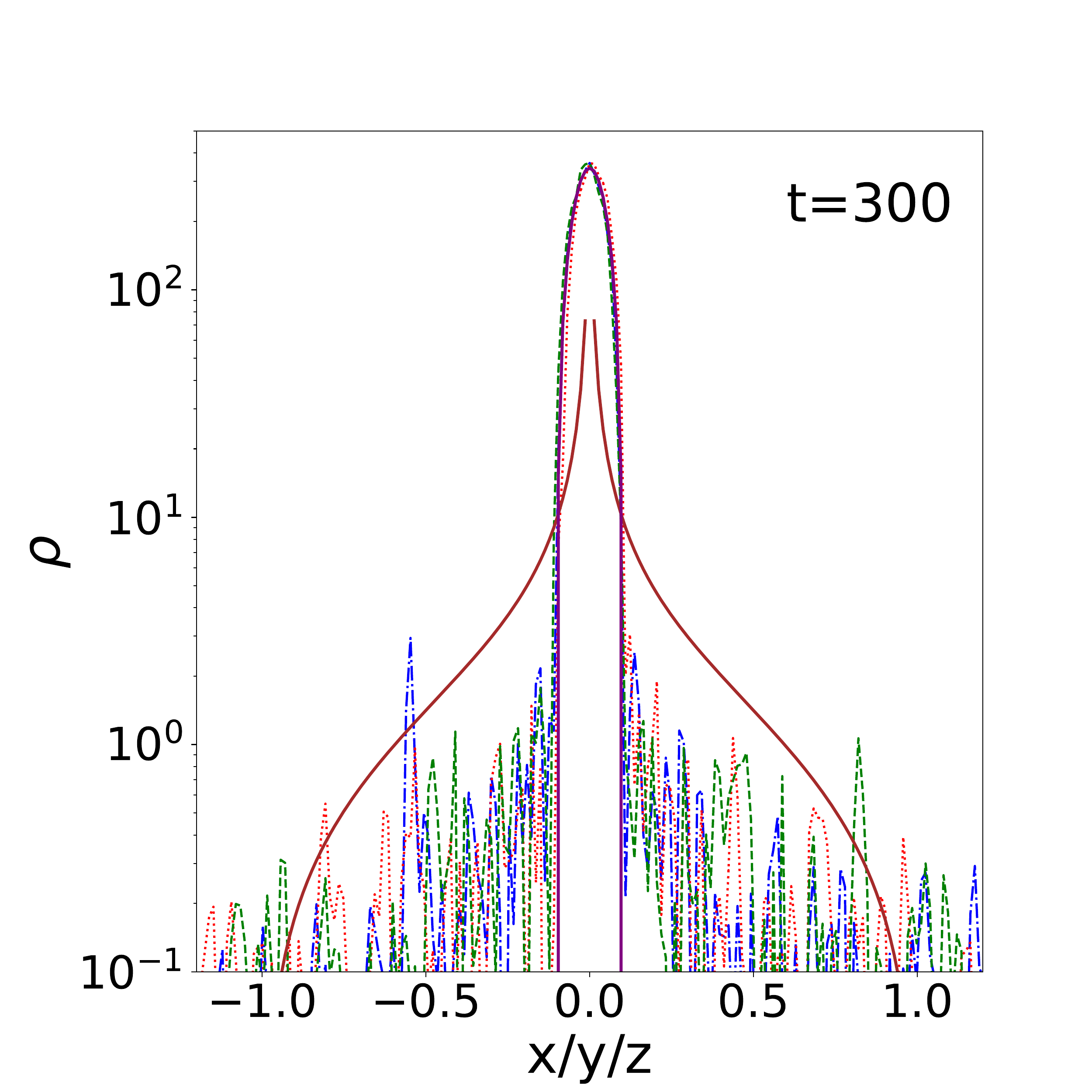}
\includegraphics[height=4.cm,width=0.25\textwidth]{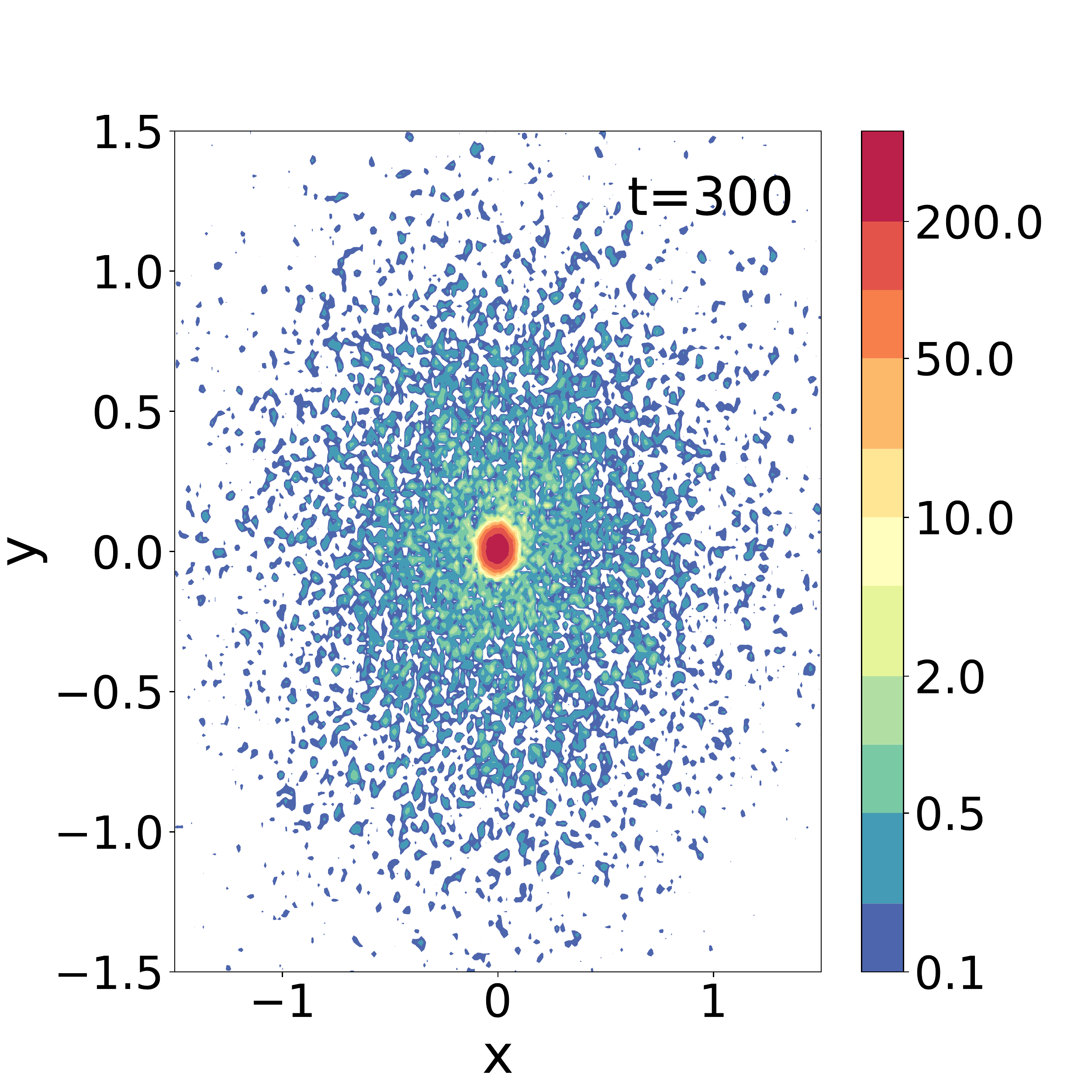}\\
\includegraphics[height=4.2cm,width=0.32\textwidth]{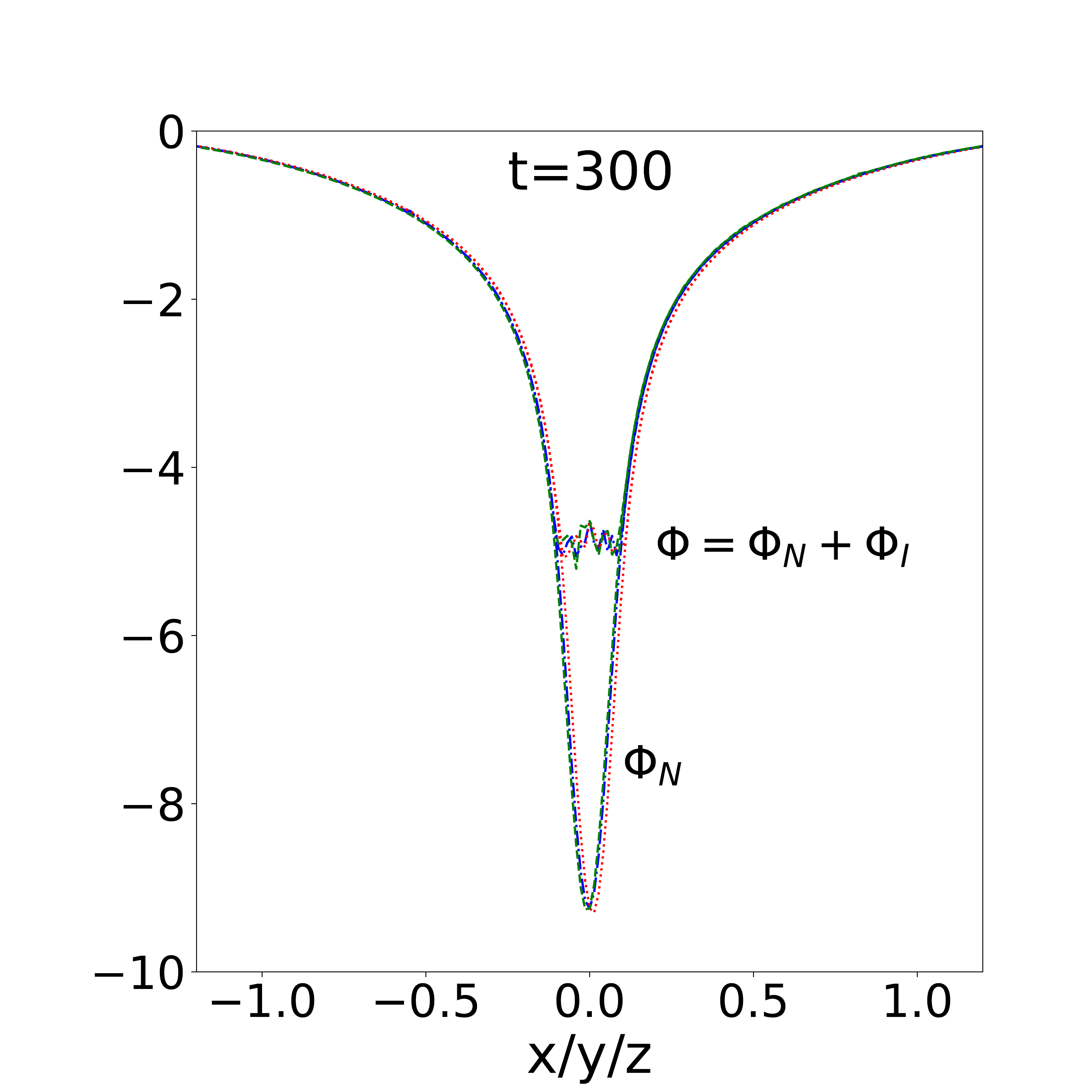}
\includegraphics[height=4.2cm,width=0.28\textwidth]{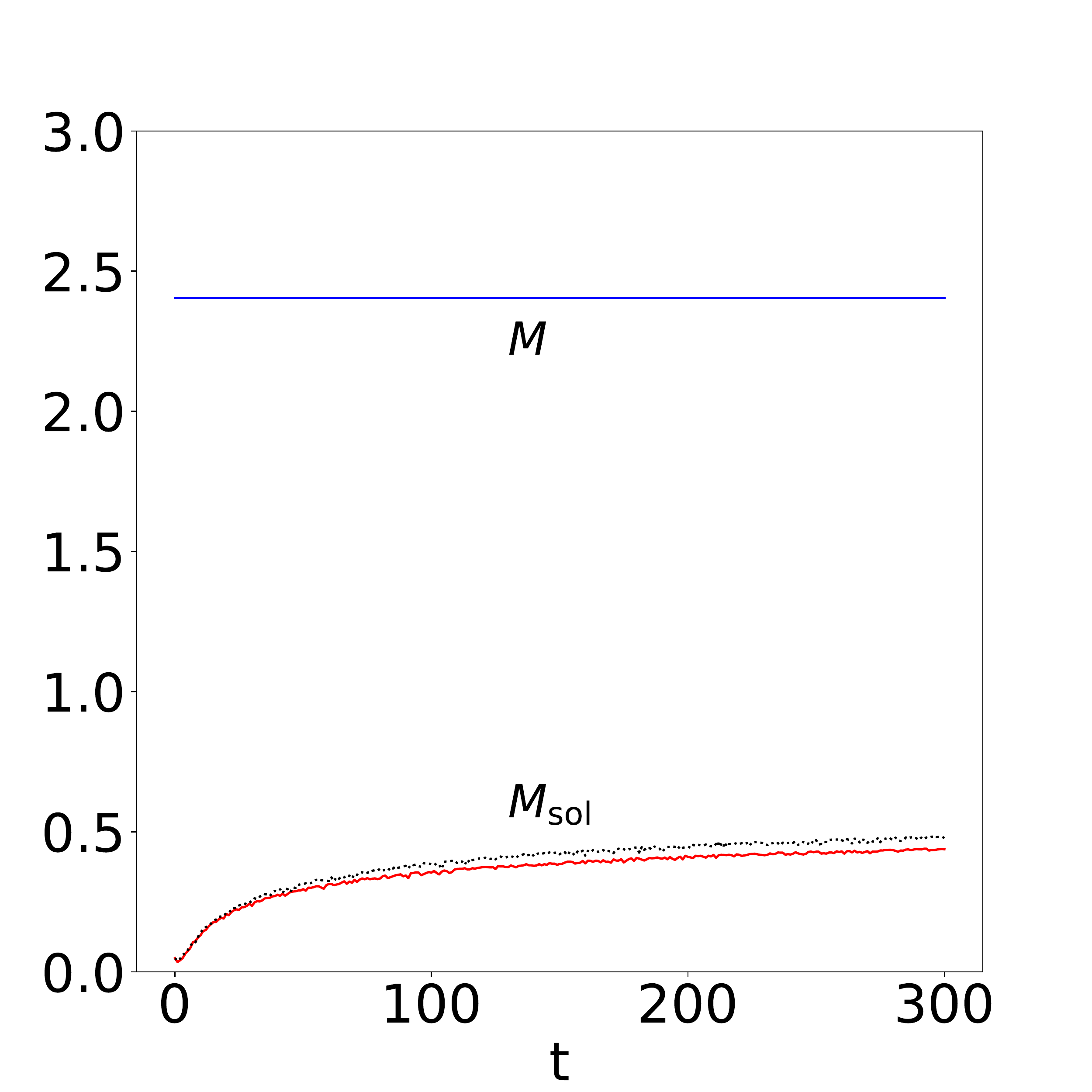}
\includegraphics[height=4.2cm,width=0.28\textwidth]{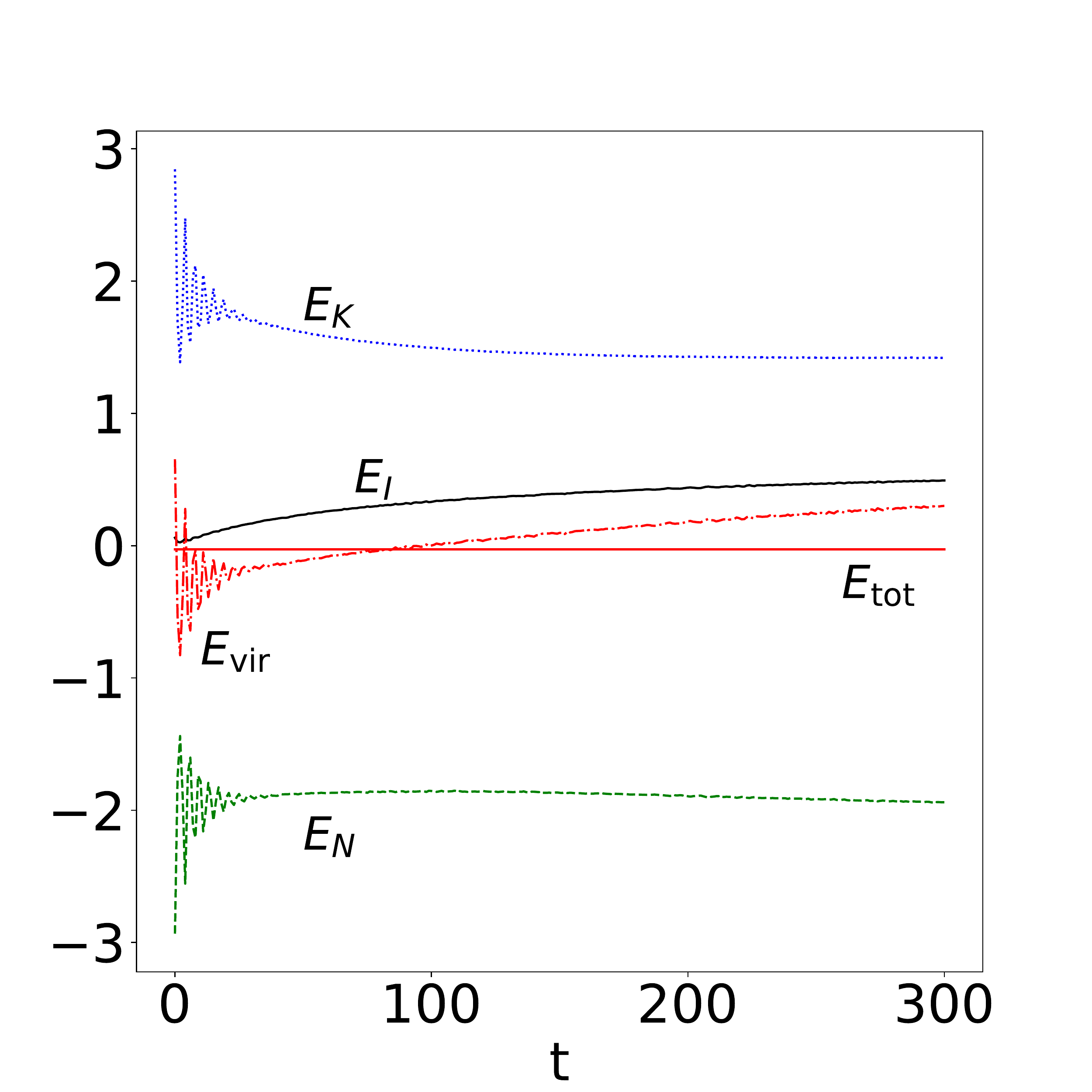}
\caption{
Evolution of a cuspy system with $R_{\rm sol}=0.1$, $\rho_{0{\rm sol}}=0$.
}
\label{fig:rho-plane-cos-R0p1-rho0}
\end{figure*}

\begin{figure*}[ht]
\centering
\includegraphics[height=4.cm,width=0.24\textwidth]{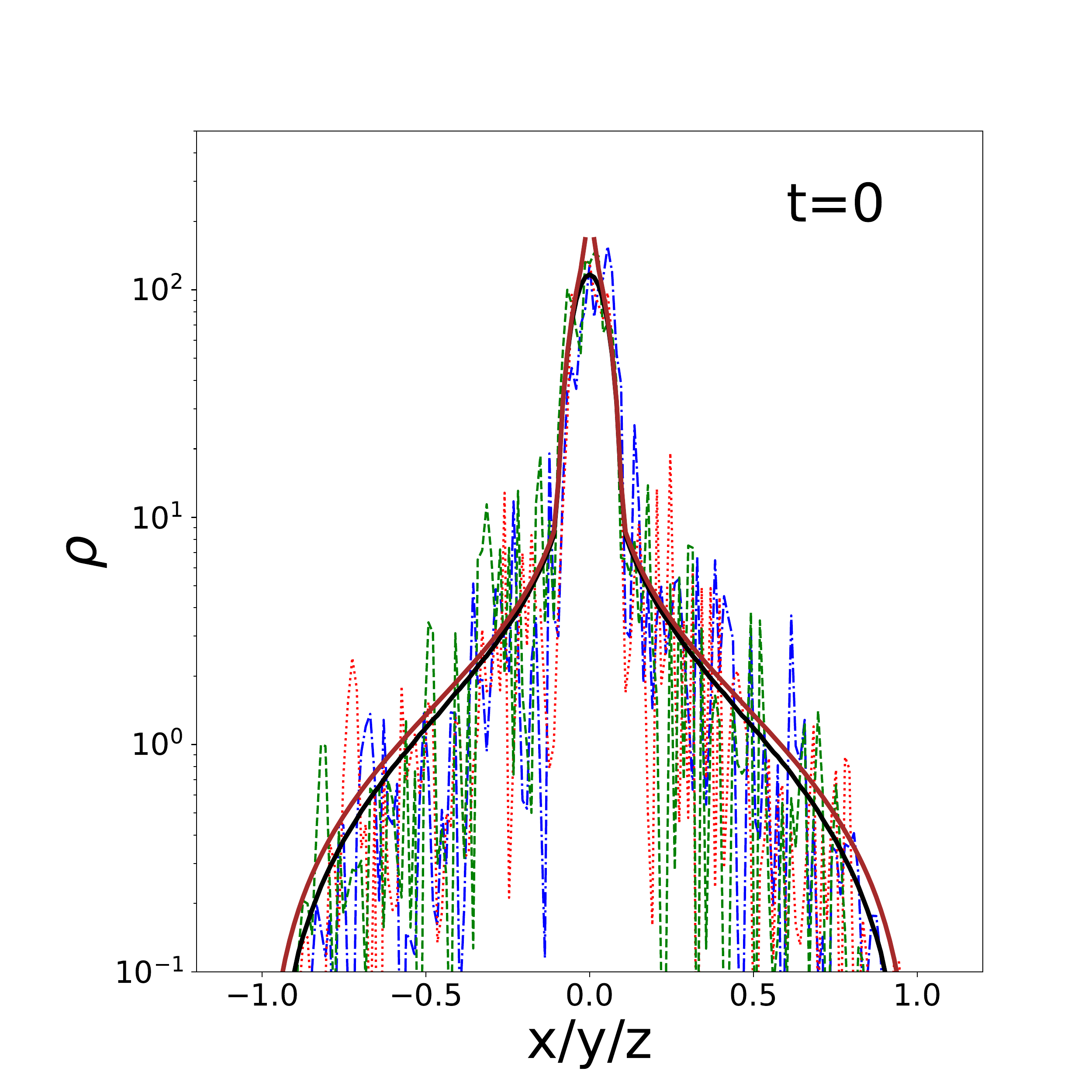}
\includegraphics[height=4.cm,width=0.24\textwidth]{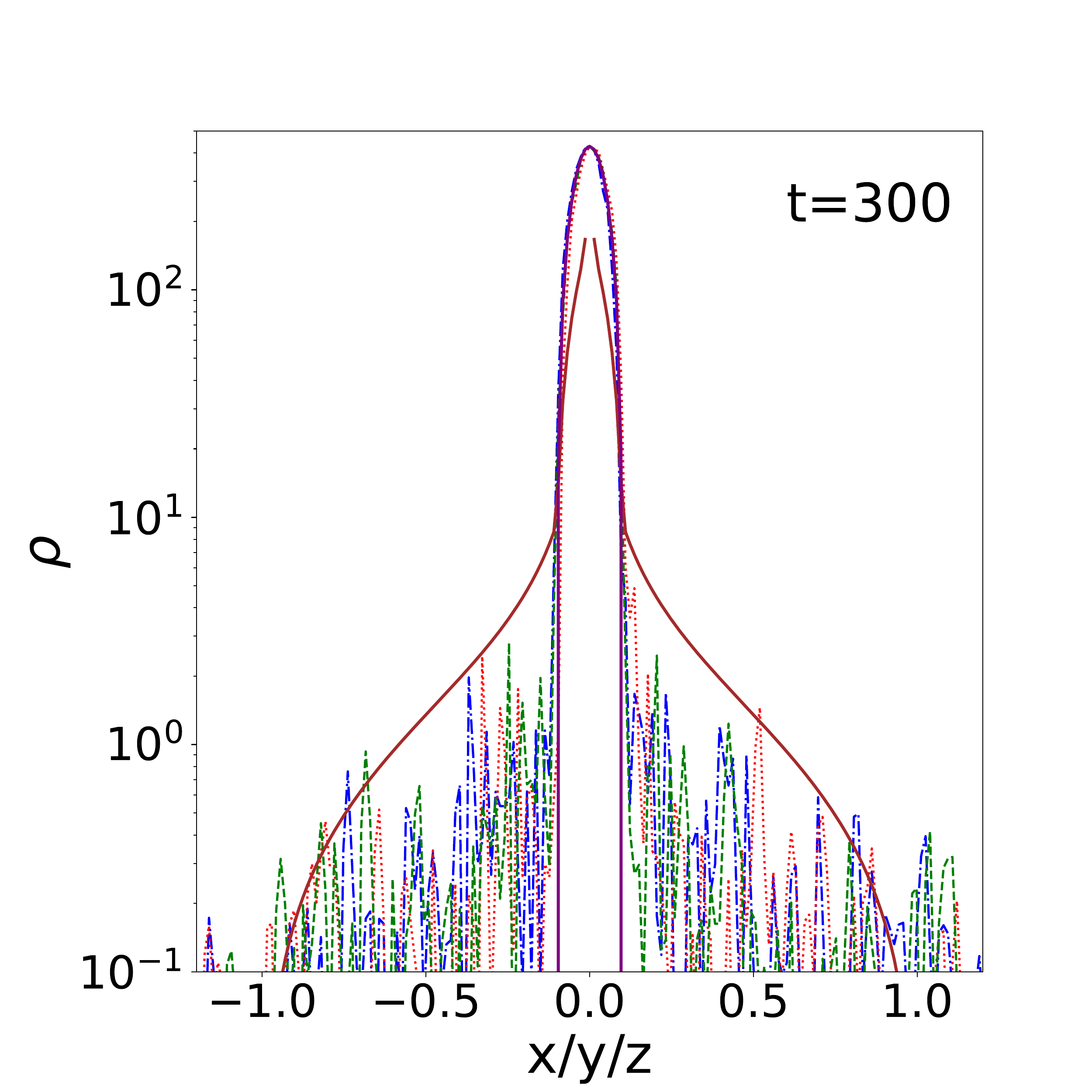}
\includegraphics[height=4.cm,width=0.25\textwidth]{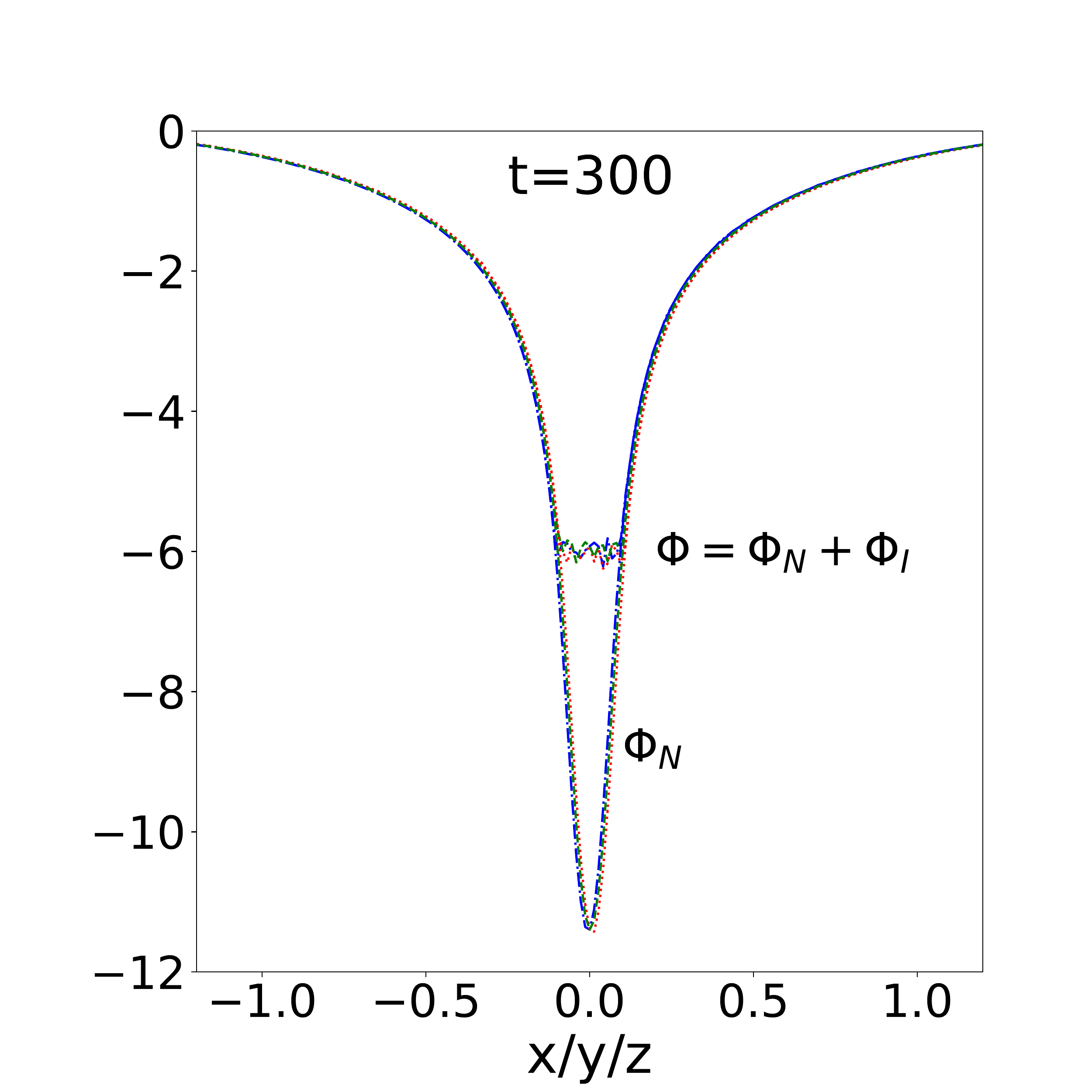}
\includegraphics[height=4.cm,width=0.24\textwidth]{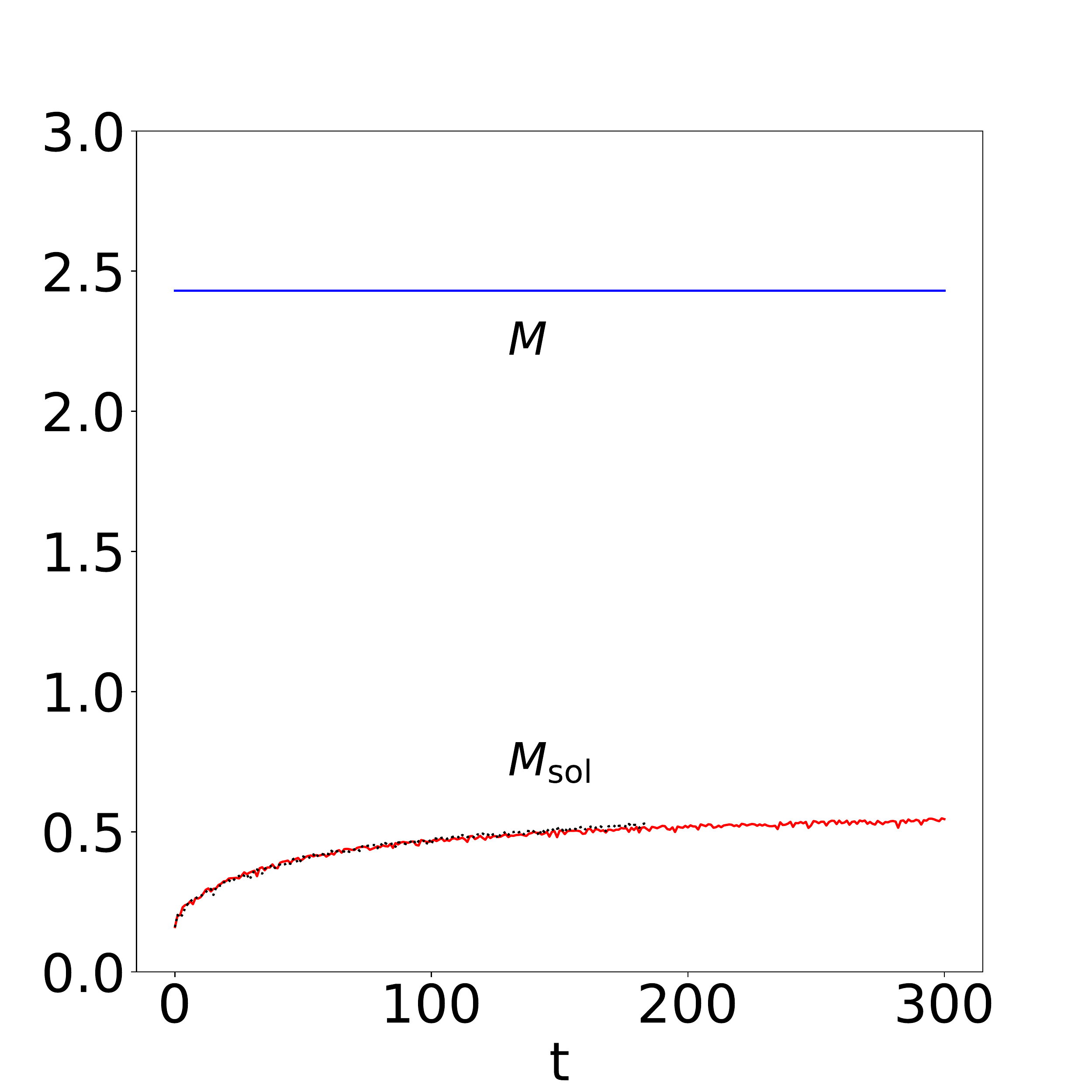}
\caption{
Evolution of a cuspy system with $R_{\rm sol}=0.1$, $\rho_{0{\rm sol}}=100$.
}
\label{fig:rho-plane-cos-R0p1-rho100}
\end{figure*}

\subsection{Halo eigenmodes}

We now study the dynamics of cuspy halos, as found in cosmological numerical
simulations of both CDM and FDM models. For simplicity, we consider an initial
target density profile 
\be
0 \leq r \leq 1: \;\;  \bar\rho(r) = \rho_0 \frac{\cos(\pi r/2)}{r} .
\label{eq:rho-cos}
\ee
This shows a cuspy profile $\rho \simeq \rho_0/r$ in the inner region, as for standard NFW and 
Hernquist profiles, while again having a finite radius $R_{\rm halo}=1$, which is convenient for
our periodic boundary conditions.
In the numerical computations we truncate this density profile at $r_{\rm cut}=0.05$, with a constant
density within $r_{\rm cut}$.
The classical phase-space distribution is obtained from the numerical integration of
Eq.(\ref{eq:Eddington}). This again gives the initial occupation numbers 
from (\ref{eq:a(E)-f(E)}).

We compute the eigenmodes and the energy levels by the same method as for the flat core
potential described in Sec.~\ref{sec:flat-core-eigenmodes}.
We show these energy levels and some radial eigenmodes in Fig.~\ref{fig:E-levels-cos}.
Because of the divergent density at the center, the gravitational potential well is deeper than for the
flat core profile studied in Sec.~\ref{sec:flat-core-eigenmodes}. This leads to a greater number of bound
states, as can be seen by comparing the energy levels in Figs.~\ref {fig:E-levels-sinc} and
\ref{fig:E-levels-cos}. 
There are $62$ energy levels for $\ell=0$ and we find bound states until $\ell_{\max}=105$.
Again, high-$n$ modes probe small scales, down to $\Delta r \sim \epsilon=0.01$, while high-$\ell$
modes probe large radii.

\subsection{Large soliton radius, $R_{\rm sol}=0.5$}

We show in Fig.~\ref{fig:rho-plane-cos-R0p5-rho0} the dynamics of a cuspy halo
(\ref{eq:rho-cos}) in the case $R_{\rm sol}=0.5$ without initial soliton.
As for the flat halo shown in Fig.~\ref{fig:rho-plane-sinc-R0p5-rho0}, we can see
in panel (a) that the WKB approximation for the coefficients $a_{n\ell m}$
provides a reasonably good agreement between the averaged density $\langle\rho_{\rm halo}\rangle$
and the target density (\ref{eq:rho-cos}). 
Again, the interferences between the different modes $\psi_{n\ell m}$ lead
to strong relative density fluctuations of order unity, in agreement with (\ref{eq:Proba-rho-halo}).

As in the flat halo case of Fig.~\ref{fig:rho-plane-sinc-R0p5-rho0}, we can see that in a few
dynamical times, $t \lesssim 4$, a central soliton of radius $R_{\rm sol}=0.5$ forms and
contains about $40\%$ of the total mass. This relaxation depletes the halo that also diffuses
beyond its initial unit radius, as the process occurs in a rather fast and violent manner.
The growth of the soliton mass obtained by the larger box simulation is close to the result of our fiducial simulation.
The shape and size of this soliton, governed by the
self-interactions, can be clearly seen in the final density profiles and in the final total potential $\Phi$,
which is flat over the extent of the soliton. However, in addition to the small wiggles associated
with high-energy modes that run across the center of the system, there remains a very high-density
spike on top of the soliton. Its width $\Delta x \sim \epsilon$ shows that it is not
supported by the self-interactions, but it is a small-scale peak on the de Broglie wavelength
that wanders over the extent of the former soliton.
Thus, in the central region there is a co-existence of the two types of features, a large smooth soliton
where gravity is balanced by the self-interactions, and one or a few high-density spikes on the de Broglie
wavelength that are far from hydrostatic equilibrium.
This is embedded within a halo of fluctuating high-energy modes.
However, it is possible that all such narrow spikes are eventually absorbed by the smooth soliton
on timescales beyond our simulation time.

As for the flat case of Fig.~\ref{fig:rho-plane-sinc-R0p5-rho5}, the virial quantity $\tilde{E}_{\rm vir}$ is closer
to zero than $E_{\rm vir}$. Again, this is due to the large value of the kinetic energy $E_K$, associated
with the fluctuations of the wave functions over the soliton and in the outer halo.
This makes the system look far from equilibrium, as seen in terms of $E_{\rm vir}$, even though the
density profile has already converged to the soliton shape in the mean.

We show in Fig.~\ref{fig:rho-plane-cos-R0p5-rho8} the initial and final density profiles when
we start with a soliton of density $\rho_{0{\rm sol}}=8$ on top of the halo. As for the case of a flat halo
shown in Fig.~\ref{fig:rho-plane-sinc-R0p5-rho5}, the system quickly reaches an equilibrium close to the
initial state, where the soliton has slightly increased its mass and depleted the halo.
The most striking result is that, as for the flat case shown in Fig.~\ref{fig:rho-plane-sinc-R0p5-rho5},
the random fluctuations inside this soliton have been significantly damped.
Thus, the soliton appears to be an attractor, damping stochastic perturbations.
Here there no longer remains high-density spikes of width $\sim \epsilon$ inside the soliton, which suggests
that this is a transient feature.
Again, the larger box simulation gives the same results as our fiducial simulation.

\subsection{Small soliton radius, $R_{\rm sol}=0.1$}

We now consider cases where the radius $r_a$ associated with the self-interactions
is much smaller than the halo radius, taking again $R_{\rm sol}=0.1$
as for the flat case.

\subsubsection{No initial soliton}

We first consider an initial profile without soliton, shown in Fig.~\ref{fig:rho-plane-cos-R0p1-rho0}.
Because of the high density at the center, $\rho_{\rm halo} \simeq \rho_0/r$, the self-interaction
$\Phi_I = \lambda\rho$ is large in the central region. This leads to the formation of a central soliton
supported by the self-interaction in a few dynamical times, $t \lesssim 2$.
This again depletes somewhat the halo, which diffuses slightly beyond its initial radius, while
the fluctuations inside the soliton are damped.
In contrast with Fig.~\ref{fig:rho-plane-cos-R0p5-rho0} 
there is no narrow density spike, supported by the quantum pressure, inside this soliton.
This is presumably because the hierarchy of scale between the de Broglie wavelength and the
self-interaction soliton is not so large, only a factor ten instead of fifty.

The larger box simulation gives the same soliton growth rate as our fiducial
simulation. 
As for the flat case shown in Fig.~\ref{fig:rho-plane-sinc-R0p1-rho0}, the energy components show
damped collective oscillations at early stages while $E_{\rm vir}$ remains close to zero as the system
is dominated by the quasi-stationary halo. Nevertheless, $E_{\rm vir}$ again shows a slow growth at later
times, because of the growth of the soliton and the remaining fluctuations that still provide a significant
kinetic energy.

\subsubsection{Small initial soliton}

We show in Fig.~\ref{fig:rho-plane-cos-R0p1-rho100} the case where there is an initial soliton
of density $\rho_{0{\rm sol}}=100$. Again, the central soliton density grows somewhat with time and damps
the central fluctuations, while the halo diffuses slightly beyond its initial radius.
The larger box simulation results for the soliton growth rate are very close to
our fiducial simulation and can hardly be distinguished in the figure.

Thus, as for the case of a flat halo, we find that the solitons governed by the balance between gravity and
the self-interactions are robust and always form, either in a few dynamical times if the initial density
is high enough, or after small-scale density fluctuations are grown large enough by a slow secular process
to trigger an instability and a fast soliton formation.

\section{Kinetic theory}
\label{sec:kine}

\subsection{Kinetic equation}

We now derive a kinetic equation for the evolution with time of the system, that is,
for the occupation numbers of the central soliton and of the halo eigenstates.
A similar approach was presented in \citep{Chan:2022bkz,Jain:2023ojg} for the
formation of FDM solitons inside an homogeneous background, which can be decomposed
over plane waves.
We go beyond these results by taking into account the self-interactions and
the non-homogeneous background. The latter can no longer be decomposed over plane
waves (i.e., we can no longer use Fourier analysis). However, as described below,
it is possible to derive simple kinetic equations by decomposing the background
over eigenmodes of a reference potential, in a fashion similar to the description
of the halo in Sec.~\ref{sec:Eigenmodes}.
In addition, as one follows the evolution along this time-dependent background, one needs
to separate this smooth background from the stochastic fluctuations that drive the
dynamics.

The equation of motion (\ref{eq:Schrod-eps}) is the Schr\"odinger equation in a self-force
potential $\Phi=\Phi_N+\Phi_I$, sourced by the system self-gravity and self-interaction.
For the quartic scalar-field model considered in this paper, $\Phi$ is quadratic over the wavefunction
$\psi$, and the equations of motion read
\be
i \epsilon \frac{\partial\psi}{\partial t} = - \frac{\epsilon^2}{2} \nabla^2\psi + \Phi \psi ,
\label{eq:Schrod-Phi}
\ee
with
\be
\Phi = (4\pi \nabla^{-2}+\lambda) \psi^\star \psi .
\label{eq:Phi-def}
\ee
If the potential $\Phi$ is fixed, $\psi(\vec x,t)$ can be decomposed as usual in energy eigenmodes
with the simple time dependence $e^{-i E t/\epsilon}$.
In the semiclassical limit, the system behaves like a collection of classical particules and the Husimi
phase-space distribution $f_H(\vec x,\vec v,t)$ \citep{Husimi1940} defined from $\psi(\vec x,t) $ approximately follows
the Vlasov equation that governs the dynamics of the classical distribution $f(\vec x,\vec v,t)$ \citep{Uhlemann2014,Mocz2018}.
As described in Sec.~\ref{sec:halo-eigenfunctions}, in this limit $\epsilon \ll 1$, we can build approximate
equilibrium configurations by choosing the eigenmode coefficients $a_{n\ell m}$ in correspondance
with a classical phase-space equilibrium solution, as in Eqs.(\ref{eq:a-E}) and
(\ref{eq:a(E)-f(E)}).
This procedure would give true equilibria if the potential $\Phi$ were only sourced by the
average density $\langle \rho \rangle$, which neglects interferences between different eigenmodes
as in Eq.(\ref{eq:rho-average-nl}).
However, as shown in Eq.(\ref{eq:Proba-rho-halo}) and in the plots of the initial conditions displayed
in the previous sections, the interference terms lead to significant fluctuations of the density profile.
They have a relative magnitude of order unity but a spatial width that decreases as $\epsilon$.
Hence they only become small in a coarse-graining sense.
These random fluctuations mean that even if we start with an equilibrium configuration in this averaged
sense, the system will not be exactly stationary as the potential $\Phi$ deviates from $\langle \Phi \rangle$.
To describe this system, we therefore split the potential $\Phi$ in an average spherically symmetric part
$\bar\Phi$ and a fluctuating part $\delta\Phi$,
\be
\bar\Phi(\vec x,t) = \bar \Phi(r) + \delta \Phi(\vec x,t) .
\label{eq:Phi-split}
\ee
Within an adiabatic approximation, we have in mind that the smooth potential $\bar\Phi$
slowly evolves on long time scales whereas the incoherent stochastic fluctuations 
$\delta \Phi$ evolve on short times and drive the averaged dynamics, as would do an
external noise for instance.
The potential $\bar\Phi$ defines the energy eigenmodes $\psi_j$,
\be
\psi_j(\vec x,t) = e^{-i E_j t/\epsilon} \hat\psi_j(\vec x) , \;\;\; \hat\psi_{n\ell m}(\vec x)
= {\cal R}_{n\ell}(r) Y_{\ell m}(\theta,\varphi) ,
\ee
where the index $j$ denotes $\{n,\ell,m\}$ and for future convenience we use the real spherical harmonics
$Y_{\ell m}$ (also called tesseral spherical harmonics) instead of the more usual complex harmonics,
\bea
&& m < 0: \;\; Y_{\ell m} = \sqrt{2} \; {\rm Im} Y_\ell^{|m|} , \;\;\; Y_{\ell 0} = Y_\ell^0 , \nonumber \\
&& m> 0 : \;\; Y_{\ell m} = \sqrt{2} \; {\rm Re} Y_\ell^m .
\label{eq:Y_lm_l^m}
\eea
Therefore, the functions $\hat\psi_j(\vec x)$ are real and form a complete orthonormal basis.
We can then expand the wavefunction $\psi$ over this basis as
\be
\psi(\vec x,t) = \sum_j \sqrt{M_j(t)} e^{-i \theta_j(t)/\epsilon} \hat\psi_j(\vec x) ,
\label{eq:psi-Mj-thetaj}
\ee
where $M_j \geq 0$ and $\theta_j$ are real. The squared amplitude $M_j$ is the mass contained in the
eigenmode $j$, if we neglect interferences.
Substituting this expansion into the equation of motion (\ref{eq:Schrod-Phi}) we obtain
\bea
i \epsilon \dot M_j + 2 M_j \dot \theta_j & = & 2 M_j E_j + \sum_{j'} 2 \sqrt{M_j M_{j'}}
e^{i (\theta_j-\theta_{j'})/\epsilon} \nonumber \\
&& \times \int d\vec x \, \hat\psi_j \, \delta\Phi \, \hat\psi_{j'}
\label{eq:Ij-dot-1}
\eea
where the dots denote the derivatives with respect to time.
We define the reference potential $\bar\Phi$ as the sum of the diagonal terms,
\be
\bar\Phi = (4\pi\nabla^{-2} + \lambda) \sum_j M_j \hat\psi_j^2 ,
\label{eq:bar-Phi-def}
\ee
while the remainder $\delta\Phi$ is given by the off-diagonal interference terms,
\be
\delta\Phi = (4\pi\nabla^{-2} + \lambda) \sum_{j\neq j'} \sqrt{M_j M_{j'}} e^{i(\theta_j-\theta_{j'})/\epsilon}
\hat\psi_j \hat\psi_{j'} .
\label{eq:delta-Phi-def}
\ee
Then, we assume that $\bar\Phi$ evolves slowly with time, so that we can neglect its time dependence
(i.e., the time dependence of the eigenmodes $\hat\psi_j$) in the equation of motion
(\ref{eq:Ij-dot-1}). The evolution of the system is due to the small fluctuations $\delta\Phi$, associated
with the interferences. They perturb the occupation numbers $M_j$ of the various energy levels,
which will in turn affect the reference potential $\bar\Phi$ as the density profile slowly changes.
However, in an adiabatic approximation, the slow change of $\bar\Phi$ only leads to a change of the phase
of the eigenmodes (and of the energy levels) while keeping their occupation number fixed.
Therefore, we focus here on the driving mechanism associated with $\delta\Phi$.
Then, we can write the equation of motion (\ref{eq:Ij-dot-1}) as
\bea
&& i \epsilon \dot M_1 + 2 M_1 \dot \theta_1 = 2 M_1 E_1 + \sum_{234}^{2\neq4} 2
\sqrt{M_1 M_2 M_3 M_4}  \nonumber \\
&& \times e^{i (\theta_1+\theta_2-\theta_3-\theta_4)/\epsilon}
\int d\vec x \, \hat\psi_1 \hat\psi_3 (4\pi\nabla^{-2}+\lambda) \hat\psi_2 \hat\psi_4 , \hspace{1cm}
\label{eq:Ij-dot-2}
\eea
where the indices $\{1,2,3,4\}$ denote $\{j_1,j_2,j_3,j_4\}$.
Let us define the vertices $V_{13;24}$ as
\be
V_{13;24} = \int d\vec x \, \hat\psi_1 \hat\psi_3 (4\pi\nabla^{-2}+\lambda) \hat\psi_2 \hat\psi_4 ,
\label{eq:V-1324-def}
\ee
which are real and symmetric over $\{1\leftrightarrow 3 \}$, $\{2 \leftrightarrow 4\}$
and $\{(13) \leftrightarrow (24)\}$.
Then, separating the real and imaginary parts of Eq.(\ref{eq:Ij-dot-2}) gives
\bea
&& \hspace{-0.2cm} \epsilon \dot M_1 = 2 \sum_{234}^{2\neq4} \sqrt{M_1 M_2 M_3 M_4} V_{13;24}
\sin\!\left( \frac{\theta_1\!+\!\theta_2\!-\!\theta_3\!-\!\theta_4}{\epsilon} \right) , \nonumber \\
&& \hspace{-0.2cm} \dot \theta_1 = E_1 + \sum_{234}^{2\neq4} \sqrt{ \frac{M_2 M_3 M_4}{M_1} }
V_{13;24} \cos\!\left(\frac{\theta_1\!+\!\theta_2\!-\!\theta_3\!-\!\theta_4}{\epsilon}\right) . \nonumber \\
&&\label{eq:I1-dot-3}
\eea
To avoid secular effects, associated with trivial resonances between products of identical oscillatory
terms \citep{Nazarenko_2011}, we define the renormalized frequencies $\omega_j$ as
\be
\omega_1 = E_1 + \sum_2^{2\neq 1} M_2 \, V_{12;21}  .
\label{eq:omega_1-def}
\ee
Then, the system (\ref{eq:I1-dot-3}) also reads
\bea
&& \dot M_1 = \frac{2\gamma}{\epsilon} \sum_{234} \sqrt{M_1 M_2 M_3 M_4} \; \hat V_{13;24} \;
\sin (\theta_{12}^{34}/\epsilon) , \nonumber \\
&& \dot \theta_1 = \omega_1 + \gamma \sum_{234} \sqrt{ \frac{M_2 M_3 M_4}{M_1} } \; \hat V_{13;24} \;
\cos (\theta_{12}^{34}/\epsilon) , \hspace{0.5cm}
\label{eq:I1-dot-4}
\eea
where we introduced the notation
\be
\theta_{12}^{34} = \theta_1+\theta_2-\theta_3-\theta_4 
\ee
and the new vertices $\hat V_{13;24}$ defined as
\be
\hat V_{13;24} = V_{13;24} \;\; \mbox{except} \;\; \hat V_{13;22}= 0 , \;\; 
\hat V_{12;21} = 0 .
\label{eq:hat-V-def}
\ee
We also introduced a book-keeping parameter $\gamma=1$ that multiplies the vertices 
$\hat V$, i.e. the potential $\delta\Phi$.
We have in mind that the fluctuating part $\delta\Phi$ leads to a slow drift of the system
as compared with the orbital motions in the mean potential $\bar\Phi$. Therefore, we will develop
a perturbation theory in $\delta\Phi$, which corresponds to a perturbation theory in powers
of $\gamma$ (taking $\gamma=1$ at the end).

The system (\ref{eq:I1-dot-4}) is similar to those encountered in four-wave systems
\citep{Nazarenko_2011,Onorato_2020}.
However, the vertices $\hat V$ are no longer fully symmetric and do not contain Kr\"onecker 
symbols $\delta_{12}^{34}$ in wavenumbers.
This is because we expand around a non-homogeneous equilibrium $\bar\Phi$, with a peculiar
radial density profile $\bar\rho(r)$. This breaks the invariance over translations obeyed by wave systems
over a uniform background.

We now look for the perturbative expansion of the squared amplitudes $M_j$ and the phases $\theta_j$
in powers of $\gamma$,
\be
M_j = M_j^{(0)} + \gamma \, M_j^{(1)} + \gamma^2 M_j^{(2)} + \dots
\ee
At zeroth order we obtain
\be
M_1^{(0)}(t) = \bar M_1 , \;\;\; \theta_1^{(0)}(t) =  \bar \theta_1 + \bar\omega_1 t ,
\ee
with $\bar M_1= M_1(0)$, $\bar\theta_1 = \theta_1(0)$, setting the initial conditions of the system
at the time $t=0$.
At first order we obtain
\be
\dot M_1^{(1)} = \frac{2}{\epsilon} \sum_{234} \sqrt{\bar M_1 \bar M_2 \bar M_3 \bar M_4} \;
\hat V_{13;24} \; \sin [ (\bar\theta_{12}^{34} + \bar\omega_{12}^{34} t)  /\epsilon] ,
\label{eq:dot-I_1}
\ee
and
\bea
M_1^{(1)}(t) & = & 2 \sum_{234} \sqrt{\bar M_1 \bar M_2 \bar M_3 \bar M_4} \;
\frac{\hat V_{13;24}}{\bar\omega_{12}^{34}} \left[ \cos (\bar\theta_{12}^{34}/\epsilon) \right. \nonumber \\
&& \left. - \cos[ (\bar\theta_{12}^{34} + \bar\omega_{12}^{34} t)/\epsilon] \right] ,
\eea
\bea
\theta_1^{(1)}(t) & = & \epsilon \sum_{234} \sqrt{ \frac{\bar M_2 \bar M_3 \bar M_4}{\bar M_1} } \;
\frac{\hat V_{13;24}}{\bar\omega_{12}^{34}} \left[
\sin [ (\bar\theta_{12}^{34} + \bar\omega_{12}^{34} t)/\epsilon] \right. \nonumber \\
&& \left. - \sin (\bar\theta_{12}^{34}/\epsilon) \right] .
\eea
At second order, using trigonometric identities we obtain
\bea
&& \hspace{-0.3cm} \dot M_1^{(2)} = \sum_{234} \hat V_{13;24} \sum_{m=1}^4 \sum_{567}
\frac{\hat V_{m6,57}}{\bar\omega_{m5}^{67}}
 \sqrt{ \frac{\bar M_1 \bar M_2 \bar M_3 \bar M_4 \bar M_5 \bar M_6 \bar M_7}{\bar M_m} } \;  \nonumber \\
&& \times  \frac{2}{\epsilon} \left[ \sin[ (\bar\theta_{12}^{34} + \bar\omega_{12}^{34} t
- \sigma_m \bar\theta_{m5}^{67} ) /\epsilon] \right. \nonumber \\
&& \left. - \sin[ ( \bar\theta_{12}^{34} + \bar\omega_{12}^{34} t - \sigma_m \bar\theta_{m5}^{67} -
\sigma_m \bar\omega_{m5}^{67} t ) /\epsilon ]  \right] ,
\label{eq:dot-I_2}
\eea
where we introduced $\sigma_1=\sigma_2=1$, $\sigma_3=\sigma_4=-1$.

At zeroth order we have $\dot M_1^{(0)}=0$. At first order we obtain from Eq.(\ref{eq:dot-I_1})
\be
\langle \dot M_1^{(1)} \rangle = 0 ,
\ee
assuming that the initial phases $\bar\theta_j$ are uncorrelated and uniformly distributed over
$[0,2\pi[$, as in (\ref{eq:a-E}). Here we used the properties (\ref{eq:hat-V-def}) of the non-symmetric
vertex $\hat V$.
At second order we obtain from Eq.(\ref{eq:dot-I_2})
\bea
&& \hspace{-0.3cm}  \langle \dot M_1^{(2)} \rangle = \frac{2}{\epsilon} \sum_{234} \bar M_1 \bar M_2
\bar M_3 \bar M_4 \biggl \lbrace \frac{ \sin(\bar\omega_{12}^{34} t/\epsilon) }{\bar\omega_{12}^{34}}
\hat V_{13;24} \nonumber \\
&& \hspace{-0.3cm} \times \left[ \frac{\hat V_{13;24}+\hat V_{14;23}}{\bar M_1} + \frac{\hat V_{23;14}
+\hat V_{24;13}}{\bar M_2}
- \frac{\hat V_{31;42}+\hat V_{32;41}}{\bar M_3} \right. \nonumber \\
&& \hspace{-0.3cm} \left. - \frac{\hat V_{41;32}+\hat V_{42;31}}{\bar M_4} \right]
+ \frac{ \sin(\bar\omega_{1}^{3} t/\epsilon) }{ \bar\omega_{1}^{3} } \hat V_{12;23}
\left[ \frac{\hat V_{14;43}}{\bar M_1} - \frac{\hat V_{34;41}}{\bar M_3} \right] \nonumber \\
&& \hspace{-0.3cm} +  \frac{ \sin(\bar\omega_{2}^{4} t/\epsilon) }{ \bar\omega_{2}^{4} } \hat V_{23;34}
\frac{ \hat V_{14;21} - \hat V_{12;41} }{\bar M_2} \biggl \rbrace  ,
\label{eq:dot-I_2-a}
\eea
where we used the properties and symmetries of the vertices $\hat V$ and $V$.
In usual four-wave systems over an homogeneous background, with a symmetric vertex $\hat V$,
the last two terms vanish and the first term simplifies as $\sin(\bar\omega_{12}^{34} t/\epsilon)
(2\hat V_{1234}^2 / \bar\omega_{12}^{34}) ( 1/\bar M_1+1/\bar M_2 - 1/\bar M_3 - 1/\bar M_4 )$.
In our case, the inhomogeneous background leads to the more complicated expression
(\ref{eq:dot-I_2-a}).

\subsection{Soliton ground state and halo excited states}

We are interested in hydrostatic solitons embedded within a halo formed by a quasi-continuum of
excited states, as described in Sec.~\ref{sec:halo-eigenfunctions}.
As shown in the figures in the previous sections,
in the limit $\epsilon \ll 1$ the central soliton follows the density profile (\ref{eq:rho-soliton}) with a flat potential
$\Phi=E_{\rm sol}$ over its extent, determined by the hydrostatic equilibrium (\ref{eq:hydrostatic-alpha}).
This is the ground state $j=0$ of the system. Higher-energy states correspond in the classical limit
to particles that orbit up to a radius $r_j^{\max} > R_{\rm sol}$, with a higher energy
$E_j = \frac{v^2}{2} + \Phi \geq \Phi(r_j^{\max}) > \Phi(R_{\rm sol})$.
The soliton contains a macroscopic mass, that can make up a significant fraction of the system,
whereas the higher-energy states that build the halo form a quasi continuum, with a mass of the order of
$\epsilon^3 \ll 1$ as in Eq.(\ref{eq:a(E)-f(E)}) and energy levels separation $\Delta E \sim \epsilon$
as in (\ref{eq:quanta-k-n}).

Therefore, we look for the evolution of $M_{\rm sol} = M_0$ and we separate the contributions of the
soliton from those of the halo quasi-continuum in the sums in the right-hand side in Eq.(\ref{eq:dot-I_2-a}).
We also consider times much longer than the orbital periods, using
\be
\lim_{t\to\infty} \frac{\sin(t x)}{x} = \pi \, \delta_D(x) .
\ee
This gives
\bea
&& \dot M_0 = \frac{2 \pi}{\epsilon} \sum_{12} M_0^2 M_1 M_2
\biggl \lbrace \delta_D(\omega_{00}^{12}) 4 V_{01;02}^2  \left( \frac{1}{M_0} - \frac{1}{M_1} \right)
\nonumber \\
&& + \delta_D(\omega_{0}^{1}) \frac{V_{02;21} V_{00,01}}{M_0} \biggl \rbrace
+ \frac{2 \pi}{\epsilon} \sum_{123} M_0 M_1 M_2 M_3
\biggl \lbrace \delta_D(\omega_{01}^{23})  \nonumber \\
&& \times \frac{1}{2} ( V_{02;13} + V_{03;12} )^2 \left( \frac{1}{M_0}
+ \frac{1}{M_1} - \frac{1}{M_2} - \frac{1}{M_3} \right) \nonumber \\
&& + \delta_D(\omega_{0}^{1}) V_{02;21} V_{03;31} \left( \frac{1}{M_0} - \frac{1}{M_1} \right)  \biggl \rbrace  ,
\label{eq:dot-I_0-a}
\eea
where the sums only run over the halo excited states $j \neq 0$ (and at least one is transformed into an
integral in the continuum limit).
Here we dropped the overbars for simplicity and we replaced $\hat V$ by $V$ as we discarded
the constraints (\ref{eq:hat-V-def}) in the sums over the halo excited states, as each of them only
contains a mass of the order of $\epsilon^3$.

\subsection{Renormalized frequencies $\omega_j$}
\label{sec:frequencies-omega-j}

We also separate the soliton from the quasi-continuum of halo excited states
in the expression (\ref{eq:omega_1-def}) of the renormalized frequencies $\omega_j$.
Thus, we write $\omega_j = E_j + \Delta E_j$ with
\be
\Delta E_0 = \sum_1 V_{01;10} M_1 , \;\;\; \Delta E_1 = V_{10;01} M_0 + \sum_2 V_{12;21} M_2 ,
\label{eq:Delta-E-j}
\ee
where the indices $1$ and $2$ stand for halo excited states.
As in Eq.(\ref{eq:a(E)-f(E)}) for the initial halo configuration, we assume that the squared amplitudes
$M_j$ only depend on the energy $E_j$, and hence on the quantum numbers $n$ and $\ell$, and
are independent of the azimuthal number $m$,
\be
M_j = a_j^2 = (2 \pi \epsilon)^3 f(E_{n,\ell}) .
\label{eq:I_j-f(E)}
\ee
In particular, as in Eq.(\ref{eq:rho-average-nl}), we obtain from (\ref{eq:psi-Mj-thetaj}) the averaged
halo density as
\be
\langle \rho_{\rm halo} \rangle = \sum_j M_j \hat\psi_j^2 = \sum_{n\ell} \frac{2\ell+1}{4\pi} \, M_{n\ell}
\, {\cal R}_{n\ell}^2 ,
\label{eq:rho-average-nl-I}
\ee
where we used again the assumption that the initial phases $\bar\theta_j$ are uncorrelated.

The vertices $V_{13;24}$ defined in Eq.(\ref{eq:V-1324-def}) can be decomposed over their
self-interaction and gravitational parts
\be
V_{13;24} = V^\lambda_{13;24} + V^N_{13;24} ,
\ee
with
\be
V^\lambda_{13;24} = \lambda \int d\vec x \, \hat\psi_1 \hat\psi_3 \hat\psi_2 \hat\psi_4 ,
\label{eq:V-lambda}
\ee
and
\be
V^N_{13;24} = - \int \frac{d\vec x \, d\vec x \,'}{| \vec x - \vec x \,' |} \, \hat\psi_1(\vec x) \hat\psi_3(\vec x)
\, \hat\psi_2(\vec x \,') \hat\psi_4(\vec x \,') .
\label{eq:V-N}
\ee
Then, we obtain for the self-interaction contribution to the frequency shifts
\bea
&& \Delta E_0^\lambda = \lambda \int dr \, r^2 {\cal R}_0^2 \langle \rho_{\rm halo} \rangle , \nonumber \\
&& \Delta E_1^\lambda = V^\lambda_{1001} M_0 + \lambda \int dr \, r^2 {\cal R}_1^2
\langle \rho_{\rm halo} \rangle .
\label{eq:Delta-Ej-lambda}
\eea
This gives the order of magnitude estimates
$\Delta E_0^\lambda \sim \lambda \langle \rho_{\rm halo} \rangle_{R_{\rm sol}}
= \Phi_{I {\rm halo}}(R_{\rm sol})$ and 
$\Delta E_1^\lambda \sim \lambda M_{\rm sol} {\cal R}_1^2(R_{\rm sol}) + \Phi_{I {\rm halo}}(R_1)$,
where $R_1$ is the radial extent of the eigenmode ${\cal R}_1$.
By definition, we consider systems where the self-interaction is negligible in the halo, which is governed
by gravity and the velocity dispersion.
We also have $\lambda \ll 1$ and $M_{\rm sol} \ll 1$.
Therefore, the shifts $\Delta E_j^\lambda \ll 1$ are negligible as compared with the energies $E_j \sim 1$,
except for low-energy modes that are confined within the soliton radius. 

The gravitational contribution reads as
\bea
&& \hspace{-0.2cm} \Delta E_0^N = - \sum_1 M_1 \int \frac{d\vec x \, d\vec x \,'}{| \vec x - \vec x \,' |} \,
\hat\psi_0 \hat\psi_1 \hat\psi_1' \hat\psi_0'   \nonumber \\
&& \hspace{-0.2cm} \Delta E_1^N = V^N_{10;01} M_0 - \sum_2 M_2 \int
\frac{d\vec x \, d\vec x \,'}{| \vec x - \vec x \,' |} \, \hat\psi_1 \hat\psi_2 \hat\psi_2' \hat\psi_1' . \hspace{0.9cm}
\eea
A crude estimate, where we would replace the mixed product $\hat\psi_1 \hat\psi_2 \hat\psi_2' \hat\psi_1' $
by $\hat\psi_1^2 \hat\psi_2'^2$, would give $\Delta E_j^N \sim \Phi_{N {\rm halo}}(R_j)$.
This is much smaller than $E_0$ for the ground state $j=0$, while for halo excited states
this would give $\Delta E_j^N \sim E_j$. However, this is a significant overestimate because
the mixed product $\hat\psi_1 \hat\psi_2 \hat\psi_2' \hat\psi_1' $ means that we have significant interferences
between the two eigenmodes in the integrals over both $\vec x$ and $\vec x \,'$.
Then, for halo excited states we also have $\Delta E_j^N \ll E_j$.

Thus, we find that the frequency shifts are small, $\omega_j \simeq E_j$, except for the low
energy modes that are confined within the soliton radius where $\Delta E_j^\lambda \geq 0$ can be
significant. The soliton frequency shift is smaller than that of these low-energy halo states, 
because it does not contain the term $V^\lambda_{1001} M_0$ in Eq.(\ref{eq:Delta-Ej-lambda}).
Therefore, as we checked numerically, the soliton ground state keeps the lowest frequency,
\be
\omega_j > \omega_0 \;\; \mbox{for} \;\; j \neq 0 .
\label{eq:omega_0-omega_j}
\ee
Some of the renormalized frequencies $\omega_j$ are shown in Figs.~\ref{fig:omega-sinc}
and \ref{fig:omega-Hern} below.

\subsection{Evolution of the soliton mass}

We are interested in the evolution with time of the mass of the soliton, given by
Eq.(\ref{eq:dot-I_0-a})
Because the halo excited states have $\omega_j > \omega_0$ from (\ref{eq:omega_0-omega_j}), 
the Dirac factors $\delta_D(\omega_{00}^{12})$
and $\delta_D(\omega_{0}^{1})$ vanish and Eq.(\ref{eq:dot-I_0-a}) simplifies as
\bea
\dot M_0 & = & \frac{\pi}{\epsilon} \sum_{123} M_0 M_1 M_2 M_3 \, \delta_D(\omega_{01}^{23}) \,
( V_{02;13} + V_{03;12} )^2 \nonumber \\
&& \times \left( \frac{1}{M_0} + \frac{1}{M_1} - \frac{1}{M_2} - \frac{1}{M_3} \right) .
\label{eq:dot-I_0-b}
\eea
This is actually similar to the usual kinetic equation of four-wave systems
\citep{Nazarenko_2011,Onorato_2020}, but as seen above for excited
states the kinetic equation would take the more complicated form (\ref{eq:dot-I_0-a}).

The kinetic equation (\ref{eq:dot-I_0-b}) shows at once that if we start without a central soliton,
it will be generated by the nonlinear dynamics, as we have
\be
\dot M_0 = \frac{2 \pi}{\epsilon} \! \sum_{123} \! M_1 M_2 M_3 \, \delta_D(\omega_{01}^{23}) \,
( V_{02;13} + V_{03;12} )^2 > 0
\label{eq:dot-I_0-I_0=0}
\ee
for $M_0=0$. However, this expression is not so useful as for small $M_0$ it is not
possible to distinguish the soliton from the random fluctuations in the central
region. In fact, the constraint (\ref{eq:TF-1}) shows that low-mass --i.e. low-density-- solitons 
supported by the self-interactions do not exist. Low-mass density peaks are first supported
by the quantum pressure and they need to reach a finite density threshold to make the transition
to solitons supported by the self-interaction pressure. This was discussed above in 
Sec.~\ref{sec:flat-R01-rho0} for the simulation shown in Fig.~\ref{fig:rho-plane-sinc-R0p1-rho0}.

Using the fact that the occupation numbers $M_j$ and the renormalized frequencies $\omega_j$ do not
depend on the azimuthal numbers $m_j$, we can perform the sums over $\{ m_1, m_2, m_3 \}$
in Eq.(\ref{eq:dot-I_0-b}). Using the expressions
(\ref{eq:V-lambda})-(\ref {eq:V-N}) of the vertices $V$ we obtain

\bea
&& \hspace{-0.2cm} \dot M_0 = \frac{1}{2\epsilon} \widehat{\sum}_{123} M_0 M_1 M_2 M_3 \,
\delta_D(\omega_{01}^{23}) \, \begin{pmatrix} \ell_1 & \ell_2 & \ell_3 \\ 0 & 0 & 0 \end{pmatrix}^2
\nonumber \\
&& \hspace{-0.2cm} \times (2\ell_1+1) (2\ell_2+1) (2\ell_3+1)
\left( \frac{1}{M_0} \!+\! \frac{1}{M_1} \!-\! \frac{1}{M_2} \!-\! \frac{1}{M_3} \right) \nonumber \\
&& \hspace{-0.2cm}  \times \left[ \frac{\lambda}{2\pi} \int dr \, r^2 R_0 R_1 R_2 R_3 - \int dx \, x^2
\frac{R_0 R_2}{2\ell_2\!+\!1} \int dx' \, x'^2 \right. \nonumber \\
&& \hspace{-0.2cm} \left. \times R_1' R_3'  \frac{x_<^{\ell_2}}{x_>^{\ell_2+1}} - \int dx \, x^2
\frac{R_0 R_3}{2\ell_3\!+\!1} \int dx' \, x'^2 R_1' R_2'  \frac{x_<^{\ell_3}}{x_>^{\ell_3+1}} \right]^2 \nonumber \\
&&
\label{eq:dot-M_0-c}
\eea
where $\widehat{\sum}$ denotes that we only sum over the quantum numbers $n_j$ and
$\ell_j$, $x_< = \min(x,x')$, $x_> = \max(x,x')$, $R_j'$ denotes $R_j(x')$,
and we used the expansion
\be
\frac{1}{|\vec x - \vec x\,'|} = \sum_{\ell,m} \frac{4\pi}{2\ell+1} \frac{x_<^\ell}
{x_>^{\ell+1}} Y_\ell^m(\vec x)^* Y_\ell^m(\vec x \,') .
\ee

\subsection{Halo with a flat density profile}
\label{sec:kinetic-flat}

\begin{figure}[ht]
\centering
\includegraphics[height=5.9cm,width=0.4\textwidth]{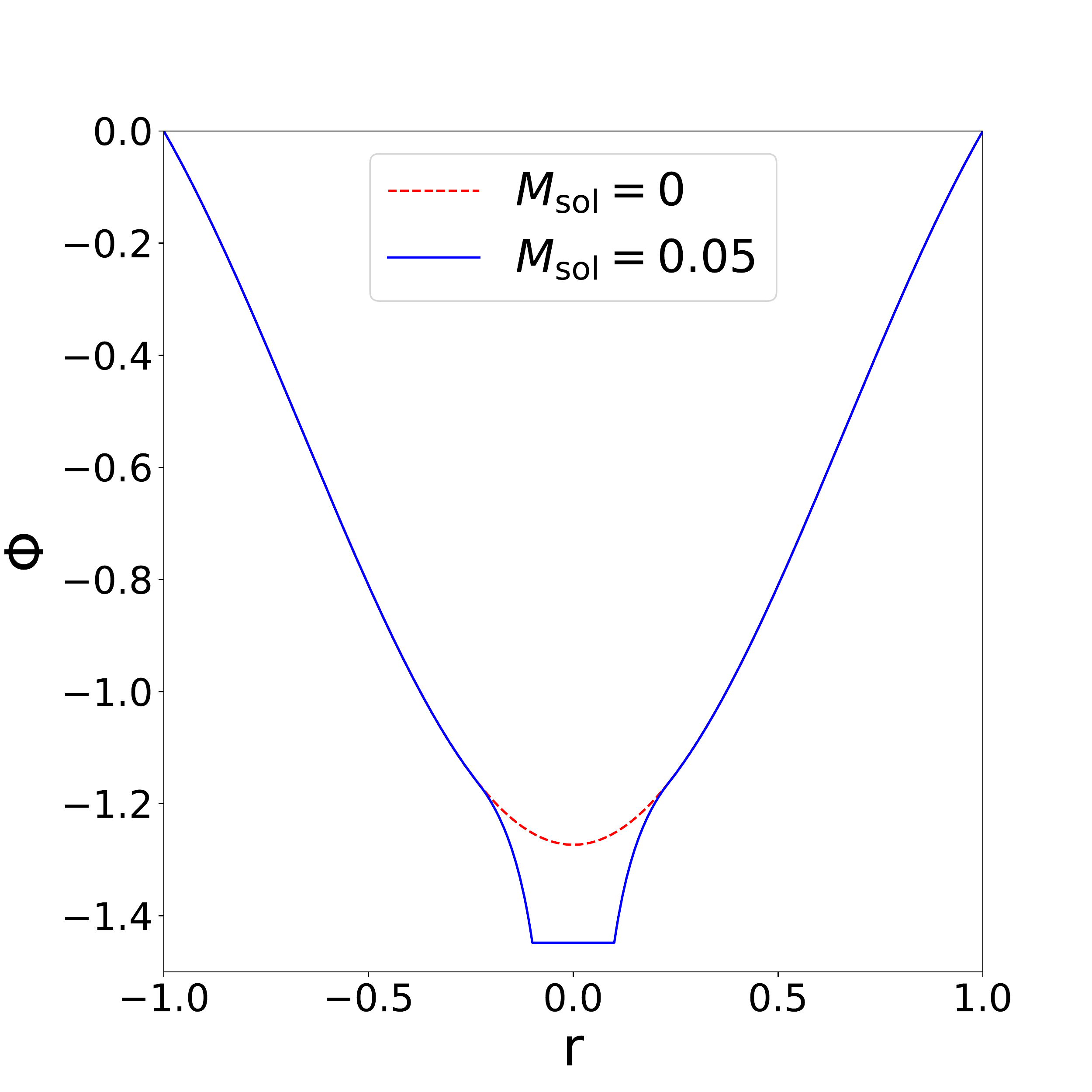}\\
\includegraphics[height=5.9cm,width=0.4\textwidth]{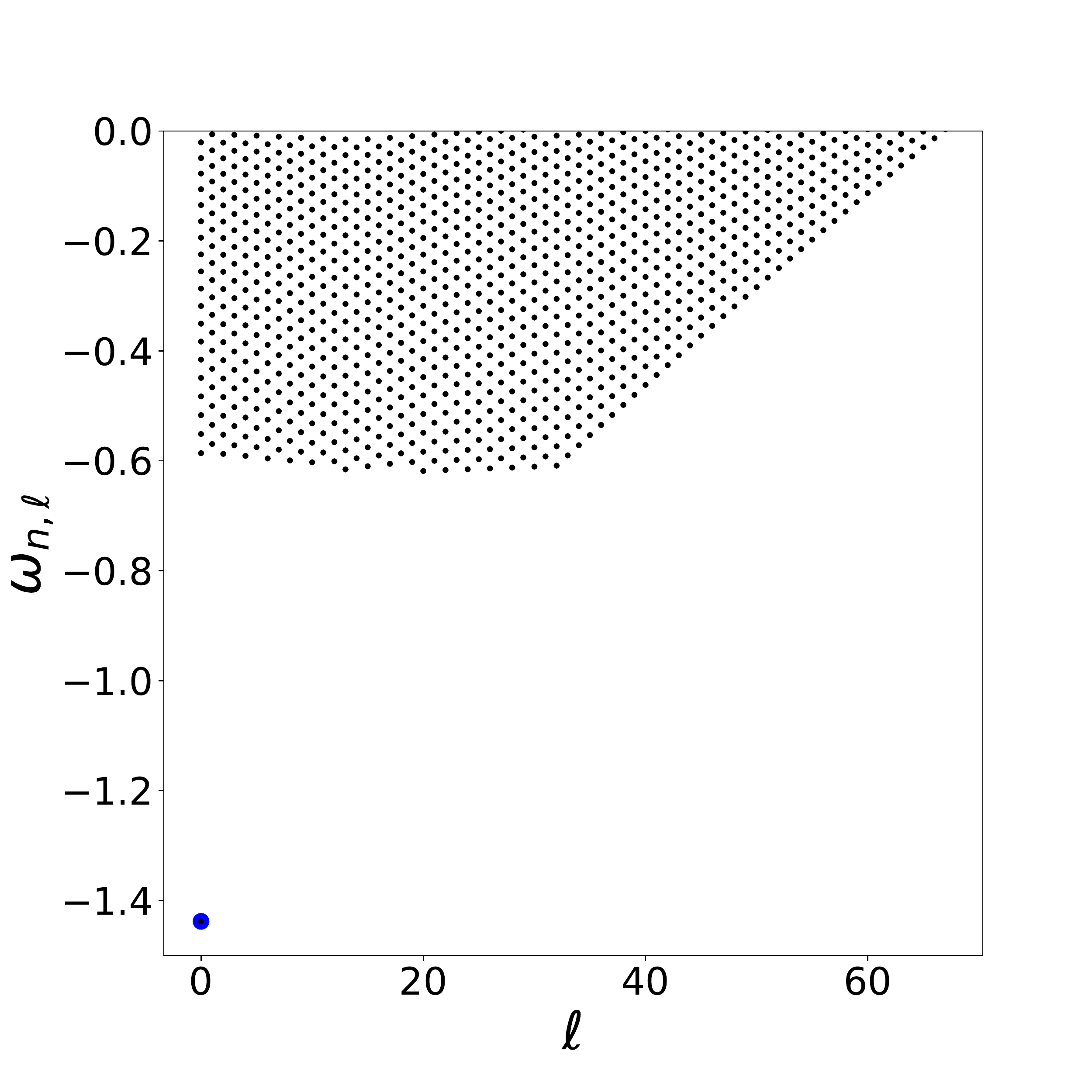}
\caption{{\it Upper panel:} potential $\bar\Phi$ without a soliton (red dashed line)
and with a soliton of mass $M_{\rm sol}=0.05$ (blue dotted line).
{\it Lower panel:} renormalized frequencies $\omega_j$ such that $E_j > E_{\rm coll}$.
The soliton ground state frequency $\omega_0$ is shown by the lower left blue point.
}
\label{fig:omega-sinc}
\end{figure}

We consider in this section the growth of the central soliton inside the flat halo studied
in Sec.~\ref{sec:flat-core}

\subsubsection{Modified potential and approximate energy cutoff}
\label{sec:energy-cutoff}

As the central soliton grows, it modifies the shape of the potential $\Phi$.
Indeed, as seen in the previous sections, inside the soliton $\Phi$ is roughly
constant, in agreement with the hydrostatic equilibrium
(\ref{eq:hydrostatic-alpha}). At radii slightly beyond the soliton radius
$R_{\rm sol}$, $\Phi$ is dominated by the gravitational potential
$\Phi_{N{\rm sol}}=-M_{\rm sol}/r$ of the soliton. Finally, at large radii
$\Phi$ is dominated by the gravitational potential $\Phi_{N{\rm halo}}$
of the halo.
In principle, we should follow simultaneously the evolution with time of the
potential $\bar\Phi$, the halo occupation numbers $M_j$, and the soliton mass
$M_0$. However, in this article we investigate a simplified approach where
we only use the kinetic equation (\ref{eq:dot-M_0-c}) to follow the soliton
growth rate and use instead approximate explicit models for the potential $\bar\Phi$
and the halo occupation numbers $M_j$.

We approximate the potential $\bar\Phi(r)$ by
\bea
&& \hspace{-0.2cm} r \!>\! R_{\rm coll} : \bar\Phi(r) = \Phi_{N{\rm halo}}(r) ,
\nonumber \\
&& \hspace{-0.2cm} R_{\rm sol} \!<\! r \!<\! R_{\rm coll} : \bar\Phi(r)
= - \frac{M_{\rm sol}}{r} + \frac{M_{\rm sol}}{R_{\rm coll}}
+ \Phi_{N{\rm halo}}(R_{\rm coll}) , \nonumber \\
&& \hspace{-0.2cm} r \!<\! R_{\rm sol} : \;\;\; \bar\Phi(r) =
- \frac{M_{\rm sol}}{R_{\rm sol}} + \frac{M_{\rm sol}}{R_{\rm coll}}
+ \Phi_{N{\rm halo}}(R_{\rm coll}) ,
\label{eq:Phi-approx-sinc}
\eea
where $\Phi_{N{\rm halo}}$ is the initial gravitational potential
(\ref{eq:rho-halo-sinc}) of the halo and $R_{\rm coll}$ is the radius where the
initial enclosed mass is equal to the soliton mass,
$M_{\rm halo}(<R_{\rm coll}) = M_{\rm sol}$.
This simple approximation provides a reasonably good description of the potential
$\Phi$ displayed in Figs.~\ref{fig:rho-plane-sinc-R0p1-rho0} and
\ref{fig:rho-plane-sinc-R0p1-rho5}, except in the outer parts as it does not capture
the diffusion of the halo somewhat beyond its initial radius.
It describes the three regimes where the total potential is i) dominated by the halo gravity
at large distance, ii) dominated by the soliton gravity closer to the soliton, iii)
constant in the soliton thanks to the balance between gravity and self-interactions.
This potential $\bar\Phi$ defines in turns the eigenmodes $\psi_j$.

Instead of using the kinetic equations (\ref{eq:dot-I_2-a}) to follow the
occupation numbers of the halo excited states, we assume an adiabatic evolution
with $M_j = M_j(t=0)$, where $M_j(0)= (2\pi\epsilon)^3 f[E_j(0)]$ are the initial
halo occupation numbers as in Eq.(\ref{eq:a(E)-f(E)}).
Then, to take into account the transfer of mass from the halo to the central
soliton, we assume that the soliton mostly builds from the lowest energy modes.
Therefore, we take $M_j = 0$ for all modes with $E_j < E_{\rm coll}$,
where the threshold $E_{\rm coll}$ is such that the mass associated with all these
modes is equal to the increase of the soliton mass,
\be
\sum_j^{E_j < E_{\rm coll}} (2\ell+1) M_{n\ell}(0) = M_{\rm sol} - M_{\rm sol}(0) ,
\label{eq:Ecoll-def}
\ee
where $M_{\rm sol}(0)$ is the initial soliton mass at time $t=0$.

We show in Fig.~\ref{fig:omega-sinc} the potential $\bar\Phi$ given by the
approximation (\ref{eq:Phi-approx-sinc}) and the renormalized frequencies $\omega_j$
for the case of a small soliton $M_{\rm sol}=0.05$.
The soliton creates a flat potential over $R_{\rm sol}$, which is deeper than the
initial halo potential because of the central overdensity.
The new energy levels $E_j$ are close to the initial energy levels $E_j(0)$
of the unperturbed halo for $E>E_{\rm coll}$ but are significantly lowered for $E<E_{\rm coll}$,
because of the increased depth of $\bar\Phi$ in the central region.
This is why the ground state (soliton) level $E_0 = \bar\Phi(0) \simeq -1.4$
is below the initial energy level $E_0 \simeq - 1.2$ shown in Fig.~\ref{fig:E-levels-sinc}.
In agreement with the analysis in Sec.~\ref{sec:frequencies-omega-j},
the shifts $\Delta E_j$ that give the renormalized frequencies $\omega_j$ are small, except
for the low-energy states that are confined within the soliton radius.
However, these states do not appear in the lower panel in Fig.~\ref{fig:omega-sinc},
because they are removed by the energy cutoff (\ref{eq:Ecoll-def}).
Nevertheless, the small but nonzero shifts $\Delta E_j$ for higher energy levels
explain why the constant-energy cutoff $E_{\rm coll}$ gives a cutoff for $\omega_{n\ell}$ that is not
completely constant with $\ell$, as seen in Fig.~\ref{fig:omega-sinc}.

We can see in the figure that for the small
mass $M_{\rm sol}=0.05$ there is already a large gap between $\omega_0$
and the remaining halo frequencies $\omega_j$.
In fact, we have $|\omega_j| < |\omega_0|/2$ for all halo modes with
$E_j > E_{\rm coll}$. This means that the Dirac factor $\delta_D(\omega_{01}^{23})$
in Eq.(\ref{eq:dot-M_0-c}) is always zero.
Therefore, the soliton growth rate $\Gamma_{\rm sol}$, defined by
\be
\Gamma_{\rm sol} = \frac{\dot M_0}{M_0} ,
\label{eq:Gamma_sol-def}
\ee
vanishes within the approximation (\ref{eq:Ecoll-def}).
This means that this approximation is not sufficient to predict the soliton growth
rate in this configuration. We need to follow more precisely the evolution with
time of the low-energy occupation numbers $M_j(t)$ as the growth rate
$\Gamma_{\rm sol}$ is very sensitive to the distribution at low energies,
for halo modes that have a significant overlap with the soliton central region
so that the kernal $V_{02;13}$ are not negligible.

\subsubsection{Growth of the soliton mass}

\begin{figure}[ht]
\centering
\includegraphics[height=6.cm,width=0.4\textwidth]{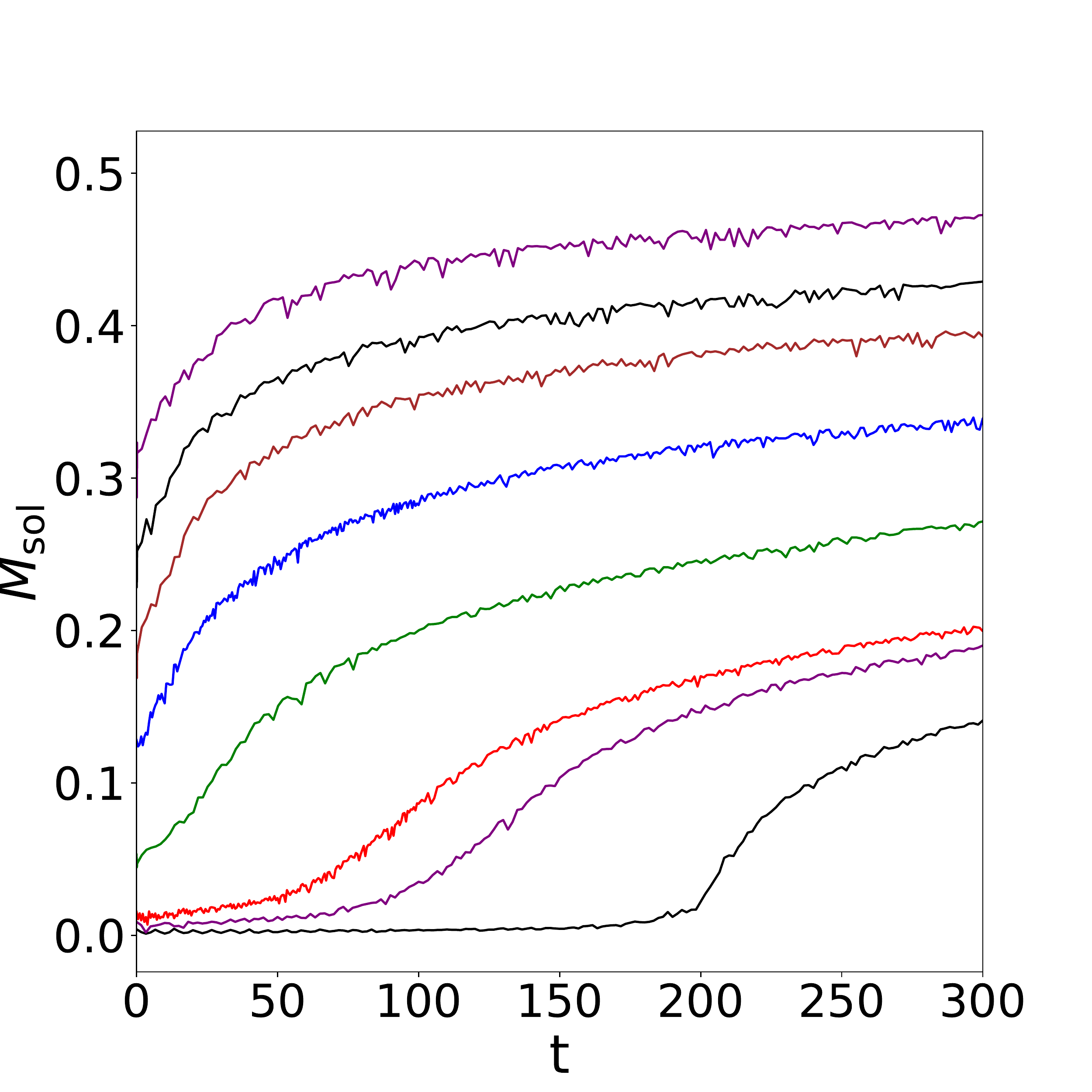}
\includegraphics[height=6.cm,width=0.4\textwidth]{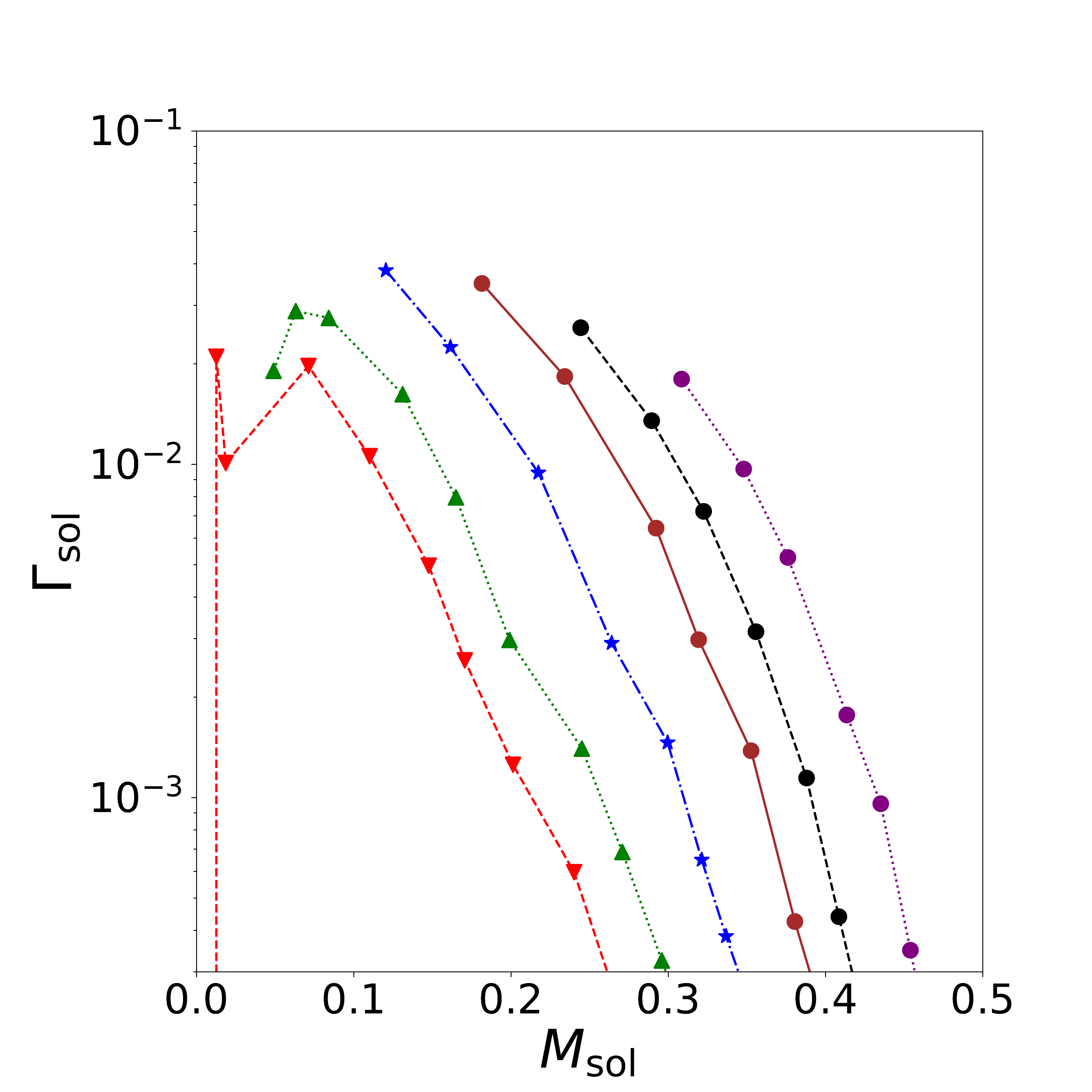}
\caption{
{\it Upper panel:} growth with time of the soliton mass $M_{\rm sol}(t)$,
for a set of simulations with different initial conditions.
{\it Lower panel:} growth rate $\Gamma_{\rm sol}$ from these simulations
shown as a function of $M_{\rm sol}$.
}
\label{fig:M-t-sinc-R0p1}
\end{figure}

We show in Fig.~\ref{fig:M-t-sinc-R0p1} the growth with time of the
soliton mass $M_{\rm sol}(t)$, for numerical simulations of the
Schr\"odinger equation (\ref{eq:Schrod-eps}) with different initial soliton
density $\rho_{0{\rm sol}}$.
Again we measure $M_{\rm sol}(t)$ by counting the total mass within the radius
$R_{\rm sol}$ from the highest density peak in the system.
This provides a good proxy for $M_{\rm sol}$ as soon as a well-characterized soliton
forms at the center of the halo.

We can see in the upper panel that, when we start with an initial soliton
$M_{\rm sol}(0) \gtrsim 0.05$, $M_{\rm sol}(t)$ typically shows an early fast
growth over a few dynamical times and next grows at a much
slower rate. The initial stage presumably corresponds to a violent relaxation,
where the low-energy levels of the halo are depleted as they mostly merge with the
soliton, while the late stage corresponds to a slow accretion limited by the
low occupation numbers of these low halo energy levels.
The three bottom curves, where there is no initial soliton or a very small overdensity at the center,
show the behavior found in Fig.~\ref{fig:rho-plane-sinc-R0p1-rho0}.
Until a long time, $t \sim 200$ in the case without initial central overdensity, there is no
soliton supported by the self-interaction but only narrow stochastic peaks.
However, they slowly grow and when one peak reaches the density threshold
(\ref{eq:Thomas-Fermi-2}) a broad soliton supported by the self-interaction appears
and next follows a similar evolution to that displayed by the other cases.

We show in the lower panel the growth rate $\Gamma_{\rm sol}(t)$ as a function
of the soliton mass. To compute $\Gamma_{\rm sol}$ we first fit the simulation curve
$M_{\rm sol}(t)$ with splines and next we compute the time
derivative (\ref{eq:Gamma_sol-def}) from this smooth curve.
We plot the result as a function of $M_{\rm sol}(t)$, to see whether the dynamics reach a scaling
regime where the growth rate only depends on the soliton mass (which also defines the halo
mass as $M_{\rm halo} = M_{\rm tot} - M_{\rm sol}$). We can see that this is not the case and
the growth rate at late times still depends on the initial conditions.
This is thus different from the scaling regime found in numerical simulations \cite{Chan:2022bkz}
for FDM (i.e. without self-interactions).
Another difference is that the solitons displayed in Fig.~\ref{fig:M-t-sinc-R0p1} 
always grow, whereas in \cite{Chan:2022bkz} small solitons evaporate.
Note that in our simulations the self-interactions indeed dominate in the central region.
However, all cases follow the same pattern. The growth rate steadily decreases with time (while
$M_{\rm sol}$ grows increasingly slowly). This falloff may be understood from the increasingly large
gap between the soliton frequency $\omega_0$ and the halo frequencies $\omega_j$
above the increasingly large cutoff $E_{\rm coll}$, shown in Fig.~\ref{fig:omega-sinc},
and the low occupation numbers of the lower energy states where resonances with the soliton
are possible. 
The leftmost red-dashed curve, which starts with a low central overdensity and mass, 
starts with a very low growth rate and oscillations, before reaching the same pattern as the other cases.
As explained above, this is because before the threshold (\ref{eq:Thomas-Fermi-2}) is reached
there is no self-interaction supported soliton and the central region is dominated for a long time
by narrow stochastic peaks, with a size set by the de Broglie wavelength.

\subsection{Halo with a cuspy density profile}
\label{sec:kinetic-cuspy}

\begin{figure}[ht]
\centering
\includegraphics[height=5.9cm,width=0.4\textwidth]{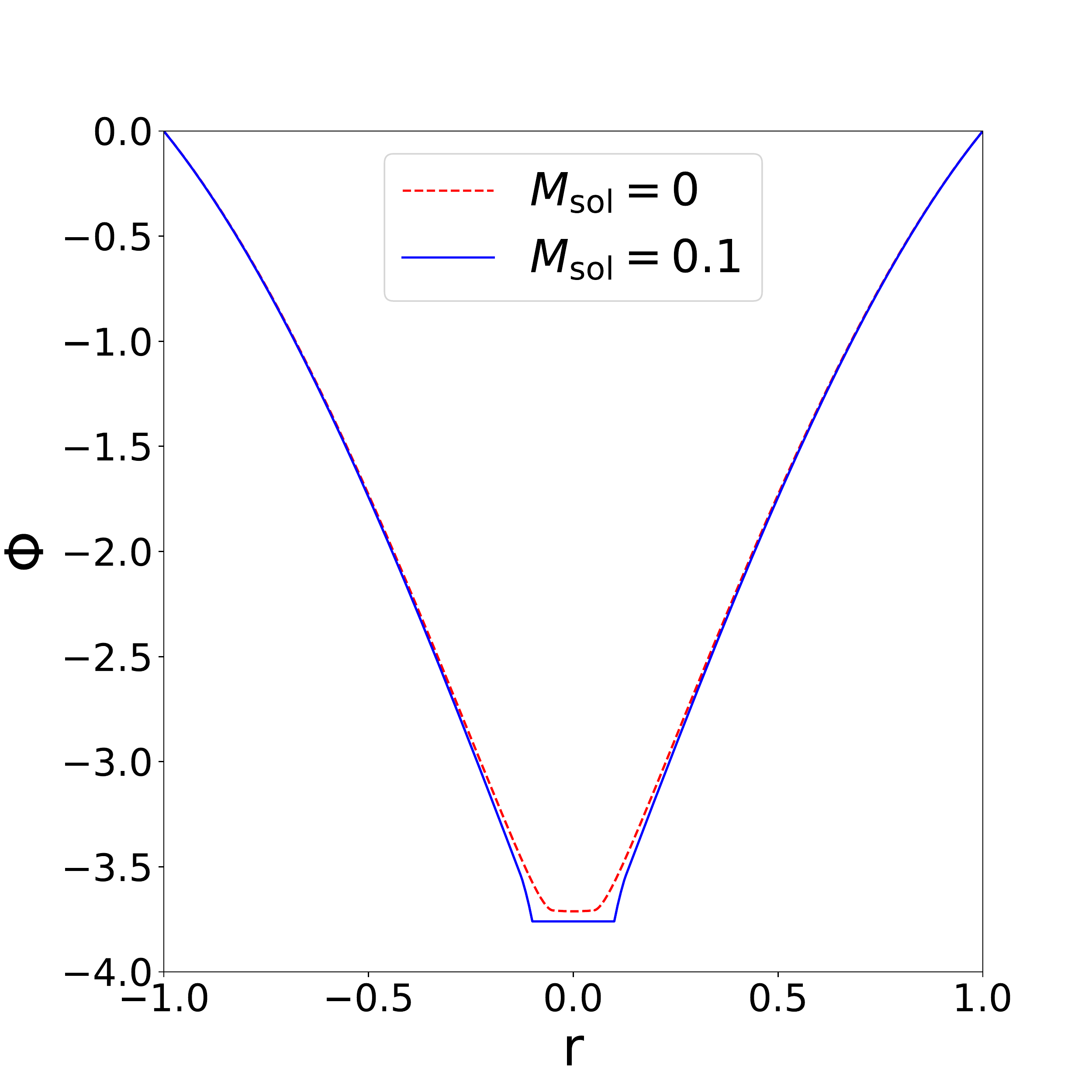}
\includegraphics[height=5.9cm,width=0.4\textwidth]{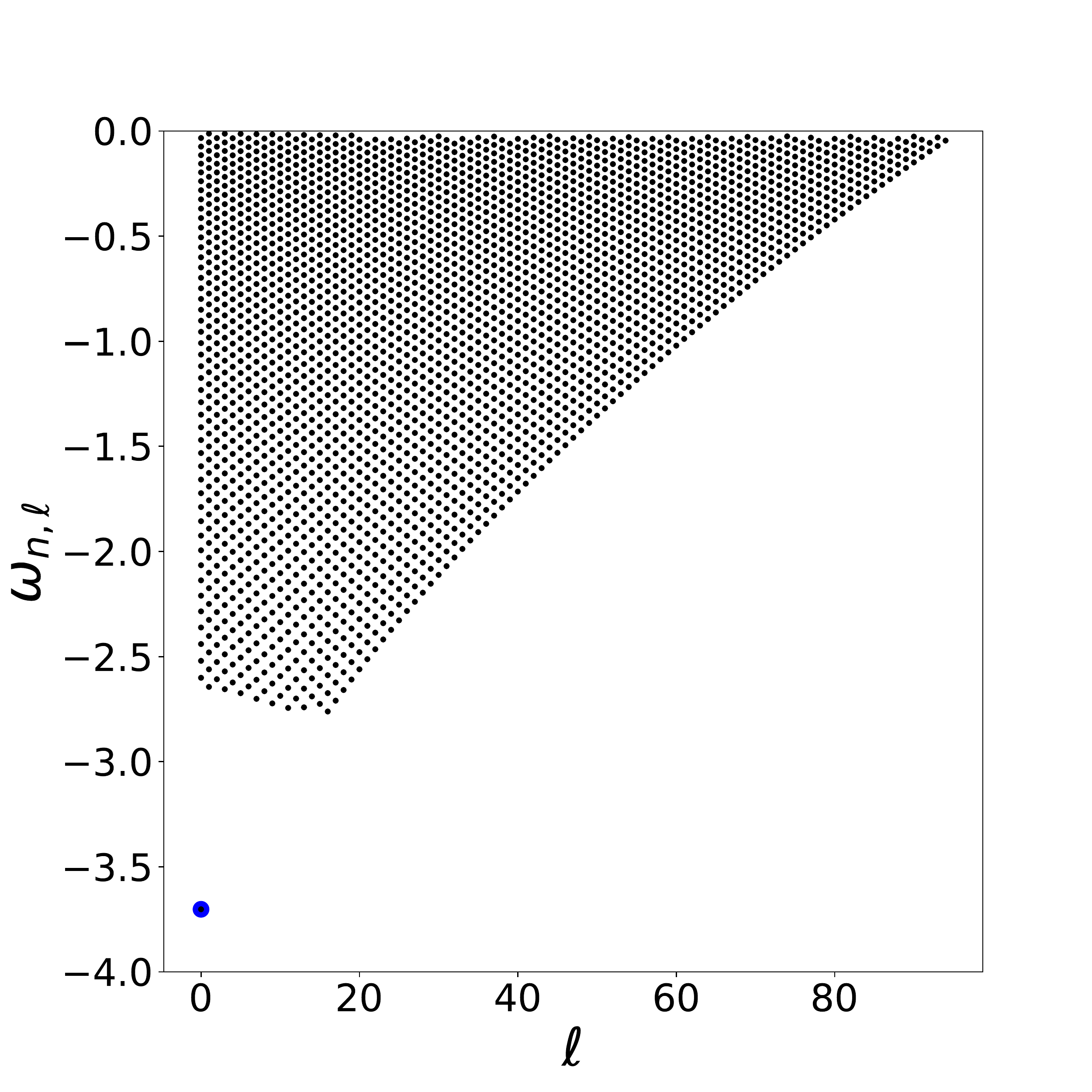}
\caption{
Potential $\bar\Phi$ and renormalized frequencies $\omega_j$ as in Fig.~\ref{fig:omega-sinc},
but for a cuspy halo and a soliton mass $M_{\rm sol}=0.1$.
}
\label{fig:omega-cos}
\end{figure}

\begin{figure}[ht]
\centering
\includegraphics[height=6.cm,width=0.4\textwidth]{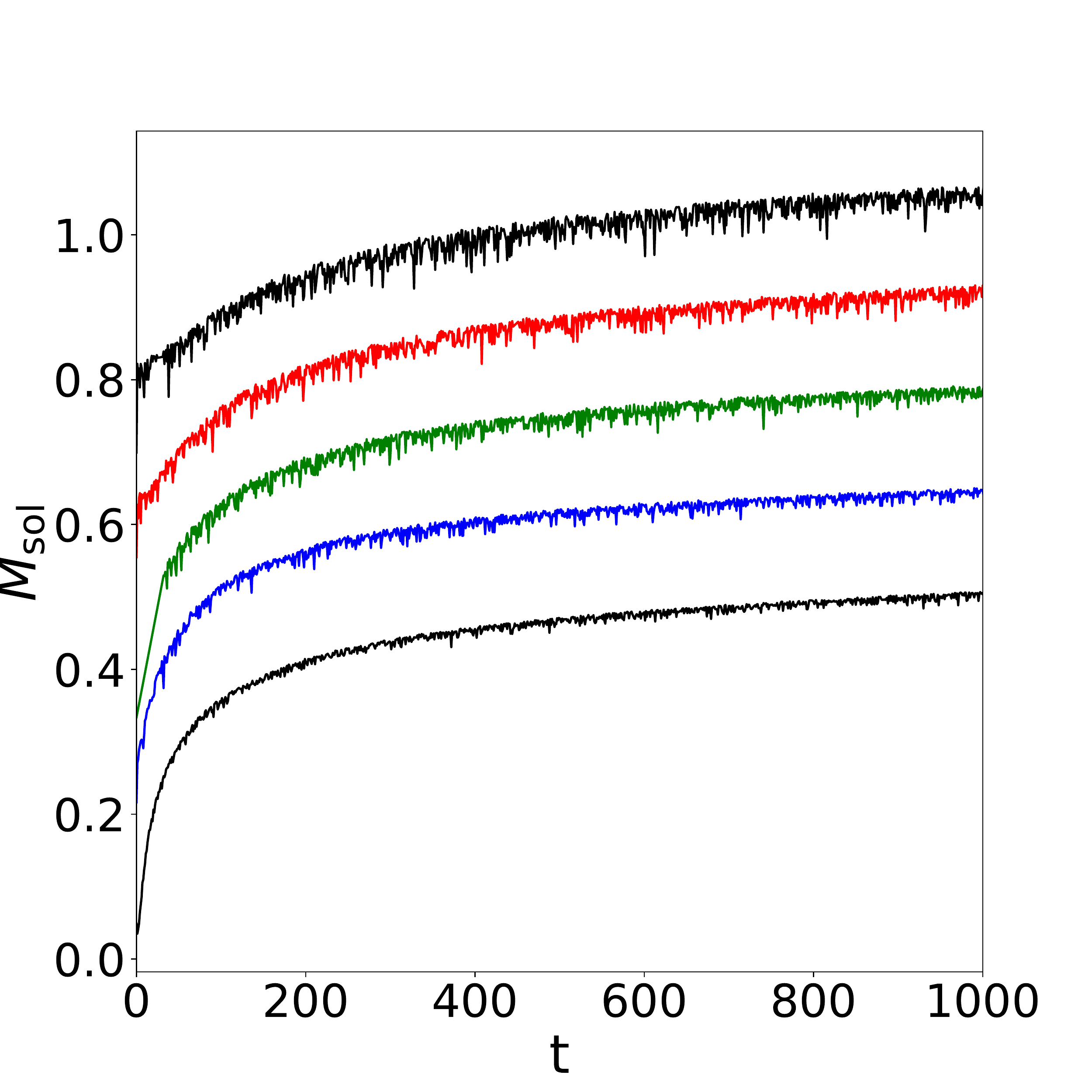}
\includegraphics[height=6.cm,width=0.4\textwidth]{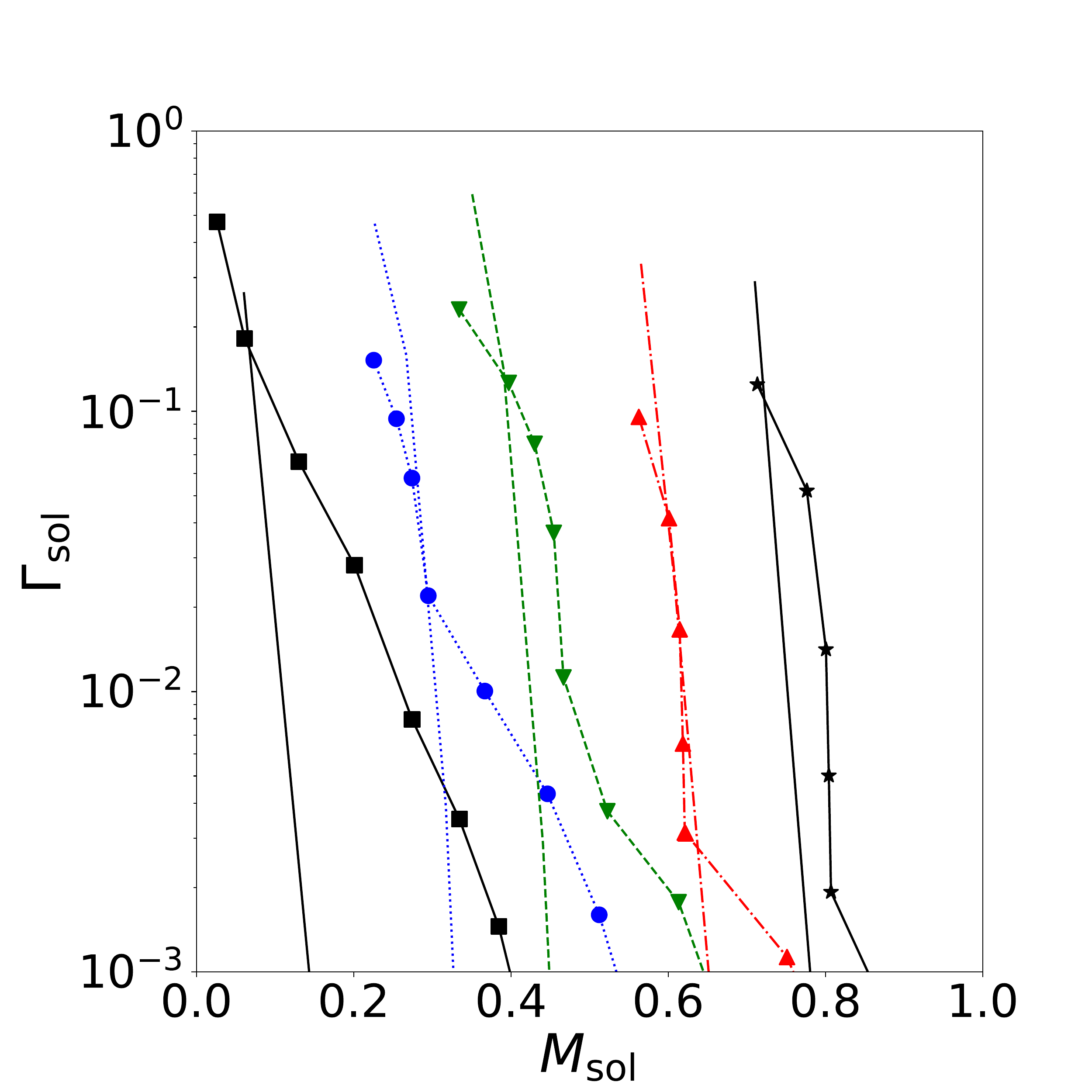}
\caption{
Growth with time of the soliton mass $M_{\rm sol}(t)$ (upper panel) and growth rate $\Gamma_{\rm sol}$
as a function of $M_{\rm sol}$ (lower panel), as in Fig.~\ref{fig:M-t-sinc-R0p1} but for a cuspy halo.
The lines with symbols are the numerical simulations while the simple lines are the theoretical predictions. 
}
\label{fig:M-t-cos}
\end{figure}

We now consider the growth of the central soliton in the case of the cuspy halo studied in
Sec.~\ref{sec:cuspy}.
We use the same approximation (\ref{eq:Phi-approx-sinc}) for the potential $\bar\Phi$
and the energy cutoff (\ref{eq:Ecoll-def}) for the removal of the low energy levels.

As seen in Fig.~\ref{fig:omega-cos}, we recover the flat potential $\bar\Phi$ over the extent of the soliton,
somewhat deeper than the initial halo potential.
The gap between the ground-state frequency $\omega_0$ and the lowest levels $\omega_j$
above the threshold $E_{\rm coll}$ is not as large as in Fig.~\ref{fig:omega-sinc},
even though the ratio $M_{\rm sol}/M_{\rm halo} \simeq 0.05$ is about the same.
This is because the cuspy initial density profile means that low-energy levels initially contain a greater
relative fraction of the total mass of the system. Therefore, a smaller fraction of them is needed to make
up the growing soliton mass.
As seen in Fig.~\ref{fig:omega-cos}, the lowest energy levels above the threshold $E_{\rm coll}$
now have $|\omega_j| > |\omega_0|/2$. Therefore, the Dirac factor $\delta_D(\omega_{01}^{23})$
in Eq.(\ref{eq:dot-M_0-c}) is no longer always zero.
The soliton mass can grow through the interaction with two low-energy levels
$-2.7 \lesssim \omega_2 , \omega_3 \lesssim -1$ and a high energy level $\omega_1 \simeq 0$.
However, for higher solition mass $E_0$ decreases while more low-energy levels are depleted,
within the approximation (\ref{eq:Ecoll-def}), so that the gap increases and eventually the
Dirac factor $\delta_D(\omega_{01}^{23})$ always vanishes.

We show in Fig.~\ref{fig:M-t-cos} the growth with time of the soliton mass.
As for the flat-core halos displayed in Fig.~\ref{fig:M-t-sinc-R0p1}
the soliton always grows, with a growth rate that decreases with time.
The total mass of the system is $M \simeq 2.5$, so that the upper curve corresponds to a central
soliton that makes about $44\%$ of the total mass.
Therefore, the numerical simulations suggest that the central soliton can slowly grow until it makes
a large fraction of the total mass of the system, of the order of $40\%$ at least.

We show in the lower panel the growth rate as a function of the soliton mass.
As for the flat-core halos displayed in Fig.~\ref{fig:M-t-sinc-R0p1}, there is no clear sign of a scaling regime, 
as the growth rate still depends on the initial conditions at late times.
The simple lines are the theoretical predictions from Eqs.(\ref{eq:dot-M_0-c}) and (\ref{eq:Ecoll-def}),
for the different initial conditions, while the lines with symbols are the results from numerical simulations.
In agreement with the lower panel in Fig.~\ref{fig:omega-cos}, at early times when the soliton has
not grown too much, the gap between the renormalized soliton and halo frequencies $\omega_0$
and $\omega_j$ is not too large and the simple energy-cutoff ansatz (\ref{eq:Ecoll-def}) allows for some 
resonances in the theoretical prediction (\ref{eq:dot-M_0-c}).
This gives a positive growth rate that shows a fast decrease with $M_{\rm sol}$ and vanishes
beyond some mass threshold as the frequency gap becomes too large to allow for resonances.
This provides a reasonably good agreement with the results from the numerical simulations,
except close to this mass threshold and beyond. There, our ansatz underestimate the growth rate,
which remains positive but steadily decreasing in the numerical simulations.
As for the flat-core halos displayed in Fig.~\ref{fig:M-t-sinc-R0p1}, this means that the halo
low-energy levels are partially refilled by the interactions between higher-energy states.
This cannot be captured by the simple ansatz (\ref{eq:Ecoll-def}) and is beyond the scope of this paper.
We leave a detailed study of this regime, where one needs to simultaneously follow the evolution
of all halo excited states, to future works.

\section{Conclusion}
\label{sec:Conclusion}

In this paper we have discussed the emergence of solitons in self-interacting scalar dark matter models. In doing so, we have first
chosen specific initial conditions for the initial halo in the form of a decomposition in eigenmodes of the Schr\"odinger equation in the
presence of the Newtonian gravity due to the halo. This allows us to solve for the eigenmodes in the WKB approximation and then construct an initial state
whose projection on this basis depends on random phases. The modulus of the coefficients of the decomposition reproduce the halo profile
whilst the random phases create strong fluctuations in the initial wavefunction. We then let the system evolve under the influence of gravity and the self-interaction and
solve the nonlinear Schr\"odinger equation.

The WKB approximation for the coefficients $a_{n\ell m}$ of the eigenmodes of the halo
provides a reasonably good approximation of the target density profile by the averaged density
$\langle\rho_{\rm halo}\rangle$.
This could be expected as we focus on the semiclassical limit $\epsilon \ll 1$.
However, the actual density profile $\rho_{\rm halo}$ always shows strong density fluctuations,
of the same order as the mean density $\langle\rho_{\rm halo}\rangle$, because of the interferences
between the different eigenmodes.
The amplitude of these fluctuations does not decrease with $\epsilon$, but their spatial width
decreases as $\Delta x \propto \epsilon$.

When halos form on a scale of the order of the length $r_a$ associated with the self-interactions
(the cases $R_{\rm sol} = 0.5$ in our units), without initial soliton, a unique central soliton supported
by the self-interactions quickly forms in a few dynamical times. It contains $30 - 50\%$ of the total mass.
It also damps the initial density fluctuations, associated with interferences, within its radius.
This holds whether we start with a flat or a cuspy halo profile.
However, in the cuspy case, we find that narrow and very high-density spikes can survive and wander
inside the large central soliton for a long time.
If there is an initial soliton, it grows in a few dynamical times to reach a quasi-stationary state where
initial fluctuations are also damped.

Next, we considered halos with a size much greater than the self-interaction scale ($R_{\rm sol}=0.1$
in our units). If the halo has a flat density profile, with density fluctuations of order unity, it takes a long
time for a central soliton supported by the self-interactions to appear, until the small-scale spikes
on sizes of the order of the de Broglie wavelength grow and  reach densities that are high enough to trigger
the self-interactions.
In contrast, if the halo has a cuspy density profile,
the high density at the center leads at once to significant self interactions.
This gives rise in a few dynamical times to a central soliton supported by the self-interactions.
Again, the fluctuations are damped within this soliton.
If there is an initial soliton, it slowly grows for many dynamical times.

We developed a kinetic theory to follow the evolution with time of the system for arbitrary profile
(i.e., going beyond plane waves in an homogeneous system).
For the quartic self-interaction $\lambda_4 \phi^4$ that we consider in this paper, which leads to an effective
quadratic pressure $P \propto \rho^2$ in the nonrelativistic limit, we obtain a kinetic equation that is similar
to the kinetic equation of four-wave systems.
To estimate the soliton growth rate, we further simplify the theory by taking a simple ansatz for
the halo excited modes, assuming that they keep their initial occupation numbers in an adiabatic fashion,
except for the low-energy levels that are depleted below a threshold $E_{\rm coll}$ to build the soliton.
This allows us to compute the soliton growth rate at once, for a given soliton and halo mass,
without following the precise evolution with time of all occupation numbers.
For a cuspy halo, this provides a reasonably good prediction for the growth rate $\Gamma_{\rm sol}$
at early times. This simple ansatz breaks down for large $M_{\rm sol}$, and for a flat halo,
because it does not follow the replenishing of the low-energy excited states and predicts an abrupt
end of the soliton growth as there are no more possible resonances.

To improve this theoretical prediction, we would need to go beyond the energy-threshold ansatz
and use the kinetic theory to follow the simultaneous evolution of all occupation numbers.
We leave such a task to future works.

All solitons that are lighter than $40 \%$ of the total mass of the system keep slowly growing until the
end of our numerical simulations, albeit at an increasingly slow rate.
Therefore, our results suggest that the soliton mass observed at a given time depends on the
past history of the system and can make up a significant fraction of the total mass of the system.

In a cosmological context, these results suggest that, in scalar-field dark matter scenarios with repulsive
self-interactions, a soliton with about half of the total mass forms when overdense regions first collapse
just above the Jeans mass. These solitons should then survive as the halos grow by accretion or mergings.
The solitons should also grow in the process by accretion or direct mergings of solitons.
The absence of clear relation between the halo and soliton masses suggests that the complex hierarchical
formation process of cosmological halos will lead to a large scatter for the mass of the soliton at fixed halo
mass, depending on the assembly history of the system.
We leave a detailed investigation of this point to future works, using cosmological simulations.

\acknowledgments

R.G.G. was supported by the CEA NUMERICS program, which has received funding from the European Union's Horizon 2020 research and innovation program under the Marie Sklodowska-Curie grant agreement No 800945.
This work was granted access to the CCRT High-Performance Computing (HPC) facility under the Grant CCRT2023-valag awarded by the Fundamental Research Division (DRF) of CEA. 

\bibliography{ref}

\end{document}